\documentclass[a4paper,11pt]{article}
\pdfoutput=1 
\usepackage{jcappub} 
\usepackage[T1]{fontenc} 

\usepackage{physics}            
\usepackage{enumitem}           
\usepackage[normalem]{ulem}     
\usepackage{xcolor}             
\usepackage{graphicx}           
\usepackage[export]{adjustbox}  
\usepackage{caption}            
\usepackage{subcaption}         
\usepackage{booktabs}           
\usepackage{colortbl}           
\usepackage{orcidlink}          

\usepackage[nameinlink,noabbrev]{cleveref}
\newcommand\pcref[1]{(\cref{#1})}
\crefname{equation}{Eq.}{Eqs.}
\crefname{section}{Section}{Sections}
\crefname{figure}{Figure}{Figures}
\crefname{table}{Table}{Tables}
\crefname{appendix}{Appendix}{Appendices}
\Crefname{figure}{Figure}{Figures}
\Crefname{equation}{Equation}{Equations}
\Crefname{section}{Section}{Sections}
\Crefname{table}{Table}{Tables}

\usepackage{lineno}

\newcommand{\fnl}[0]{f_{\rm NL}^{\rm loc}}
\newcommand{\ra}[0]{\mathrm{R.A.}}
\newcommand{\dec}[0]{\mathrm{Dec.}}

\newcommand{\ie}[0]{\textit{i.e.} }

\newcommand{\referee}[1]{#1} 

\title{\boldmath Constraining primordial non-Gaussianity with DESI 2024 LRG and QSO samples.}

\emailAdd{echaussidon@lbl.gov}

\author[1,2]{{E.~Chaussidon}\orcidlink{0000-0001-8996-4874},}
\author[2]{{C.~Yèche}\orcidlink{0000-0001-5146-8533},}
\author[2]{{A.~de~Mattia}\orcidlink{0000-0003-0920-2947},}
\author[2]{{C.~Payerne}\orcidlink{0000-0002-1818-929X},}
\author[1]{{P.~McDonald}\orcidlink{0000-0001-8346-8394},}
\author[3,4,5]{{A.~J.~Ross}\orcidlink{0000-0002-7522-9083},}
\author[6]{{S.~Ahlen}\orcidlink{0000-0001-6098-7247},}
\author[7]{{D.~Bianchi}\orcidlink{0000-0001-9712-0006},}
\author[8]{{D.~Brooks},}
\author[2]{{E.~Burtin},}
\author[1]{{T.~Claybaugh},}
\author[9]{{A.~de la Macorra}\orcidlink{0000-0002-1769-1640},}
\author[8]{{P.~Doel},}
\author[1,10]{{S.~Ferraro}\orcidlink{0000-0003-4992-7854},}
\author[8,11]{{A.~Font-Ribera}\orcidlink{0000-0002-3033-7312},}
\author[12,13]{{J.~E.~Forero-Romero}\orcidlink{0000-0002-2890-3725},}
\author[14,15,16]{{E.~Gaztañaga},}
\author[17,14,18]{{H.~Gil-Mar\'in}\orcidlink{0000-0003-0265-6217},}
\author[1]{{S.~Gontcho A Gontcho}\orcidlink{0000-0003-3142-233X},}
\author[19]{{G.~Gutierrez},}
\author[1]{{J.~Guy}\orcidlink{0000-0001-9822-6793},}
\author[3,20,5]{{K.~Honscheid}\orcidlink{0000-0002-6550-2023},}
\author[21]{{C.~Howlett}\orcidlink{0000-0002-1081-9410},}
\author[22,23]{{D.~Huterer}\orcidlink{0000-0001-6558-0112},}
\author[24]{{R.~Kehoe},}
\author[25]{{D.~Kirkby}\orcidlink{0000-0002-8828-5463},}
\author[1]{{T.~Kisner}\orcidlink{0000-0003-3510-7134},}
\author[1]{{A.~Kremin}\orcidlink{0000-0001-6356-7424},}
\author[26]{{L.~Le~Guillou}\orcidlink{0000-0001-7178-8868},}
\author[1]{{M.~E.~Levi}\orcidlink{0000-0003-1887-1018},}
\author[27,11]{{M.~Manera}\orcidlink{0000-0003-4962-8934},}
\author[28]{{A.~Meisner}\orcidlink{0000-0002-1125-7384},}
\author[29,11]{{R.~Miquel},}
\author[30]{{J.~Moustakas}\orcidlink{0000-0002-2733-4559},}
\author[31]{{J.~ A.~Newman}\orcidlink{0000-0001-8684-2222},}
\author[32,33]{{G.~Niz}\orcidlink{0000-0002-1544-8946},}
\author[2,1]{{N.~Palanque-Delabrouille}\orcidlink{0000-0003-3188-784X},}
\author[34,35,36]{{W.~J.~Percival}\orcidlink{0000-0002-0644-5727},}
\author[37]{{F.~Prada}\orcidlink{0000-0001-7145-8674},}
\author[38]{{I.~P\'erez-R\`afols}\orcidlink{0000-0001-6979-0125},}
\author[39]{{C.~Ravoux}\orcidlink{0000-0002-3500-6635},}
\author[40]{{G.~Rossi},}
\author[41]{{E.~Sanchez}\orcidlink{0000-0002-9646-8198},}
\author[1]{{D.~Schlegel},}
\author[22,23]{{M.~Schubnell},}
\author[42]{{H.~Seo}\orcidlink{0000-0002-6588-3508},}
\author[28]{{D.~Sprayberry},}
\author[23]{{G.~Tarl\'{e}}\orcidlink{0000-0003-1704-0781},}
\author[9]{{M.~Vargas-Maga\~na}\orcidlink{0000-0003-3841-1836},}
\author[28]{{B.~A.~Weaver},}
\author[43]{{C.~Zhao}\orcidlink{0000-0002-1991-7295},}
\author[44]{{H.~Zou}\orcidlink{0000-0002-6684-3997},}

\affiliation{
\noindent \hangindent=.5cm $^{1}${Lawrence Berkeley National Laboratory, 1 Cyclotron Road, Berkeley, CA 94720, USA} \\
\noindent \hangindent=.5cm $^{2}${IRFU, CEA, Universit\'{e} Paris-Saclay, F-91191 Gif-sur-Yvette, France} \\ \vspace{-2mm}

Rest of the affiliations are in Appendix \ref{sec:affiliations}.}

\abstract{We analyse the large-scale clustering of the Luminous Red Galaxy (LRG) and Quasar (QSO) sample from the first data release (DR1) of the Dark Energy Spectroscopic Instrument (DESI). In particular, we constrain the primordial non-Gaussianity (PNG) parameter $\fnl$ via the large-scale scale-dependent bias in the power spectrum using $1,631,716$ LRGs ($0.6 < z < 1.1$) and $1,189,129$ QSOs ($0.8 < z < 3.1$). This new measurement takes advantage of the enormous statistical power at large scales of DESI DR1 data, surpassing the latest data release (DR16) of the extended Baryon Oscillation Spectroscopic Survey (eBOSS). For the first time in this kind of analysis, we use a blinding procedure to mitigate the risk of confirmation bias in our results. \referee{We improve the model of the radial integral constraint proposing an innovative technique allowing the correction through the window matrix convolution.} We also carefully test the mitigation of the dependence of the target selection on the photometry qualities by incorporating an angular integral constraint contribution to the window function, and validate  our methodology with the blinded data. Finally, combining the two samples, we measure $\fnl = {-3.6}_{-9.1}^{+9.0}$ at $68\%$ confidence, where we assume the universality relation for the LRG sample and a recent merger model for the QSO sample about the response of bias to primordial non-Gaussianity. Adopting the universality relation for the PNG bias in the QSO analysis leads to $\fnl = 3.5_{-7.4}^{+10.7}$ at $68\%$ confidence. \referee{Due to restricted selection in the LRG sample, the inclusion of the LRGs allows for $10\%$ improvement.} This measurement is the most precise determination of primordial non-Gaussianity using large-scale structure to date, surpassing the latest result from eBOSS by a factor of $2.3$.}

\begin{document}
\setcounter{tocdepth}{2}
\maketitle
\flushbottom
\clearpage \newpage

\section{Introduction}
Since its introduction in the early 80s, inflation \cite{Guth1981, Albrecht1982, Linde1982a} is still the leading paradigm for describing the early Universe. Without direct observation of this epoch, one can only probe the properties of the primordial fields from later time, such as the tilt of the primordial scalar power spectrum, the primordial gravitational waves, or the primordial non-Gaussianity (PNG) to test different inflation models. The tild is well constrained with the latest Planck cosmic microwave background (CMB) data \cite{Akrami2020a}. The gravitational waves are gaining a growing interest with the future missions to observe B-mode polarization of the CMB \cite{Hazumi2020, Abazajian2022}. 

PNG remains still poorly constrained by current experiments \cite{Akrami2020, Mueller2021} relative to the precision needed to rule out inflationary scenarios of interest. At the same time, it is a powerful probe to distinguish the simplest models of inflation that predict a nearly Gaussian distribution of primordial fluctuations \ie a minimal amount of PNG, to more sophisticated ones like multi-field inflation \cite{Chen2010}. In particular, one can study the so-called \emph{local} primordial non-Gaussianity, quantified by the parameter $\fnl$,
\begin{equation}
    \Phi =  \Phi_g + f_{\mathrm{NL}}^{\mathrm{loc}} \left( \Phi_g^2 - \langle \Phi_g ^2\rangle \right),
\end{equation}
where $\Phi$ is the primordial gravitational potential field parametrised in terms of $\Phi_g$, a Gaussian potential field. A detection of local non-Gaussianity such that $\mathcal{O}\left(\fnl\right)~\sim~1$ could rule out slow-roll single-field inflation \cite{Creminelli2004}.

Currently, the best constraints on PNG are obtained from Planck data: $\fnl = -0.9 \pm 5.1$ at $68\%$ confidence \cite{Akrami2020}, but they are almost limited by the cosmic variance. A promising approach to circumvent this limit in CMB observations is to use the enormous statistical power in the 3D galaxy clustering, and in particular, through the tiny imprint left at large scales on the matter power spectrum by local PNG, known as the \emph{scale-dependent bias} \cite{Dalal2008, Slosar2008}. The best constraint using this imprint is from the latest data release of the extended Baryon Oscillation Spectroscopic Survey (eBOSS) \cite{Ahumada2020} using the quasar sample and measuring $-23 < \fnl < 21$ at $68\%$ confidence \cite{Cagliari2023}. 

Despite a significant effort to mitigate the residual dependence of the targets on the properties of the imaging survey used to select them \cite{Rezaie2019, Rezaie2021}, imaging uncertainties were the main systematic in this measurement. This effect, known as the \emph{imaging systematics}, will still be a crucial systematic for the upcoming galaxy survey that can bias the measurement of $\fnl$ \cite{Rezaie2023}. To avoid this systematic \cite{Krolewski2023, Bermejo-Climent2024} cross-correlate the galaxy field with CMB lensing, but  the statistical power is lower than the autocorrelation of the galaxy field although not biased by this systematic. \referee{Recent works try also to incorporate high-order correlation functions and to include also the small scales \cite{MoradinezhadDizgah2021, DAmico2022, Cabass2022, Brown2024}.}
 
Here, we will analyse, for the first time, the large-scale modes of the Luminous Red Galaxy and the Quasar power spectra from the first data release (DR1) \cite{Schlafly2023,DESI2024I} of the Dark Energy Spectroscopic Instrument (DESI). DESI is a robotic, fiber-fed, highly multiplexed spectroscopic surveyor that operates on the Mayall 4-meter telescope at Kitt Peak National Observatory \cite{Levi2013,DESICollaboration2016,Abareshi2022}. DESI can obtain simultaneous spectra of almost 5000 objects over a  3° field  \cite{DESICollaboration2016a, Silber2023, Miller2024}, and is currently conducting a five-year survey of about a third of the sky. The data used here correspond to the first year and half of the main survey.

This first data release of DESI is already the most extensive catalog from a spectroscopic galaxy survey for galaxy clustering measurement and provides the best constraints on baryon acoustic oscillations (BAO) \cite{DESI2024III, DESI2024IV} and on redshift space distortion (RSD) measurements \cite{DESI2024V}, leading to some of the most precise constraint today, when combine it with Planck 2018 result \cite{Planck18}, on the cosmological parameters describing the Universe \cite{DESI2024VI, DESI2024VII}. Note that Early DESI Data Release \cite{EDR}, used for the survey validation phase \cite{SV} is already publicly available.

This analysis is the natural follow-up of the latest measurements performed with the eBOSS data \cite{Castorina2019, Rezaie2021, Mueller2021, Cagliari2023} but improves it on several points. First, with the first data release of DESI, we use the most extensive data set available to date. Then, we forward model a multiplicative correction to deal with the radial integral constraint and compute the angular integral constraint associated with the imaging systematic weights. Finally, for the first time for such a scale-dependent-bias PNG measurement, we conduct a complete blinded analysis, enabling us to validate the systematic imaging mitigation carefully. The paper is organised as follows: \cref{sec:theory} describes the theoretical model used, \cref{sec:data} the data from the first DESI data release, \cref{sec:model} gives the geometrical effect from the survey and tests it with simulations. Finally, the blinded analysis is performed in \cref{sec:blinded_analysis}, the unblinded constraints is given in \cref{sec:unblinded_analysis}, and we conclude in \cref{sec:conclusion}.

\section{Theory} \label{sec:theory}
The presence of local primordial non-Gaussianity imprints scale-dependent bias on the spatial distribution of biased tracers, impacting the power spectrum of biased tracers as follows \citep{Dalal2008, Slosar2008}:
\begin{equation} \label{eqn:scale_depedent_bias}
    P(k, z) = \left(b_1(z) + \dfrac{b_{\Phi}(z)}{T_{\Phi \rightarrow \delta}(k, z)} \fnl \right)^2 \times P_{\mathrm{lin}}(k, z),
\end{equation}
where $b_1$ is the linear bias of the tracer and $b_{\Phi}$ is the PNG bias given the response to the presence of local PNG of the tracer, $P_{\mathrm{lin}}$ is the linear matter power spectrum and $T_{\Phi \rightarrow \delta}(k, z)$ is the transfer function between the primordial gravitational field $\Phi$ and the matter density perturbation. It can be computed directly from \texttt{CLASS}\footnote{We are using the user-friendly Python wrapper: \url{https://github.com/cosmodesi/cosmoprimo}.} \cite{Blas2011} by:
\begin{equation}
    T_{\Phi \rightarrow \delta}(k, z) = \sqrt{\dfrac{P_{\rm lin}(k, z)}{P_{\Phi}(k)}} \quad \text{ with }  \quad P_{\Phi}(k) = \dfrac{9}{25} \dfrac{2 \pi^2}{k^3} A_s \left(\dfrac{k}{k_{\rm pivot}}\right)^{n_s-1},
\end{equation}
where $P_{\Phi}$ is the primordial potential\footnote{$\Phi$ is normalised to $3/5 \mathcal{R}$ to match the usual definition of \cite{Slosar2008}.} power spectrum, $n_s$ is the spectral index and $A_s$ the amplitude of the initial power spectrum at $k_{\rm pivot} = 0.05\;\text{Mpc}^{-1}$. Hence, with the Poisson equation, $T_{\Phi \rightarrow \delta}(k, z)$ has the well-known scale dependency \cite{Dalal2008}:
\begin{equation}
T_{\Phi \rightarrow \delta}(k, z) \propto  k^2 \times T_{\Phi \rightarrow \Phi}(k, z),
\end{equation}
where $T_{\Phi \rightarrow \Phi}(k, z)$ is the usual transfer function, oftenly denoted $T$.
    
In addition, as in the latest eBOSS measurement~\cite{Castorina2019, Mueller2021, Cagliari2023}, we model the redshift space distortion \cite{Sargent1977} effect with a simple model including the Kaiser effect \cite{Kaiser1987} and a damping factor for small scales\footnote{The model is available: \url{https://github.com/cosmodesi/desilike/blob/hmc/nb/png_examples.ipynb}}
\begin{equation} \label{eqn:pk_theo}
    P(k, \mu)=\dfrac{\left[b_1(z_{\rm eff}) + \dfrac{b_{\phi}(z_{\rm eff})}{T_{\Phi \rightarrow \delta}(k, z_{\rm eff})} \fnl + f(z_{\rm eff}) \mu^2\right]^2}{\left[1+\frac{1}{2}\left(k \mu \Sigma_s\right)^2\right]^{2}} \times P_{\rm lin}(k, z_{\rm eff}) + s_{n,0},
\end{equation}
where the different redshift-dependent quantities are fixed or measured at the effective redshift $z_{\rm eff}$ of the tracer sample, see \cref{sec:effective_redshift} for how we estimate it. $f$ is the linear growth rate and $\Sigma_s$ the amount of damping at small scales. Although the shot noise contribution is always removed from our power spectrum measurements, we include also a potential residual shot noise $s_{n, 0}$ which should be close to 0.

Finally, the power spectrum is expanded in Legendre multipoles:
\begin{equation} \label{eqn:pk_legendre}
    P_{\ell}(k) = \dfrac{2 \ell+1}{2} \int_{-1}^1 \dd \mu P(k, \mu) \mathcal{L}_{\ell}(\mu).
\end{equation}
In the following, we use only the monopole ($\ell=0$) and the quadrupole ($\ell=2$) since the statistical errors on the hexadecapole are too big at large scales. The statistical gain on $\fnl$ when adding the quadrupole is detailed in \cref{sec:gain_with_quadrupole}.

The theoretical prescription of the PNG bias $b_{\Phi}$ is a widely discussed topic \cite{Biagetti2017, Barreira2020, Barreira2021, Barreira2022, Fondi2023, Sullivan2023, Hadzhiyska2024, Adame2024}, but it will not be discussed here. We simply follow \cite{Slosar2008}, assuming the usual relation 
\begin{equation} \label{eqn:b_phi_with_p}
    b_{\Phi}(z) = 2 \delta_c \times (b_1(z) - p)
\end{equation}
where $\delta_c = 1.686$ is the critical density for spherical collapse and $p$ quantifies the merger history of the tracer. For the QSOs, by default, we assume a recent merger model \ie following \cite{Slosar2008}, we use $p=1.6$. Note that with this choice we assume that all the quasars have a recent merger history which it is not known \cite{Breiding2024}. This choice leads to an increase of the statistical uncertainty on $\fnl$ and shift the measured value of $\fnl$ compared to using $p=1.0$ as in the universality relation. For the LRGs, we use the universality relation ($p=1.0$) \ie we assume that their halo occupation distribution (HOD) depends only on halo mass. Note that \cite{Barreira2020a} suggests that stellar mass selected samples could have a PNG bias described by $p=0.55$ which would increase the statistical power of the LRG sample. Measuring and validating the description of the PNG bias is deferred to future work and will represent a crucial upgrade for upcoming analyses.

In addition, we give also the assumption-free constraint on $b_{\Phi} \times \fnl$ in order to circumvent this discussion. However, this constraint cannot be given in the case where we combine the LRG and the QSO measurements. In practice, we fit either $(\fnl, b_1, s_{n,0}, \Sigma_s)$ assuming \cref{eqn:b_phi_with_p} or $(b_{\Phi}\fnl, b_1, s_{n,0}, \Sigma_s)$. Note that in the following, we fix the $\Lambda$CDM parameters to the values of the Plank 2018 cosmology \cite{Planck18} values, thus neglecting the uncertainty on the shape of the power spectrum.

\section{Data} \label{sec:data}
In this section, we present the two samples from the first DESI data release \cite{DESI2024I} that are used to constrain the local primordial non-Gaussianity, the power spectrum estimator, and the optimal weights that we are using, and finally, how we generate simulations as realistic as possible to mimic these two samples.

\subsection{DESI DR1 Samples}
\subsubsection{Luminous Red Galaxies (LRG) and Quasars (QSO)} \label{sec:lrg_qso_data}
\begin{figure}
     \begin{subfigure}{0.49\textwidth}  
         \centering
         \includegraphics[scale=0.68, center]{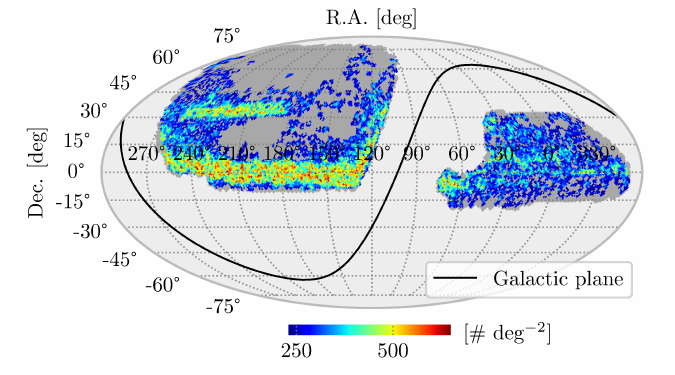}
         \caption{DESI DR1 LRG.}
     \end{subfigure}
     \begin{subfigure}{0.49\textwidth}
         \centering
         \includegraphics[scale=0.68, center]{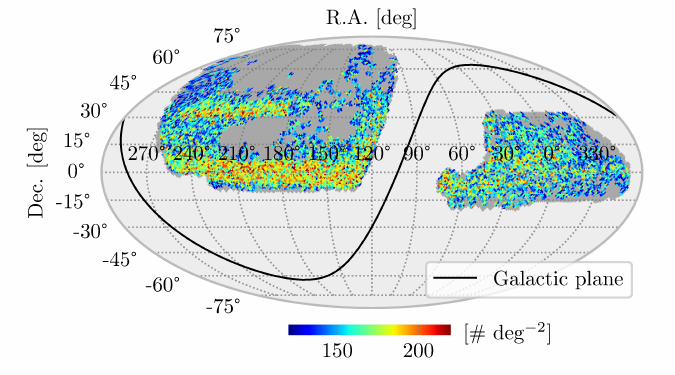}
         \caption{DESI DR1 QSO.}
     \end{subfigure} \newline
     \begin{subfigure}{0.49\textwidth}
        \centering
        \includegraphics[scale=0.68, center]{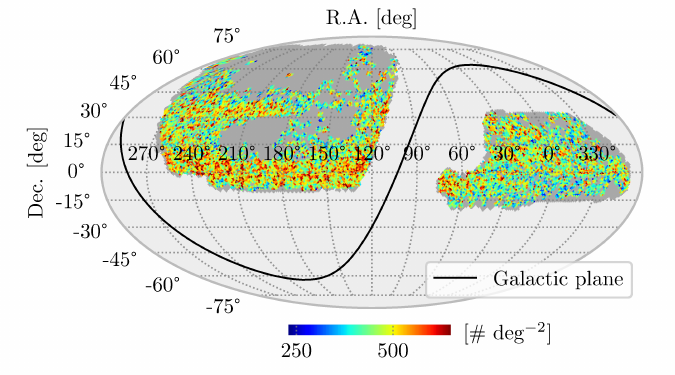}
        \caption{LRG corrected by the completeness.}
    \end{subfigure}
    \begin{subfigure}{0.49\textwidth}
        \centering
        \includegraphics[scale=0.68, center]{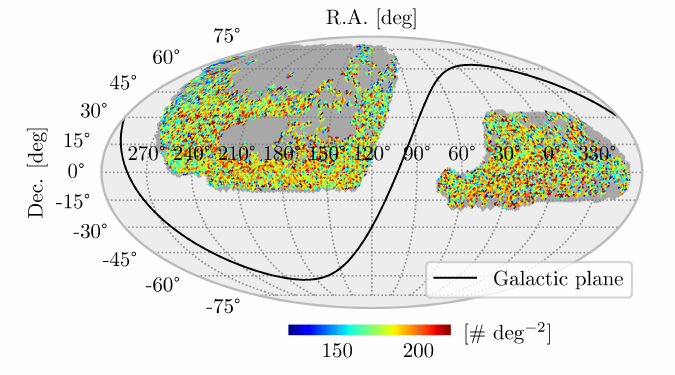}
        \caption{QSO corrected by the completeness.}
    \end{subfigure}
       \caption{Angular density distribution for the DESI DR1 LRG/QSO on the left/right corrected by the completeness of the observation (some regions of the sky have yet to be fully observed) on the bottom and not on the top. The comparison between top and bottom panel shows from which region the statistical information comes from. Although both appear almost isotropic, one must correct for the anisotropy due to the dependence of the target selection on the image quality. The dark gray region represents the expected final DESI footprint.}
     \label{fig:angular_density_LRG_QSO}
\end{figure}

In this analysis, we use the LRG \cite{Zhou2023} and QSO \cite{Chaussidon2023} samples from the first DESI data release. Compared to the baseline of the DESI clustering measurements used in \cite{DESI2024III, DESI2024VI}, we consider each tracer in its full range and include quasars with a higher redshift ($z>2.1$): $0.4 < z < 1.1$ for the LRGs and $0.8 < z < 3.1$ for the QSOs, to enhance the measurement of the large-scale modes of the power spectrum. Hence, this work analyses the clustering of 2,130,621 LRGs and 1,189,129 QSOs, improving the size of the sample by a factor 8 and 2.5 times larger compared to the latest measurement performed in eBOSS \cite{Ross2013, Mueller2021}. The angular density distributions of these two samples are displayed in \cref{fig:angular_density_LRG_QSO}. Although both appear isotropic at first order, the angular fluctuations of the number of densities due to the alteration of the target selection by the quality of the imaging survey heavily contaminate the large-scale modes of the power spectrum. This effect is known as imaging systematics and is the primary source of systematics in this analysis; see \cref{sec:blinded_analysis}.

\begin{figure}
    \centering
    \includegraphics[scale=1]{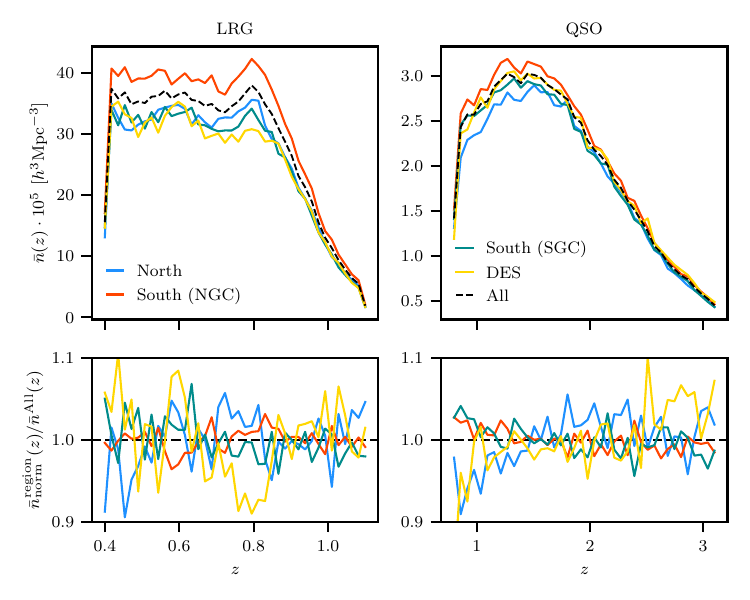}
    \caption{On the top, the redshift distributions for the LRGs (left) from 0.4 to 1.1 and QSOs (right) from 0.8 to 3.1 not corrected by the completeness of the observation, split according to the four distinct photometric regions: North in blue, South (NGC) in red, South (SGC) in green and DES in gold. The black dashed line is for the four regions combined. The difference in amplitudes highlights the difference of completeness between the different footprints. There are more objects in South (NGC) because it is the area that is the most complete, see \cref{fig:angular_density_LRG_QSO}. On the bottom, the ratio between the redshift distributions of the different photometric regions and the ones from the four regions combined. At the bottom panel, redshift distributions are normalized to have the same amplitude than the ones from the four regions combined. The redshift distributions are remarkably similar for the LRGs; some differences in the North are visible for the QSOs.}
    \label{fig:redshift_distribution}
\end{figure}

The target selection of the LRGs and QSOs are based on the DESI Legacy Imaging Survey\footnote{\url{https://www.legacysurvey.org}} \cite{Dey2019} that contains four different photometric regions with different image qualities, see Fig. 1 of \cite{Chaussidon2022} for instance. These photometric regions are called:
\begin{itemize}[noitemsep,topsep=1pt]
    \item \textbf{\emph{North}} is the northern part of the footprint in the North Galactic cap (NGC) ($\dec > 32.375~\rm{deg}$) corresponding to the part of the sky covered by BASS \cite{Zou2019}, and MzLS \cite{Silva2016}.
    \item \textbf{\emph{DES}} is the region covered by the Dark Energy Survey \cite{DarkEnergySurveyCollaboration2016} and is significantly deeper than the rest of the footprint.
    \item \textbf{\emph{South (NGC)}} is the rest of the footprint in NGC, which was also collected by DECam \cite{Flaugher2015} but is less deep than DES.
    \item \textbf{\emph{South (SGC)}} is similar to South (NGC) but in the South galactic cap (SGC). We split South (NGC) and South (SGC) although they have similar photometry because the two regions are spatially disjointed, and thus are never used in the same time when we compute the power spectrum either on the full NGC or SGC.
\end{itemize}
The target selection for the LRGs is the same across each region \cite{Zhou2023}; however, to adapt to each region specificities, the target selection for the QSOs is adapted for each region \cite{Chaussidon2023}. The discrepancy in either the target selection or imaging quality leads to different mean densities in these different photometric regions, as given in \cref{tab:mean_density}, and slightly different redshift distributions as displayed in \cref{fig:redshift_distribution}. In particular, DES has more high-$z$ quasars than the others because the imaging is deeper, and DES and North have less low-$z$ quasars than the South since the PSF is better resolved, such that low-$z$ quasars are preferentially detected with a non-PSF morphology and therefore do not pass the cut on PSF-like objects imposed by the QSO target selection, see Fig.11 in \cite{Chaussidon2023}. Therefore, in the following, these regions are always treated separately and mutually renormalised before computing the power spectrum over the entire NGC or SGC, see \cref{sec:pk_normalization}.

\begin{table}
    \centering
    \caption[]{Mean density of the LRG and QSO samples (corrected by the completeness of the observation) in each photometric region.}
    \label{tab:mean_density}
    \begin{tabular}{lcc}
      \toprule
      & LRG $[\rm{deg}^{-2}]$ & QSO $[\rm{deg}^{-2}]$ \\ 
      \midrule
      North & $531.7$ & $184.5$ \\
      South (NGC) & $535.3$ & $186.6$ \\
      South (SGC) & $532.6$ & $187.2$ \\
      Des   & $519.5$ & $191.7$ \\
      \bottomrule
    \end{tabular}
\end{table}

The construction of the catalogues for the data and the randoms are described in \cite{Ross2024}, and \cite{Guy2023,Bailey2023,Brodzeller2023} describe the spectroscopic reduction pipeline, as well as, the redshift estimation from the spectra. We only discuss, in the following, the correction of the imaging dependence of the target selection, see \cref{sec:imaging_systematic_data_blind}. During this analysis, we only use ten files of randoms\footnote{Each randoms file has a density of $2,500$ randoms per $\mathrm{deg}^2$. So, one needs to compare $10 \times 2,500$ to the densities given in \cref{tab:mean_density}, leading to a  randoms/data ratio of 50 for the LRGs and 125 for the QSOs.}  instead of 18 as in the fiducial BAO \cite{DESI2024III} and RSD \cite{DESI2024V} analysis, and after comparison, we denote no impact on the scales of the power spectrum that we use.

Note that at this stage of the DESI DR1 analysis and with the standard treatment, the large-scale modes of the Emission Line Galaxy (ELG) sample \cite{Raichoor2023} are not yet fully reliable despite a significant effort and development of new tools \cite{Rosado-Marin2024} and are therefore not used here. See, in particular, the large-scale modes of the ELGs power spectrum displayed in \cref{sec:impact_imaging_systematic_weights} that exhibit an unexpected excess of power about one magnitude. One could only analyse smaller scales ($k>0.008~h\text{Mpc}^{-1}$) where the power spectrum seems more reasonable, but due to the low linear bias of this tracer, no interesting constraints can be extracted at present.

\subsubsection{Correct for observational systematics} \label{sec:observational_weights}
As described in \cite{Ross2024}, one needs to correct the data for several observational effects by weighting them with
\begin{equation} \label{eqn:observational_weights}
    w_{\rm tot} = w_{\rm comp} \times w_{\rm sys} \times w_{\rm zfail}.
\end{equation}
The different contributions in \cref{eqn:observational_weights} are for
\begin{itemize}[noitemsep,topsep=1pt]
    \item Completeness ($w_{\rm comp}$): The complex geometry of the survey is taken into account via the randoms, uniform distribution of objects without any clustering which occupies the same volume as the data, and does not represent any difficulties. On the other hand, targets that are unobserved because there are no free fibers to observe them \referee{during the fiber assignment step \citep{Schlafly2023},} lead to a biased estimation of the power spectrum. The randoms cannot take these missing targets into account since the survey observes the geometrical regions associated with these missing targets. This impacts both large and small scales. The large scales can be easily corrected by the \emph{completeness} weights $w_{\mathrm{comp}}$\footnote{As described in \cite{Ross2024}, the completeness weights were split into two parts: 1 / FRACZ\_TILELOCID and 1 / FRAC\_TLOBS\_TILES, where 1 / FRAC\_TLOBS\_TILES is applied to the randoms (multiply the weights after the shuffling method by FRAC\_TLOBS\_TILES) instead of applying to the data directly. This choice has no impact on our scales of interest.} that simply overweight the observed objects to match the number of total targets in the patrol radius of a fiber. \referee{The impact of the fiber assignment is shown in \cref{sec:fiber_assignment}.}


    \item Imaging systematics ($w_{\rm sys}$): Imaging systematics can be defined as the dependence of the target selection on the properties of the used photometric survey and create, unfortunately, an excess of correlation at large scales (see \cref{sec:impact_imaging_systematic_weights}). Since the imaging systematics are the major bias in our analysis, the mitigation of this contribution is carefully validated in \cref{sec:imaging_systematic_data_blind}. 

    \item Spectroscopy efficiency ($w_{\rm zfail}$): Classification and redshift determination depend on the quality of the observation. Poor weather, noise in the CDDs, or dust in the sky can indeed impact the spectra collected. This effect is corrected but has a minor impact, as reported in \cite{Yu2024, Krolewski2024}. Note that any residual redshift determination errors are naturally be taken into account in our theoretical model, given in \cref{eqn:pk_theo}, by the parameter $\Sigma_s$ representing the typical damping velocity dispersion.
\end{itemize}
The completeness and redshift failures weights are the same as the ones summarized in \cite{DESI2024II}, while the imaging systematic weights are fully re-determined for this analysis, see \cref{sec:wsys_description}. 

Since the randoms should reproduce the same redshift distribution as the data, we randomly draw the redshifts and associated weights from the data for the randoms, such that randoms have similar weights ($w_{\mathrm{tot},r}$). This procedure is known as the \emph{shuffling} method \cite{Ross2012} and impacts the measurement as described in \cref{sec:ric}.

\subsubsection{Blinding the data} \label{sec:blinding}
To avoid any confirmation bias during our analysis, we have developed a blinding scheme that reproduces the scale-dependent bias in the power spectrum. This blinding is described and validated in \cite{Chaussidon2024}. Note that this is the first time the scale-dependent bias has been measured with a blinding strategy. 

Similar to the blinding schemes applied to conceal the BAO and RSD signal \cite{Andrade2024}, this blinding is applied at the catalog level. The one for PNG is a set of weights reproducing a value of $f_{\rm NL}^{\rm blind} \in [-15, 15]$ randomly chosen, such that the amount of PNG measured from the blinded data, $f_{\rm NL}^{\rm obs}$, is
\begin{equation}
    f_{\rm NL}^{\rm obs} = f_{\rm NL}^{\rm loc} + f_{\rm NL}^{\rm blind}.
\end{equation}
These weights are multiplied by the completeness weights, preventing anyone from breaking the blinding since the large-scale modes of the power spectrum cannot be recovered without correcting for completeness.

As described in \cite{Chaussidon2024}, the blinding is applied coherently across all the samples such that one can compare the large-scale modes of the power spectrum measured from sub-part of the sample, allowing us internal consistency validation, see \cref{sec:consistency}. Although the blinding value $f_{\rm NL}^{\rm blind}$ is the same for the different tracers, the weights were generated with a value of $b_{\Phi}$ computed for $p=1.0$ in each situation so that the apparent blinding value for the QSO when analysing with $p=1.6$ should be larger in absolute value.


\subsubsection{Linear bias evolution}

\begin{table}[t]
    \centering
    \caption[]{Value of the parameters $(a, b)$ used to describe the redshift evolution of the linear bias $b_1$ as given in \cref{eqn:linear_bias_evolution}. These values are estimated from the unblind DESI DR1 LRG and QSO samples.}
    \label{tab:alpha_beta_linear_bias}
    \begin{tabular}{lcc}
      \toprule
          & $a$ & $b$ \\ 
      \midrule
      LRG & $0.209 \pm 0.025$ & $1.415 \pm 0.076$ \\
      QSO & $0.237 \pm 0.010$ & $0.771 \pm 0.070$ \\
      \bottomrule
    \end{tabular}
  \end{table}

The evolution of the linear bias can modeled by 
\begin{equation} \label{eqn:linear_bias_evolution}
    b_1(z) = a (1 + z)^2 + b,
\end{equation}
where $a, b$ are given for LRGs and QSOs in \cref{tab:alpha_beta_linear_bias}. These two parameters are measured from the unblind DESI DR1 LRG and QSO samples. The measurement is detailed in \cref{sec:linear_bias}. \referee{The evolution of the linear bias will be used to compute the optimal weighting scheme described in \ref{sec:optimal_weight}.}

\subsection{Measuring the power spectrum from a spectroscopic survey} \label{sec:power_spectrum}
\subsubsection{Power spectrum estimator} \label{sec:power_spectrum_estimator}
In the following, the multipoles of the power spectrum are estimated through the so-called \emph{Yamamoto estimator} \cite{Yamamoto2006}. The estimation of the average power spectrum $\hat{P}\left(\vb{k}_\mu\right)$, in a phase-space volume $V_{\vb{k}_\mu}$ corresponding to the binning used in $\vb{k}_\mu$ space for the measurement, is given by
\begin{equation} \label{eqn:yamamoto-midpoint}
    \hat{P}_{\ell}\left(k_\mu\right)=\dfrac{2 \ell+1}{A V_{k_\mu}} \int_{V_{k_\mu}} \dd \vb{k} \int \dd \vb{x}_1 \int \dd \vb{x}_2 \, e^{i \vb{k} \cdot\left(\vb{x}_2-\vb{x}_1\right)}\mathcal{F}\left(\vb{x}_1\right) \mathcal{F}\left(\vb{x}_2\right) \mathcal{L}_{\ell}\left(\hat{\vb{k}} \cdot \hat{\vb{x}}_1\right)\\
    - \mathcal{N}_{\ell} ~,
\end{equation}
where we use the first-point $\vb{x}_1$ as the line-of-sight instead of the \emph{midpoint} line-of-sight to speed up the computation (see below), $F(\vb{x})$ is the FKP field \cite{Feldman1994} estimated from the data and the random catalogues
\begin{equation} \label{eqn:FKP_field}
    \mathcal{F}(\vb{x})= n_g(\vb{x}) - \alpha n_r(\vb{x}) \quad \text{ with } \quad \alpha = \dfrac{\int \dd \vb{x} \, n_g(\vb{x})}{\int \dd \vb{x} \, n_r(\vb{x})} ~,
\end{equation}
where $n_g$ is the galaxy weighted density, $w_g = w_{\mathrm{FKP}} \times w_{\rm tot}$, and $n_r$ the randoms' one, $w_r = w_{\mathrm{FKP}} \times w_{\mathrm{tot},r}$. The randoms are used to sample the survey geometry, more specifically the survey selection function $W(\vb{x})$, which is the ensemble average of the galaxy density: 
\begin{equation} \label{eqn:selection_function}
    W(\vb{x}) \equiv \left\langle n_g(\vb{x})\right\rangle = \alpha \left\langle n_r(\vb{x})\right\rangle. 
\end{equation}

The FKP weights\footnote{Note that contrary to equation (7.2) of \cite{Ross2024}, we do not use the dependence of the number of overlapping tiles for the density $\bar{n}(z)$, since this refinement was developed in particular for the emission line galaxy sample.} \cite{Feldman1994}, $w_{\mathrm{FKP}}$, are  weighting scheme that improve the power spectrum measurement by minimising the expected errors of $\hat{P}\left(\vb{k}_\mu\right)$:
\begin{equation} \label{eqn:fkp_weights}
    w_{\mathrm{FKP}}(\vb{x}, \vb{k}) = \dfrac{1}{1+\bar{n}_g(\vb{x}) P(\vb{k})} ~,
\end{equation}
where $P(\vb{k})$ is fixed as about the maximal amplitude measured in the data around $k_{\rm eq}$ which are the scales of interest\footnote{This is a different choice than in the BAO analysis \cite{DESI2024II,DESI2024III} which uses the value of the power spectrum at $k\sim0.14~h\mathrm{Mpc}^{-1}$.}: $P(\vb{k})= P_{0} = 3 \times 10^4~\left(\mathrm{Mpc}/h\right)^3$ for the QSO and $5 \times 10^4~\left(\mathrm{Mpc}/h\right)^3$ for the LRG. In what follows, FKP weights are computed independently in each of the four photometric regions using the redshift distributions displayed in \cref{fig:redshift_distribution}.

Additionally, the normalization factor used in \cref{eqn:yamamoto-midpoint} is given by\footnote{See \url{https://pypower.readthedocs.io/en/latest/api/api.html\#pypower.fft_power.normalization} for its numerical derivation.}
\begin{equation} \label{eqn:normalization_factor}
        A = \int \dd \vb{x} \, \bar{n}_g(\vb{x})^2,
\end{equation}
and the shot noise contribution is removed with
\begin{equation}
        \mathcal{N}_{\ell} = \dfrac{\delta_{\ell 0}}{A} \int \dd \vb{x} \, W(\vb{x})\left[w_g(\vb{x}) + \alpha w_r(\vb{x}) \right] .
\end{equation}

Finally, the choice of $\vb{x}_1$ as a line-of-sight coordinate reduces the computation time of \cref{eqn:yamamoto-midpoint}, by splitting the double integral \cite{Yamamoto2006},
\begin{equation} \label{eqn:yamamoto_estimator}
    \hat{P}_{\ell}\left(k_\mu\right)=\dfrac{2 \ell+1}{A V_{k_\mu}} \int_{V_{k \mu}} \dd \vb{k} \, F_0(\vb{k}) F_{\ell}(-\vb{k}) - \mathcal{N}_{\ell} ~,
\end{equation}
where we have introduced\footnote{\cref{eqn:F_l} can be written as a sum of Fourier transforms \cite{Hand2017}, by decomposing the Legendre polynomials $\mathcal{L}_\ell$ into spherical harmonics $Y_{\ell m}$:
\begin{equation}
    \mathcal{L}_{\ell}(\hat{\vb{x}} \cdot \hat{\vb{k}})=\frac{4 \pi}{2 \ell+1} \sum_{m=-\ell}^{m=\ell} Y_{\ell m}(\hat{\vb{x}}) Y_{\ell m}^{\star}(\hat{\vb{k}}),
\end{equation}
\cref{eqn:F_l} becomes
\begin{equation}
    F_{\ell}(\vb{k}) = \dfrac{4 \pi}{2 \ell+1} \sum_{m=-\ell}^{m=\ell} Y_{\ell m}^{\star}(\hat{\vb{k}}) \int d^3 x \, e^{-i \vb{k} \cdot \vb{x}} F(\vb{x}) Y_{\ell m}(\hat{\vb{x}}),
\end{equation}
and requires the computation of \emph{only} $2\ell + 1$ Fast Fourier Transforms for each multipole $\ell$. 
}
\begin{equation} \label{eqn:F_l}
    F_{\ell}(\vb{k})=\int \dd \vb{x} \, e^{i \vb{k} \cdot \vb{x}} \mathcal{F}(\vb{x}) \mathcal{L}_{\ell}(\hat{\vb{k}} \cdot \hat{\vb{x}}).
\end{equation}
Note that this choice of line-of-sight leads to the so-called \emph{wide-angle effects} that is described in \cref{sec:window_function}.

In the following, all power spectra are computed with \texttt{pypower}\footnote{\url{https://github.com/cosmodesi/pypower}} using the Triangular Shaped Cloud (TSC) sampling and interlacing at order $n=3$ to mitigate the aliasing. For both LRG and QSO samples, we use a physical box sizes of $16000~h^{-1}\mathrm{Mpc}$ with a grid cell size of $6 ~h^{-1}\mathrm{Mpc}$, leading to a Nyquist frequency of $k_{N} \sim 0.5~h\mathrm{Mpc}^{-1}$.

\subsubsection{Optimal quadratic estimator for the scale-dependent bias} \label{sec:optimal_weight}
The FKP weights introduced above miss the redshift dependence of the PNG signal that we want to measure, such that introduce this dependence into the FKP weights provides a more optimal way to extract the scale-dependent bias signal in the power spectrum. This can be achieved by using optimised redshift weights that are inspired from the optimal quadratic estimator (OQE) for $\fnl$ \cite{Mueller2019, Castorina2019}. In the following, we follow \cite{Castorina2019} who propose to weight each galaxy, which is more natural to compute the FKP field $\mathcal{F}$, instead of weighting pairs of galaxies as in \cite{Mueller2021}. 

The optimal estimator for extracting $f_{\mathrm{NL}}^{\mathrm{loc}}$ has the same form as \cref{eqn:yamamoto_estimator} but with a different weighting scheme:
\begin{equation}
    \hat{P}_{\ell}\left(k_\mu\right) = \dfrac{2 \ell+1}{A_{\ell} V_{k_\mu}} \int_{V_{k \mu}} \dd \vb{k} \, \tilde{F}(\vb{k}) F_{\ell}(-\vb{k}) - \mathcal{N}_{\ell},
\end{equation}
where 
\begin{equation}
    \left\{
    \begin{array}{ll}
        \tilde{F}(\vb{k}) &= \displaystyle\int \dd \vb{x} \, e^{i \vb{k} \cdot \vb{x}} \tilde{w} \left(\vb{x}\right) \mathcal{F}(\vb{x}) \\[3pt]
        F_{\ell}(\vb{k})  &= \displaystyle\int \dd \vb{x} \, e^{i \vb{k} \cdot \vb{x}} w_{\ell}\left(\vb{x}\right) \mathcal{F}(\vb{x}) \mathcal{L}_{\ell}(\hat{\vb{k}} \cdot \hat{\vb{x}})
    \end{array}
    \right., 
\end{equation}
and the shot noise contribution is 
\begin{equation}
    \mathcal{N}_{\ell} = \dfrac{\delta_{\ell 0}}{A} \int \dd \vb{x} \, W(\vb{x})  \left[w_g(\vb{x}) + \alpha w_r(\vb{x}) \right] \tilde{w}\left(\vb{x}\right)w_{\ell}\left(\vb{x}\right).
\end{equation}
Similarly, the normalization factor in \cref{eqn:normalization_factor} becomes
\begin{equation} 
A_{\ell} = \int \dd \vb{x} \, \bar{n}_g(\vb{x})^2  \tilde{w}\left(\vb{x}\right)w_{\ell}\left(\vb{x}\right).
\end{equation}

The optimal weights\footnote{Here, we assume that the density distribution is isotropic and only depends on the redshift.} $\tilde{w}$, $w_{0}$ and $w_2$ for the quadratic estimator are
\begin{equation} \label{eqn:oqe_weights}
    \left\{
    \begin{array}{ll}
        \tilde{w}(z) &= \left[b(z) - p \right] \\[3pt]
        w_0(z) &= D(z) \left[ b(z)+f(z) / 3 \right] \\[3pt]
        w_2(z) &= 2 / 3 D(z) f(z)
    \end{array} 
    \right., 
\end{equation}
where $p$ is the parametrization used for $b_{\Phi}$ in \cref{eqn:b_phi_with_p}, $f$ is the growth rate and $D$ is the growth factor.

\begin{figure}[t]
    \centering
    \includegraphics[scale=1]{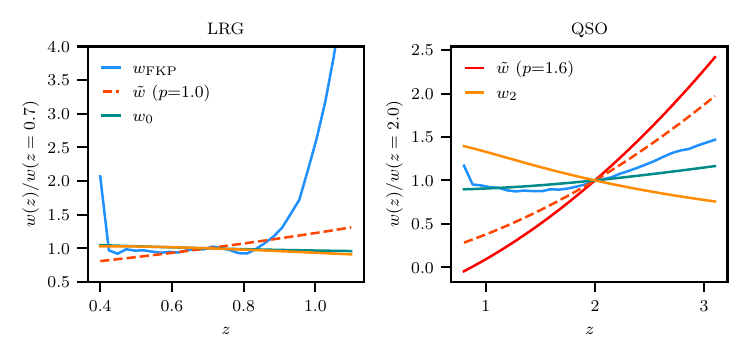}
    \caption{Optimal weighting scheme used to measure $\fnl$ for LRG sample (left) and QSO sample (right) as a function of redshift. The usual FKP weights are displayed in blue. For the quasars, the most important weight is $\tilde{w}$, which follows the redshift evolution of the linear bias. For comparison, we normalize the weight to one at $z=0.7$ for the LRGs and $z=2.0$ for the QSOs.}
    \label{fig:optimal_weights}
\end{figure}

These weights used in this analysis are displayed in \cref{fig:optimal_weights} for the LRG and QSO samples. For the LRGs (left panel), the shapes of $\tilde{w}$, $w_{0}$ and $w_2$ are very similar to $w_{\rm FKP}$ such that we do not expect substantial improvement by using these optimal weights compared to the traditional FKP ones. The FKP weight shape, increasing a lot around $z\sim1$, comes from the decrease of the density $n(z)$ of this sample in this region. 

For the QSOs (right panel), the optimal weights in \cref{eqn:oqe_weights} overweight the objects at high redshift, naturally increasing the effective redshift of the sample. The first effect is to increase the value of $b_1$ and $b_{\Phi}$ in our weighted sample such that the precision on $\fnl$ is improved. Due to this effective redshift modification, it is hard to quantify precisely how these weights improve the measurement compared to the standard FKP weights.

Practically speaking, computing the power spectrum with the OQE weights is just the cross-power spectrum between one FKP field weighted by $\tilde{w}$ and another one weighted by $w_{\ell}$. Therefore, the computation of the power spectrum causes no problem. 

Since the bias vanishes in the hexadecapole ($\ell=4$), no specific weights are needed for this multipole. Due to a lack of statistical significance, we do not include the hexadecapole in our analysis. The gain by using the quadrupole ($\ell=2$) in addition to the monopole ($\ell=0$) is shown in \cref{sec:gain_with_quadrupole}.

\subsubsection{Effective redshift} \label{sec:effective_redshift}
\begin{figure}
    \centering
    \hspace{-2cm}\includegraphics[scale=1]{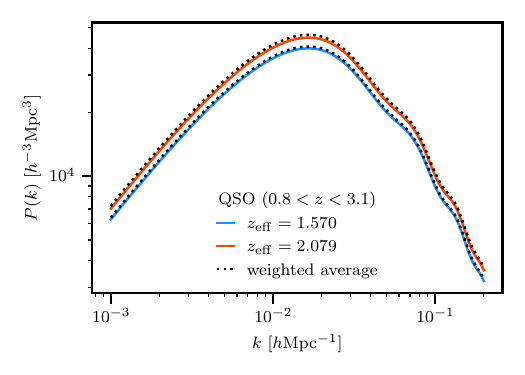}
    \caption{\referee{Comparison between the monopoles of the power spectrum at two effective redshifts for the quasar sample (FKP in blue and OQE in red) and the weighted average of the monopoles over the selection function in dotted black.}}
    \label{fig:p_zeff}
\end{figure}

During the parameter estimation step, one needs to evaluate the model \cref{eqn:pk_theo} at the \emph{effective redshift} of the data. We are following the definition used in \citep{Castorina2019} that is correct up to first order in the Taylor expansion\footnote{See, for instance, Appendix B from \cite{DeMattia2021}}:
\begin{equation} \label{eqn:effective_redshift}
    z_{\mathrm{eff}}=\frac{\int \dd z \, n(z)^2 w_a(z) w_b(z) z}{\int \dd z \, n(z)^2 w_a(z) w_b(z)},
\end{equation}
where $w_a, w_b$ are the weights of two fields that are cross-correlated. \referee{This definition of the effective redshift is validated in \cref{fig:p_zeff} where the color lines are the monopoles computed at the effective redshifts while the dotted lines are the weighted averages of the monopole over the selection function similarly to \cref{eqn:effective_redshift}}.

\Cref{tab:effective_redshift} gives the effective redshift under the different sets of weights used in the following. The redshift distribution of the DR1 LRGs and QSOs are displayed in \cref{fig:redshift_distribution}. For comparison, we also give the effective redshift without any weighting scheme and the effective redshift for the quasars when using $p=1.0$. Using OQE weights significantly increases the effective redshift for the quasar, while it is less pronounced for the LRG since the $n(z)$ is mostly flat. Using $p=1.6$ reduces the response of the scale-dependent bias to the presence of PNG such that the OQE weights increase the weight for the higher redshift part of the sample, where the signal is the most important.

Note that the use of OQE weights requires to evaluate the linear power spectrum at two different effective redshifts, one for the monopole and one for the quadrupole. In the following, when we use the OQE weights with $\ell=0, 2$, we assume that $\Sigma_s$ does not depend on the redshift while we fit $b_1$ by considering the evolution given in \cref{eqn:linear_bias_evolution}.

\begin{table}
    \centering
    \caption{Effective redshift for the DESI DR1 LRG and QSO samples computed with the completeness and spectroscopic efficiency weights. For comparison, the mean redshift of the samples is displayed in the first line, and the unweighted effective redshift in the second one. Due to the shape of the redshift distribution, using OQE weights significantly increases the effective redshift for the quasars.}
    \label{tab:effective_redshift}
    \begin{tabular}{lccc}
        \toprule
        & LRG & QSO  \\
        & 0.4 < z < 1.1 & 0.8 < z < 3.1 \rule{0pt}{2.6ex}\rule[-1.5ex]{0pt}{0pt} \\
        \midrule
        $\bar{z}$ & 0.741 & 1.768 \\
        $z_{\mathrm{eff}}$ & 0.665 & 1.573  \\
        $z_{\mathrm{eff}}$ (FKP) & 0.733 & 1.651 \\
        \midrule
        $z_{\mathrm{eff}}$ (OQE $\ell = 0$, $p=1.0$) & 0.754 & 1.926 \\
        $z_{\mathrm{eff}}$ (OQE $\ell = 2$, $p=1.0$) & 0.751 & 1.813 \\
        \midrule
        $z_{\mathrm{eff}}$ (OQE $\ell = 0$, $p=1.6$) & - & 2.082 \\
        $z_{\mathrm{eff}}$ (OQE $\ell = 2$, $p=1.6$) & - & 1.989 \\
        \bottomrule
    \end{tabular}  
  \end{table}


\subsubsection{Normalization across the different photometric regions} \label{sec:pk_normalization}
As shown in \cref{tab:mean_density} and in \cref{fig:redshift_distribution}, the LRG \cite{Zhou2023} and QSO \cite{Chaussidon2023} samples have different redshift distributions and angular densities in different photometric regions of the Legacy Imaging Surveys used for the target selection: North, South and DES. These differences may be due to slightly different target selection cuts or to different photometric properties of a specific region. For instance, the DES region is about one magnitude deeper than South \cite{Dey2019}.

Although one can compute the power spectrum independently on each of these photometric regions, one wants to compute the power spectrum simultaneously on the different regions to avoid losing any modes across the different regions and reduce the statistical uncertainty of our measurement. In the following, the power spectrum is computed on all the NGC and on all the SGC such that we need to normalize the North to the South (NGC) and DES to the South (SGC).

Here, the normalization of the randoms means that $\alpha$ in \cref{eqn:FKP_field} is set to match the corresponding data separately in each region. Thus, the randoms weights in the South (NGC) are multiplied by the normalization factor:
\begin{equation} \label{eqn:normalization_factor_region}
  f_{\textrm{norm}} = \dfrac{\sum_{i \in \textrm{North}} w_{r, i}}{\sum_{i \in \textrm{North}} w_{d, i}} \times \dfrac{\sum_{i \in \textrm{South (NGC)}} w_{d, i}}{\sum_{i \in \textrm{South (NGC)}} w_{r, i}},
\end{equation}
and similarly for the weights in the South (SGC) (South (NGC), North $\rightarrow$ South (SGC), DES).

This normalization is crucial to measure the power spectrum's large-scale modes without bias. The impact of the normalization of DES to South (SGC) is shown in \cref{sec:appendix_pk_normalization}.

\subsection{Estimating the covariance matrix}
\subsubsection{EZmocks} \label{sec:ezmocks}
In the following, the covariance matrix is estimated as the covariance between the measurements done in a large set of realistic simulations (mocks) that emulate the observations as faithfully as possible.

As in eBOSS \citep{Zhao2021}, we choose the approximate method known as \emph{EZmocks} \citep{Chuang2015} to generate our realistic simulations. This method generates a galaxy field with position and velocity that follows an input power spectrum at an effective redshift, thanks to the Zel’dovich approximation \citep{Zeldovich1970}. This approximation is enough in our situation since our analysis is focused on large scales where the linear theory holds such that the EZmocks predict the desired covariance matrix \cite{Zhang2023}. 

We use the EZmocks generated for DESI that are similar to what is described in \cite{Zhao2021}. For a full description of these simulations, we refer the reader to \cite{Zhao2024}. In particular, we use 2000 boxes of $6~\mathrm{Gpc}~h^{-1}$ side, 1000 for the NGC and 1000 for the SGC, so that we do not need duplications to cover the full volume of DESI. Boxes for the LRGs (\textit{resp.} QSOs) are generated at $z=0.8$ (\textit{resp.} $z=1.4$) using the fiducial DESI cosmology. Despite the large size of these boxes, quasars are too dispersed in redshift such that they can be emulated only up to $z_{\rm max}=3.1$ without repeating the box. It explains why we stop our analysis at this maximum redshift for the quasars, although DESI has observed quasars at higher redshifts, $\sim$32,500 with $3.1<z<3.5$.

In Section 5.7 of \cite{DESI2024IV} (see also Section 10.2 of \cite{DESI2024II}), the covariance matrix from the EZmocks is rescaled to match the analytical prediction from \texttt{RascalC} \cite{Rashkovetskyi2023}. Here, however, we do not rescale our covariance matrix. The difference arises because the EZmocks in \cite{DESI2024IV} incorporate a method to emulate the fiber assignment of DESI known as the FFA \cite{Bianchi2024} (see Section 11.2 of \cite{DESI2024II}), which results in an underestimation of the variance. In our case, we neglect the impact of the fiber assignment and we verified that our EZmocks do not under-estimate the covariance. Further investigations could be required for the upcoming data release.

\subsubsection{Generate realistic simulations}
First, the cubic EZmocks are remapped according to \cite{Carlson2010} to increase the sky coverage of these simulations, and all the coordinates are transformed into sky coordinates ($\ra, \dec, z$) after the remapped box is moved along an axis. Then, we add the redshift space distortion effect along the line-of-sight by translating the real space position to redshift space \cite{Kaiser1987}.

Next, we match these simulations to the DESI survey\footnote{All these steps can be performed with \texttt{mockfactory} (\url{https://github.com/cosmodesi/mockfactory}) an MPI-based code to generate cutsky mocks from box simulations.}. We imprint the redshift distribution (\cref{fig:redshift_distribution}) and the mean density (\cref{tab:mean_density}) independently in each of the photometric regions. At that time, we did not differentiate between South (NGC) and South (SGC), and we used the redshift distribution and density from South (NGC+SGC) for the two regions. This will be improved with the upcoming DESI data release and is neglected in the following.


\referee{As illustrated in \cref{sec:fiber_assignment}, the large-scale modes of the power spectrum are only sensitive to the global completeness of the observation, such that we do not need to apply the entire fiber assignment step in each mocks. Hence, we apply} the global completeness by downsampling the data and the randoms via an \texttt{HEALPix} map \cite{Gorski2005, Zonca2019} at $N_{\rm side}=256$ representing the fraction of the pixel that was observed in DESI DR1. We finally remove objects located in bad imaging regions as in \cite{Ross2024}. In particular, we use the LRG mask developed in \cite{Zhou2023} for the LRGs and the imaging maskbits\footnote{\url{https://www.legacysurvey.org/dr10/bitmasks/}} 1, 7, 8, 11, 12 and 13 for the QSOs. This is not exactly what it is done in \cite{DESI2024II} that, in addition, also removes pixels where the imaging properties of the photometric survey are too extreme and regions with bad hardware\footnote{\url{https://github.com/desihub/LSS/blob/main/py/LSS/globals.py\#L77}}. This represents a small fraction of the sky that is negligible for our covariance matrix estimation compared to the expected statistical errors. 


We also generate, from the same boxes, mocks that describe the expected final DESI sample after the five years of observation referred in the following as Y5. We use the exact sky coverage to match the expected observations and assume full completeness and the same redshift distribution and density for each photometric region as the DR1 sample. These Y5 mocks will be used to have an accurate forecast for the expected final DESI sample and validate our theoretical description, see \cref{sec:validation_ezmocks}.

\subsubsection{Computing the covariance matrix} \label{sec:covariance_matrix}
\begin{figure}
    \centering
    \includegraphics[scale=1]{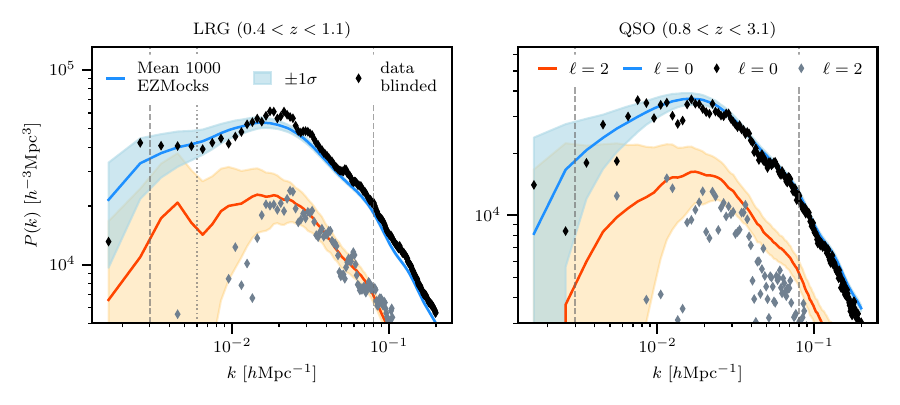}
    \caption{Comparison between the multipoles (NGC+SGC) from the mean of 1000 EZmocks (solid lines) and the blinded data (diamond points) for the LRGs (left) and the QSOs (right). The coloured dashed regions are the $\pm1\sigma$ regions from the EZmocks. Note that although the large scales are blinded for the data, there is a good agreement between the mean from the EZmocks and the data. The discrepancy at large scales on the quadrupole is from the radial integral constraints and is detailed in \cref{sec:ric}.}
    \label{fig:data_vs_ezmocks}
\end{figure}

Once each mock realization is matched to DESI DR1, the power spectrum with different sets of weights (FKP or OQE) is computed precisely in the same way as the one calculated from the data, as explained in \cref{sec:power_spectrum}. \Cref{fig:data_vs_ezmocks} shows the power spectrum of the mean of 1000 EZmocks for LRGs (left) and QSOs (right) as well as the power spectrum from the blinded data catalog. The coloured shaded regions represent the $\pm 1 \sigma$ deviation estimated from 1000 realizations. Although the EZmocks were not generated at the correct effective redshift of the data (see \cref{tab:effective_redshift}), they remain usable to estimate the covariance matrix because they match the amplitude of the power spectrum from the data, and our analysis is limited to scales that are almost linear. 

Finally, the covariance $C_{ij}$ is simply the covariance between these 1000 measured power spectra. As proposed in \cite{Hartlap2007}, we re-scale by multiplying the inverse of the covariance by the Hartlap factor to deal with the skewed nature of the inverse Wishart distribution as
\begin{equation} \label{eqn:cov_with_hartlap}
    C_{ij}^{-1} \longleftarrow \dfrac{N_m - n - 2}{N_m - 1} C_{ij}^{-1},
\end{equation}
where $N_m$ is the number of mocks, and $n$ is the number of data points. In addition, we also add the extra correction provided in \cite{Percival2014} to correct for the propagation of errors in the covariance matrix to the errors on estimated parameters, re-scaling \cref{eqn:cov_with_hartlap} by dividing it by the Percival factor
\begin{equation}
    C_{ij}^{-1} \longleftarrow \left(\dfrac{1 + B(n - n_p)}{1 + A + B(n_p + 1)}\right)^{-1} C_{ij}^{-1},
\end{equation}
with $A= 2 \left[\left(N_m - n - 1\right)\left(N_m - n - 4\right)\right]^{-1}$, $B=\left(N_m - n - 2\right)\left[\left(N_m - n - 1\right)\left(N_m - n - 4\right)\right]^{-1}$ and $n_p$ the number of estimated parameters. Typically, in the following, $N_m=1000$, $n=116$, and $n_p = 4$ such that the Hartlap factor evaluates to $\sim 0.88$ and the Percival factor to $\sim 1.12$. 

Note that in practice, all the inference is performed using the \texttt{desilike}\footnote{Publicly available: \url{https://github.com/cosmodesi/desilike}. In particular, the model that we used is presented in \url{https://github.com/cosmodesi/desilike/blob/hmc/nb/png_examples.ipynb}.} framework. The posterior profiling is performed through the \texttt{iminuit} \cite{iminuit} minimiser\footnote{\url{https://github.com/cosmodesi/desilike/blob/hmc/desilike/profilers/minuit.py}} and all the Monte Carlo Markov chains (MCMC) use the \texttt{emcee} \cite{Foreman-Mackey2013} sampler\footnote{\url{https://github.com/cosmodesi/desilike/blob/hmc/desilike/samplers/emcee.py}}.

\section{Modeling of geometrical effects} \label{sec:model}
The large-scale modes of the observed power spectrum are impacted by the geometry of the survey. In this section, we present how to modify our model in order to account for the geometry as well as correct for the integral constraints.  

\subsection{Window function and wide-angle effect} \label{sec:window_function}
Due to stars, Milky Way dust, or incompleteness of the observations, some parts of the footprint are masked or unobserved such that we do not exactly observe the full density field, but only a fraction of it. This is described by the survey selection function $W(\vb{x})$, see \cref{eqn:selection_function}. Hence, the expected value of the power spectrum estimator in \cref{eqn:yamamoto_estimator} reads as\footnote{\cite{Feldman1994} shows that $\langle \mathcal{F}(\vb{x}) \mathcal{F}(\vb{x^\prime}) \rangle = W(\vb{x})W(\vb{x^\prime})\xi(\vb{x}, \vb{x^\prime}) + W(\vb{x})\delta_D^{(3)}(\vb{x} - \vb{x^\prime})$.} 
\begin{equation} \label{eqn:expected_value_yamamoto}
    \left\langle\hat{P}_{\ell}(k)\right\rangle = \dfrac{(2 \ell+1)}{A} \int \dfrac{\dd \Omega_k}{4 \pi} \int \dd \vb{s_1} \int \dd \vb{s_2} \, e^{i \vb{k}\left(\vb{s_2} - \vb{s_1} \right)} \xi(\vb{s_1}, \vb{s_2}) W\left(\vb{s_1} \right) W\left(\vb{s_2}\right) \mathcal{L}_\ell \left(\vu{k} \cdot \vu{s_1} \right)
\end{equation}

Following \cite{Beutler2019}, the correlation function can be expanded into Legendre multipoles and under the local plane-parallel approximation limit ($s \ll s_1, s_2$ with $\vb{s} = \vb{s_1} - \vb{s_2}$), \cref{eqn:expected_value_yamamoto} becomes
\begin{equation}\label{eqn:convolved_pk}
    \begin{aligned}
    \left\langle\hat{P}_{\ell}(k)\right\rangle =& \dfrac{(2 \ell+1)}{A} \sum_p \int \dfrac{\dd \Omega_k}{4 \pi} \int \dd \vb{x} \int \dd \vb{s} \, e^{-i \vb{k} \cdot \vb{s}} W(\vb{x}) W(\vb{x}-\vb{s}) \xi_p(s) \mathcal{L}_p(\vu{x} \cdot \vu{s}) \mathcal{L}_{\ell}(\vu{k} \cdot \vu{x}) \\
    &= 4\pi (-i)^\ell (2 \ell + 1) \sum_{\ell_1, \ell_2}  \left(\begin{array}{ccc}
        \ell_1 & \ell_2 & \ell \\
        0 & 0 & 0
        \end{array}\right)^2  \int \dd s \, s^2 j_\ell(ks) \xi_{\ell_1}(s) \mathcal{W}_{\ell_2}(s) 
    \end{aligned}
\end{equation}
where we have introduced the real space \emph{window matrix}
\begin{equation}
    \mathcal{W}_\ell(s) \equiv \dfrac{(2 \ell+1)}{4\pi \times A} \int \dd \Omega_s \int \dd \vb{x} \, W(\vb{x}) W(\vb{x}-\vb{s}) \mathcal{L}_\ell(\vu{x} \cdot \vu{s}).
\end{equation}

To speed up the computation of the \cref{eqn:yamamoto-midpoint}, we have chosen the first galaxy as the line-of-sight \cite{Yamamoto2006}. This is a good choice under the local plane-parallel approximation. However, this choice creates the so-called \emph{wide-angle effect} when this approximation does not hold \cite{Castorina2018}. One can take into account this wide-angle effect by expanding the theoretical correlation function as 
\begin{equation}
    \xi\left(\mathbf{x}_1, \mathbf{x}_2\right) = \sum_{p, n}\left(\frac{s}{d}\right)^n \xi_p^{(n)}(s) \mathcal{L}_p(\hat{\mathbf{d}} \cdot \hat{\mathbf{s}}),
\end{equation}
where $\vb{s} = \vb{x_2} - \vb{x_1}$ is the pair separation and $\vb{d}$ the line-of-sight.

As described in \cite{Beutler2019}, the wide-angle effect can be easily handled with the above window matrix formalism by introducing the $(s / d)^n$ expansion in \cref{eqn:convolved_pk}, and one need to compute new window matrices
\begin{equation}
    \mathcal{W}_{\ell}^{(n)}(s)= \dfrac{2 \ell+1}{4 \pi \times A} \int \dd \Omega_s \int \dd \vb{x} \, x^{-n} W(\vb{x}) W(\vb{x}-\vb{s}) \mathcal{L}_{\ell}(\vu{x} \cdot \vu{s}).
\end{equation}
In the following, we only consider the first order of the effect ($n=1$) such that
\begin{equation}
    \left\{
    \begin{array}{ll}
    \xi_1^{(1)}(s) &= -\dfrac{3}{5} \xi_2^{(0)}(s) \\[3pt]
    \xi_3^{(1)}(s) &= \dfrac{3}{5} \xi_2^{(0)}(s) - \dfrac{10}{9} \xi_4^{(0)}(s)
    \end{array}
    \right., 
\end{equation}
and the multipoles of the correlation function $\xi_{\ell}^{(0)}$ are computed from \cref{eqn:pk_theo} and \cref{eqn:pk_legendre} using
\begin{equation}
    \xi_{\ell}^{(0)}(r)=\frac{i^{\ell}}{2 \pi^2} \int \dd k \, k^2  j_{\ell}(k r) P_{\ell}(k).
\end{equation}

Hence, the convolved power spectrum is evaluated on a finite size wavelength vector $k_i$ through a single matrix multiplication 
\begin{equation} \label{eqn:convolved_prediction}
    \left(P_{\ell}^{\rm obs}\right)_{i} = \left( \mathcal{W}_{\ell \ell^\prime} \right)_{ij} \left(P_{\ell^\prime}\right)_{j},
\end{equation}
where the summation run over $\ell^\prime$ and $j$ the indices on which the unconvolved power is evaluated.

\begin{figure}
    \centering
    \includegraphics[scale=1]{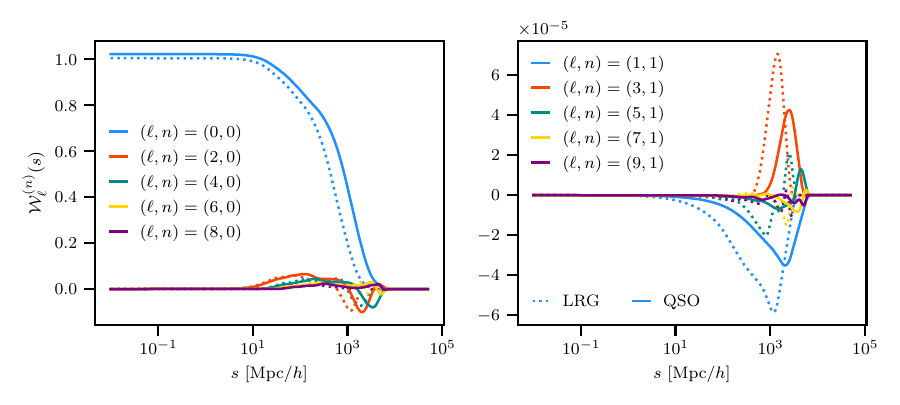}
    \caption{Real space window matrix for the DESI DR1 LRG (dashed lines) and QSO (solid lines) sample. The window matrices without the wide-angle effect ($n=0$) are displayed on the left, while the first order ($n=1$) contribution are displayed on the right.}
    \label{fig:window_function}
\end{figure}

The real space window matrix for the DESI DR1 LRG and QSO sample is displayed in \cref{fig:window_function}. One can note that, as indicated by \cite{DeMattia2019}, the real space window matrix $\mathcal{W}_0(s)$ is not normalised to 1 when $s \rightarrow 0$. The normalization factor is the same as the one used in Yamamoto's estimator \cref{eqn:yamamoto_estimator} and does not introduce a bias during the parameter estimation.

\begin{figure}
    \centering
    \includegraphics[scale=1]{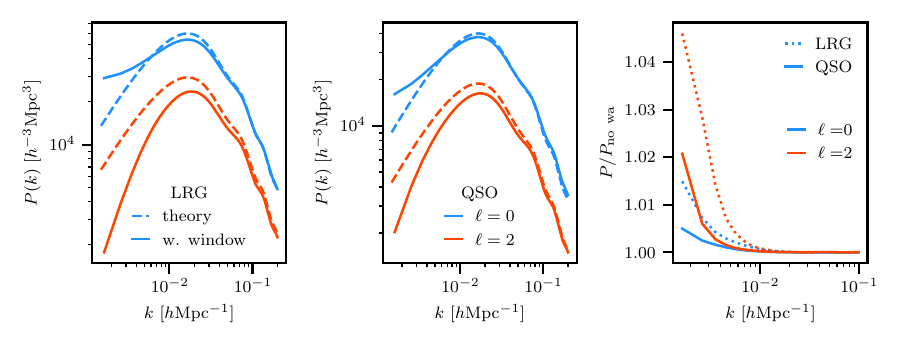}
    \caption{Multipoles of the convolved (\textit{resp.} unconvolved) power spectrum are displayed in solid (\textit{resp.} dashed) lines for the DESI DR1 LRG (left panel) and QSO (middle panel) sample. Right panel shows the ratio between the convolved power spectrum accounting for the first order wide-angle effect and not accounting for it. As expected the wide-angle effect is bigger for the LRG that are closer to us.}
    \label{fig:window_impact}
\end{figure}

The multipoles of the theoretical convolved power spectrum accounting for the geometry of the DESI DR1 LRG or QSO sample are shown in \cref{fig:window_impact}. The impact on large scales of the window matrix cannot be neglected when measuring the large-scale modes of the power spectrum. Additionally, the impact of considering the wide-angle effect is displayed in left panel. As expected, no strong effects are visible for the QSOs, since the quasars are sufficiently far away from us for the local plane-parallel approximation to hold. However, the effect is larger for the LRG but is still very small such that we do not account for high-order contribution ($n \geq 2$) of the wide-angle effect \cite{Benabou2024}. 

In practice, the window matrix is computed from the random catalog that fully described the geometry of the data, using the implementation available in \texttt{pypower}\footnote{The window matrix is from the concatenation of three windows obtained with box sizes 20x, 5x and 1x the nominal box size used for the power spectrum measurement, as shown in \url{https://github.com/cosmodesi/pypower/blob/main/nb/window_examples.ipynb}}. Then the convolution of the model with the window function is done in \texttt{desilike}\footnote{\url{https://github.com/cosmodesi/desilike/blob/hmc/desilike/observables/galaxy_clustering/power_spectrum.py\#L19}}. As in \cite{Beutler2017}, we only use the multipoles up to $\ell=4$ in \cref{eqn:convolved_pk} such that we only consider the window matrix up to $\ell=8$\footnote{Non-zero Wigner 3-j symbols must respect: $\vert l_1 - l_2 \vert < l \leq l_1 + l_2$}. 

\subsection{Validation with EZmocks} \label{sec:validation_ezmocks}
\subsubsection{Analysis setup}
One can use the mean of the power spectrum over the 1000 EZmocks as a null test, to validate the theoretical prediction given in \cref{eqn:convolved_prediction} and forecast with great accuracy the statistical errors expected for this analysis. During the MCMC and the profiling, we fit either $(\fnl, b_1, s_{n,0}, \Sigma_s)$ assuming \cref{eqn:b_phi_with_p} or $(b_{\Phi}\fnl, b_1, s_{n,0}, \Sigma_s)$ together.

First, as explained in \cref{sec:ezmocks}, the EZmocks were not generated at the correct effective redshift with the correct bias. Fortunately, the amplitudes of the power spectrum of these EZmocks match quite well the one from the data, see \cref{fig:data_vs_ezmocks}, such that they can be used without renormalization to build the covariance matrix. However, to use them as a null test, we need to match the amplitude to recover the expected bias at a specific effective redshift\footnote{Note that we have performed all these tests before the final weights for the clustering catalog were available. In particular, at that time we did not have the spectroscopic efficiency weights such that, in the following, we use a slightly different effective redshift than the one given in \cref{tab:effective_redshift}, typically lower about $\Delta z \sim 0.002$.} by renormalising the monopole and the quadrupole. This does not pose any problems, as we mainly use linear scales of the power spectrum. This can be achieved by measuring the actual bias $b$ from the mocks at $z_{\rm eff}$ and then replacing the multipoles by 
\begin{equation*} 
    \left\{
        \begin{array}{ll}
            P_{0}(k) & \longrightarrow \dfrac{b_n^2}{b^2}  \dfrac{1 + 2/3 \beta_n + 1/5 \beta_n^2}{1 + 2/3 \beta + 1/5 \beta^2} \times P_{0}(k) \\
            P_{2}(k) & \longrightarrow \dfrac{b_n^2}{b^2} \dfrac{4/3 \beta_n + 4/7 \beta_n^2}{4/3 \beta + 4/7 \beta^2} \times P_{2}(k)
        \end{array}
    \right. ,
\end{equation*}
where $b_n$ is the desired bias and $\beta_n = f(z_{\rm eff}) / b_n$ (similarly for $\beta$).

\begin{figure}
    \centering
    \includegraphics[scale=1]{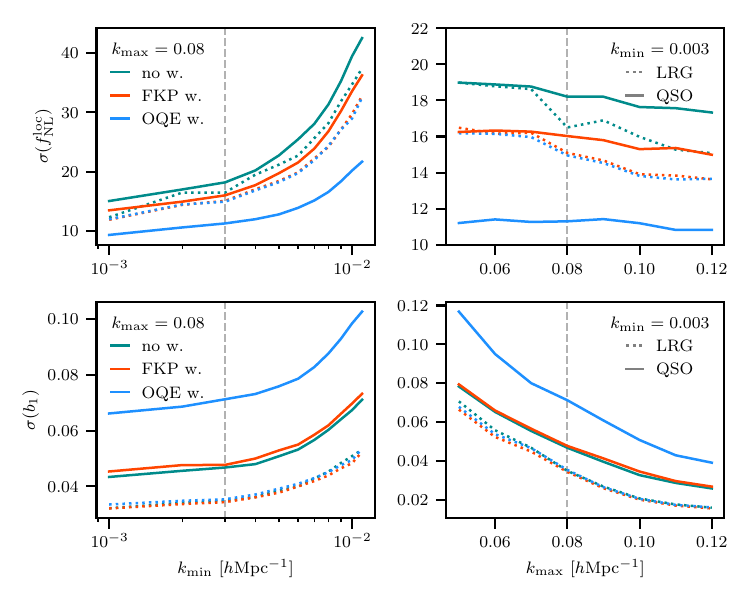}
    \caption{Dependence of the errors on $\fnl$ (top) and $b_1$ (bottom) as a function of either $k_{\rm min}$ (left) or $k_{\rm max}$ (right) for the LRGs and the QSOs and for the different weighting schemes. The fiducial values, $k_{\rm min} = 0.003~h\textrm{Mpc}^{-1}$ and $k_{\rm max} = 0.08~h\textrm{Mpc}^{-1}$, are denoted with dashed vertical lines. Note that the errors here are the standard deviation from MCMC chains, however the distribution is not symmetric for $\fnl$, see \cref{fig:fit_ezmocks}. Therefore, the errors displayed here do not perfectly match those quoted in \cref{tab:fit_ezmocks} that are the $1\sigma$ credible interval.}
    \label{fig:error_evolution}
\end{figure}

Next, we quantify the dependence of the statistical error as a function of the range that it is used during the fit. This dependence is shown in \cref{fig:error_evolution} for the different tracers and the different weighting scheme. This range is limited by two factors:  
\begin{itemize}[noitemsep,topsep=1pt]
    \item $k_{\rm min}$: At large scales, the measurement is impacted either by a mismodeling of the geometrical effects or by a imperfect correction of the imaging systematics (see \cref{sec:blinded_analysis}). Due to the statistics, the very large-scales are not the most important ones as shown in the left panel of \cref{fig:error_evolution}, and we decide to use a conservative cut for our fiducial pipeline, avoiding any bias in our measurement: $k_{\rm min} = 0.003~h\textrm{Mpc}^{-1}$. Note that the geometrical effects are still handled up to $k_{\rm min} = 0.001~h\textrm{Mpc}^{-1}$, the limiting factor here is the efficiency of the imaging systematic mitigation.
    \item $k_{\rm max}$: At small scales, the simple description that we used, see \cref{eqn:pk_theo}, cannot deal with the non-linearity. As shown in the right panel of \cref{fig:error_evolution}, there is not much to gain by increasing $k_{\rm max}$ to constrain $\fnl$. However, we still need some small-scale information to obtain a small uncertainty on $b_1$. Consequently, we choose to include the scales where the modes are mostly linear: $k_{\rm max} = 0.08~h\textrm{Mpc}^{-1}$.
\end{itemize}
Hence, unless mentioned, all the fits in the following use
\begin{equation*} 
    \left\{
        \begin{array}{ll}
            \ell_0: 0.003 < k < 0.08 \text{ with } \Delta k = 0.001 \quad h\mathrm{Mpc}^{-1} \\
            \ell_2: 0.003 < k < 0.08 \text{ with } \Delta k = 0.002 \quad h\mathrm{Mpc}^{-1}
        \end{array}
    \right..
\end{equation*}

\subsubsection{DR1 validation and Y5 forescast}
\begin{figure}
    \centering
    \begin{subfigure}[t]{0.48\textwidth}
        \centering
        \includegraphics[scale=0.97]{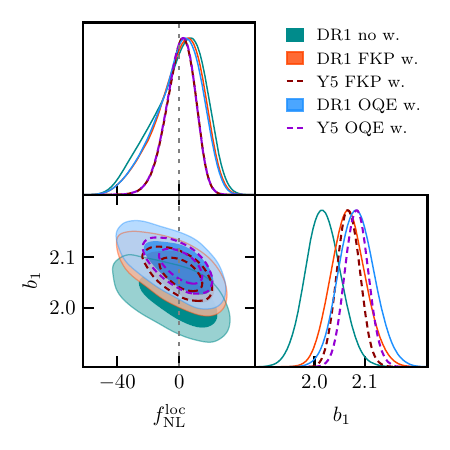}
        \caption{LRG with $b_{\Phi}(b_1) = 2 \delta_c (b_1 - 1)$.}
        \label{fig:fit_ezmocks_lrg_fnl}
    \end{subfigure}
    \begin{subfigure}[t]{0.48\textwidth}
        \centering
        \includegraphics[scale=0.97]{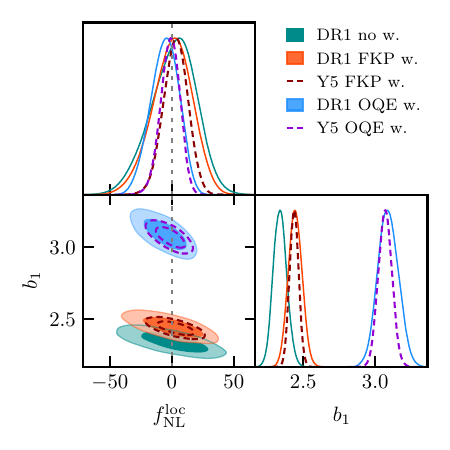}
        \caption{QSO with $b_{\Phi}(b_1) = 2 \delta_c (b_1 - 1.6)$.}
        \label{fig:fit_ezmocks_qso_fnl}
    \end{subfigure}
    \begin{subfigure}[t]{0.48\textwidth}
        \centering
        \includegraphics[scale=0.97]{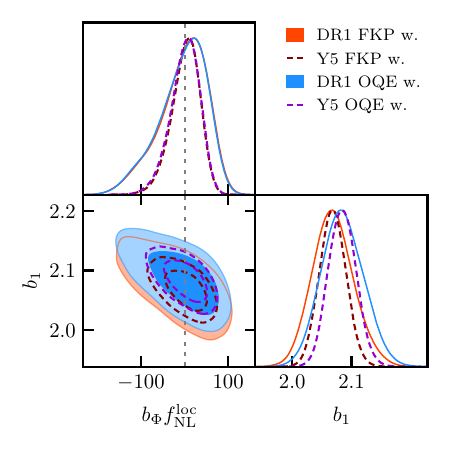}
        \caption{LRG with $b_{\Phi} \fnl$ as a free parameter.}
        \label{fig:fit_ezmocks_lrg_bphi_fnl}
    \end{subfigure}
    \begin{subfigure}[t]{0.48\textwidth}
        \centering
        \includegraphics[scale=0.97]{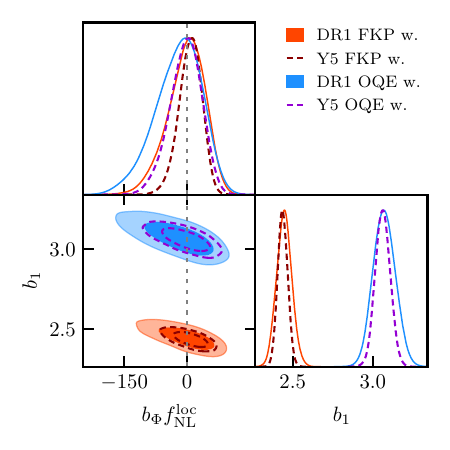}
        \caption{QSO with $b_{\Phi} \fnl$ as a free parameter.}
        \label{fig:fit_ezmocks_qso_bphi_fnl}
    \end{subfigure}
    \caption{Posteriors in the $(\fnl, b_1)$ plane with $b_{\Phi}(b_1)$ derived via \cref{eqn:b_phi_with_p} in (a) / (b) and in the $(b_{\Phi} \fnl, b_1)$ plane in (c) / (d). The parameters $s_{n,0}, \Sigma_s$ are free during the MCMC but are not shown for visibility. We fit the mean the 1000 realizations and use them to derive the covariance matrix. Filled contours are for the DR1 mocks while the dark dashed contours are for the Y5 mocks. Red colours are for the FKP weights, while the blue ones are for the OQE weights. For comparison, the case without any weighting is also included for the DR1 posteriors in green. The corresponding maximum-a-posteriori (MAP) values are displayed in \cref{tab:fit_ezmocks}.}
    \label{fig:fit_ezmocks}
\end{figure}

Using the EZmocks as a null test, we can test our model with the fiducial scale range. The posteriors for the different tracers and weighting schemes are displayed in \cref{fig:fit_ezmocks} and the best fit values in \cref{tab:fit_ezmocks}. The second row in \cref{fig:fit_ezmocks} gives the posterior when considering $b_{\Phi} \fnl$ as a single parameter\footnote{While fitting $b_{\Phi} \fnl$ with the OQE weights, we fit the quadrupole at the effective redshift computed for the monopole because we do not know the redshift evolution of $b_{\Phi} \fnl$. The gain including the quadrupole is very small and this choice has a negligible impact.} without assuming any value for $b_{\Phi}$ and gives the overall sensitivity of the two tracers for the detection of the presence of primordial non-Gaussianity. For each configuration, we also give the measurement performed with mocks describing the DESI Y5 data.

In all the configurations, we recover $\fnl = 0$ well within $1\sigma$ validating the description of the geometrical effects described in \cref{sec:window_function}. The difference between the value of the linear bias $b_1$ and the different weighting configurations is from the difference of effective redshift $z_{\rm eff}$, see \cref{tab:effective_redshift}. Although, there is a small discrepancy for the LRG DR1 EZmocks, it seems this is only a statistical fluctuation since it disappears when considering the Y5 footprint of the same realization. The systematic error contribution is discussed in \cref{sec:systematic_errors}. In addition, for the QSO case, one can notice a discrepancy between the value fitted with FKP and OQE weights; it will be discussed in \cref{sec:fkp_vs_oqe}.

As illustrated in \cref{fig:fit_ezmocks_qso_fnl}, the use of OQE weights helps to increase the value of $b_1$ since we are fitting the data with a higher effective redshift, see \cref{tab:effective_redshift}, which improves the constraint on $\fnl$. 

The constraints on $b_{\Phi} \fnl$, given in \cref{tab:fit_ezmocks}, are better with the FKP weights compared to the OQE weights. This is not surprising, as the effective redshift, and due to the redshift evolution of $b_1$, the value of $b_\Phi$, is higher when using the OQE weights. This same reasoning explains why the errors on $\fnl$ are smaller with OQE weights compared to FKP weights. Note that, without assuming any value for $b_{\Phi}$, we cannot obtain a competitive constraint with respect to Planck18 \cite{Akrami2020}.


\begin{table}
    \centering
    \caption[]{Results of the fit using the mean of the power spectrum over 1000 realizations with the corresponding covariance matrix for the LRGs and QSOs and the different weighting scheme. Central values are the best fit values from the \texttt{minuit} minimization while the errors are the $1\sigma$ credible interval from the chains that are displayed in \cref{fig:fit_ezmocks}. The second part of the Table gives the best fits using $b_{\Phi}\fnl$ as single parameter avoiding any assumption on the value of $b_{\Phi}$. The systematic error contribution is discussed in \cref{sec:systematic_errors}.}
    \label{tab:fit_ezmocks}
    \begin{tabular}{llcccc} 
        \toprule
           &           & $f_{\mathrm{NL}}^{\mathrm{loc}}$ & $b_1$ & $s_{n, 0}$ & $\Sigma_{s}$                                       \\ \midrule \vspace{1mm}
     LRG   & DR1       & $6_{-11}^{+21.}$     & $2.010_{-0.036}^{+0.033}$ & $0.048_{-0.065}^{+0.068}$  & $4.45_{-0.44}^{+0.49}$     \\ \vspace{1mm}
           & DR1 (FKP) & $5_{-11}^{+18}$      & $2.061_{-0.037}^{+0.032}$ & $0.042_{-0.061}^{+0.064}$  & $4.35_{-0.37}^{+0.52}$     \\ \vspace{1mm}
           & DR1 (OQE) & $5_{-11}^{+18}$      & $2.077_{-0.035}^{+0.035}$ & $0.040_{-0.058}^{+0.069}$  & $4.32_{-0.42}^{+0.49}$     \\ \midrule \vspace{1mm}
           & Y5 (FKP)  & $2.0_{-8.4}^{+10.7}$ & $2.068_{-0.023}^{+0.022}$ & $0.031_{-0.045}^{+0.042}$  & $4.56_{-0.28}^{+0.31}$     \\ \vspace{1mm}
           & Y5 (OQE)  & $2.2_{-7.6}^{+10.2}$ & $2.083_{-0.023}^{+0.023}$ & $0.029_{-0.044}^{+0.044}$  & $4.52_{-0.28}^{+0.33}$     \\ \midrule \vspace{1mm}   
    QSO    & DR1       & $3_{-16}^{+20}$      & $2.339_{-0.051}^{+0.042}$ & $-0.076_{-0.048}^{+0.057}$ & $3.15_{-0.58}^{+1.19}$     \\ \vspace{1mm}
           & DR1 (FKP) & $3_{-13}^{+18}$      & $2.442_{-0.045}^{+0.051}$ & $-0.010_{-0.051}^{+0.056}$ & $2.69_{-0.81}^{+1.22}$     \\ \vspace{1mm}
           & DR1 (OQE) & $-3_{-10}^{+13}$     & $3.085_{-0.080}^{+0.064}$ & $-0.045_{-0.074}^{+0.075}$ & $0_{-0.93}^{+0.25}$        \\ \midrule \vspace{1mm}
           & Y5 (FKP)  & $3.4_{-9.6}^{+9.9}$  & $2.437_{-0.033}^{+0.030}$ & $-0.292_{-0.037}^{+0.035}$ & $3.43_{-0.39}^{+0.59}$     \\ \vspace{1mm}
           & Y5 (OQE)  & $-2.4_{-7.8}^{+7.8}$ & $3.086_{-0.049}^{+0.045}$ & $-0.080_{-0.050}^{+0.053}$ & $0_{-0.67}^{+0.19}$        \\ \bottomrule \toprule 
           &           & $b_{\Phi}f_{\mathrm{NL}}^{\mathrm{loc}}$ & $b_1$ & $s_{n, 0}$ & $\Sigma_{s}$                               \\ \midrule \vspace{1mm}
    LRG    & DR1 (FKP) & ${21}_{-38}^{+68}$ & ${2.061}_{-0.035}^{+0.034}$ & ${0.041}_{-0.060}^{+0.066}$  & ${4.35}_{-0.43}^{+0.46}$ \\ \vspace{1mm}
           & DR1 (OQE) & ${19}_{-41}^{+65}$ & ${2.077}_{-0.035}^{+0.036}$ & ${0.039}_{-0.065}^{+0.065}$  & ${4.32}_{-0.37}^{+0.54}$ \\ \midrule \vspace{1mm}
           & Y5 (FKP)  & ${9}_{-26}^{+37}$  & ${2.066}_{-0.021}^{+0.023}$ & ${0.034}_{-0.041}^{+0.047}$  & ${4.55}_{-0.27}^{+0.32}$ \\ \vspace{1mm}
           & Y5 (OQE)  & ${9}_{-24}^{+42}$  & ${2.082}_{-0.024}^{+0.023}$ & ${0.032}_{-0.044}^{+0.046}$  & ${4.50}_{-0.28}^{+0.33}$ \\ \midrule \vspace{1mm}
    QSO    & DR1 (FKP) & ${7}_{-34}^{+54}$  & ${2.442}_{-0.047}^{+0.047}$ & ${-0.010}_{-0.048}^{+0.056}$ & ${2.69}_{-0.76}^{+1.28}$ \\ \vspace{1mm}
           & DR1 (OQE) & ${-1}_{-48}^{+62}$ & ${3.058}_{-0.075}^{+0.064}$ & ${-0.013}_{-0.068}^{+0.077}$ & ${0.0}_{-1.26}^{+0.39}$  \\ \midrule \vspace{1mm}
           & Y5 (FKP)  & ${10}_{-25}^{+29}$ & ${2.437}_{-0.031}^{+0.030}$ & ${-0.292}_{-0.035}^{+0.036}$ & ${3.44}_{-0.39}^{+0.61}$ \\ \vspace{1mm}
           & Y5 (OQE)  & ${4}_{-35}^{+41}$  & ${3.055}_{-0.050}^{+0.043}$ & ${-0.045}_{-0.051}^{+0.049}$ & ${0.0}_{-1.04}^{+0.34}$  \\ \bottomrule
    \end{tabular}
\end{table}

\begin{figure}
    \centering
    \includegraphics[scale=1]{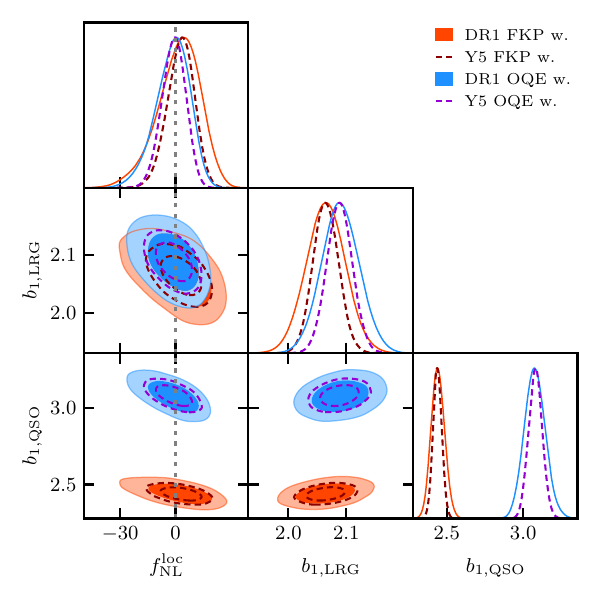}
    \caption{Posteriors in the $(\fnl, b_{1, \textrm{LRG}}, b_{1, \textrm{QSO}})$ plane of the combine fit of the mean of power spectrum $(\ell=0, \ell=2)$ over the 1000 realizations of the LRGs and QSOs. $b_{\Phi}(b_1, p)$ is derived via \cref{eqn:b_phi_with_p}, where $p=1$ for the LRGs and $p=1.6$ for the QSOs. The other parameters are allowed to vary but not shown here. The corresponding MAP values are displayed in \cref{tab:fit_fnl_EZmocks_LRG+QSO}.}
    \label{fig:fit_EZmocks_LRG+QSO}
\end{figure}

Fixing the value of $b_{\Phi}$ via the universal mass relation breaks the degeneracy between $b_{\Phi}$ and $\fnl$ such that the different tracers can be combined to increase the statistical accuracy. Note that this is the first time this is done for $\fnl$ with 3D galaxy clustering. The posterior combining the two tracers are displayed in \cref{fig:fit_EZmocks_LRG+QSO} and the best fit values are given in \cref{tab:fit_fnl_EZmocks_LRG+QSO}. Note that we assume the two tracers are independent and neglect the cross-covariance between them. The gain combining the LRGs and the QSOs is about $20\%$ in the statistical errors compared to the QSOs only, and this motivates the inclusion of the LRGs in this analysis.

Based on our mocks (\cref{tab:fit_fnl_EZmocks_LRG+QSO}), we forecast that the DESI Y5 sample will enhance the constraint on $\fnl$ by approximately $40\%$ compared to the current DR1 sample, achieving $\sigma(\fnl) \sim 6.5$ in the current setup and with the combined LRG and QSO samples.

\begin{table}
    \centering
    \caption[]{Results of the fit of the combine mean of power spectrum over 1000 realizations of the LRGs and QSOs. Central values are the best fit values from the \texttt{minuit} minimization while the errors are the $1\sigma$ credible interval from the chains that are displayed in \cref{fig:fit_EZmocks_LRG+QSO}. Combining the LRGs and the QSOs leads to a direct statistical gain about 20\% compared to the QSOs only.}
    \label{tab:fit_fnl_EZmocks_LRG+QSO}
    \resizebox{0.99\textwidth}{!}{%
    \begin{tabular}{lccccccc}
        \toprule
        & $f_{\mathrm{NL}}^{\mathrm{loc}}$ & $b_{1, \mathrm{QSO}}$ &  $s_{n, 0, \mathrm{QSO}}$ & $\Sigma_{s, \mathrm{QSO}}$  & $b_{1, \mathrm{LRG}}$ &  $s_{n, 0, \mathrm{LRG}}$ & $\Sigma_{s, \mathrm{LRG}}$  \\
        \midrule \vspace{1mm}
        DR1 (FKP) & ${5}_{-9.8}^{+13.0}$ & ${2.437}_{-0.045}^{+0.043}$ & ${-0.006}_{-0.049}^{+0.056}$ & ${2.63}_{-0.79}^{+1.36}$ & ${2.062}_{-0.035}^{+0.032}$ & ${0.039}_{-0.062}^{+0.066}$ & ${4.35}_{-0.42}^{+0.54}$ \\ \vspace{1mm}
        DR1 (OQE) & ${1.1}_{-7.9}^{+10.5}$  & ${3.068}_{-0.072}^{+0.064}$ & ${-0.031}_{-0.072}^{+0.078}$  & ${0.0}_{-0.95}^{+0.26}$ & ${2.087}_{-0.034}^{+0.032}$ & ${0.022}_{-0.065}^{+0.062}$ & ${4.35}_{-0.44}^{+0.52}$ \\ 
        \midrule \vspace{1mm}
        Y5 (FKP) & ${3.6}_{-6.8}^{+7.6}$ & ${2.439}_{-0.029}^{+0.028}$ & ${-0.293}_{-0.036}^{+0.035}$ & ${3.46}_{-0.46}^{+0.58}$ & ${2.064}_{-0.022}^{+0.022}$ & ${0.038}_{-0.044}^{+0.045}$ & ${4.55}_{-0.31}^{+0.33}$ \\  \vspace{1mm}
        Y5 (OQE) & ${-0.1}_{-6.2}^{+6.8}$ & ${3.078}_{-0.045}^{+0.044}$ & ${-0.073}_{-0.049}^{+0.054}$ & ${0.0}_{-0.70}^{+0.18}$ & ${2.088}_{-0.022}^{+0.022}$ & ${0.023}_{-0.046}^{+0.043}$ & ${4.59}_{-0.29}^{+0.34}$ \\ 
        \bottomrule
    \end{tabular}%
    }
\end{table}

Due to the shape of the redshift distribution, the OQE weights have a negligible impact on the constraint of $\fnl$ for the LRGs such that we do not use them in the following, and we only give the result for the use of FKP weights. 

\subsection{Discrepancy between FKP and OQE weights} \label{sec:fkp_vs_oqe}
As reported in \cref{tab:fit_ezmocks}, in the case of the QSO there is a discrepancy between the measured value of $\fnl$ between the use of the FKP ($\fnl \simeq 3$) or OQE ($\fnl \simeq -3$) weighting schemes. This discrepancy is statistically significant because we are fitting the mean over 1000 realizations, which reduces the expected statistical uncertainty by a factor $\sqrt{1000} \sim 31$. For the LRGs, OQE weights have a minor impact such that the discrepancy does not exist. Although shown in the following, we do not discuss it and only focus in the QSO case.

\begin{figure}
    \centering
    \includegraphics[scale=1]{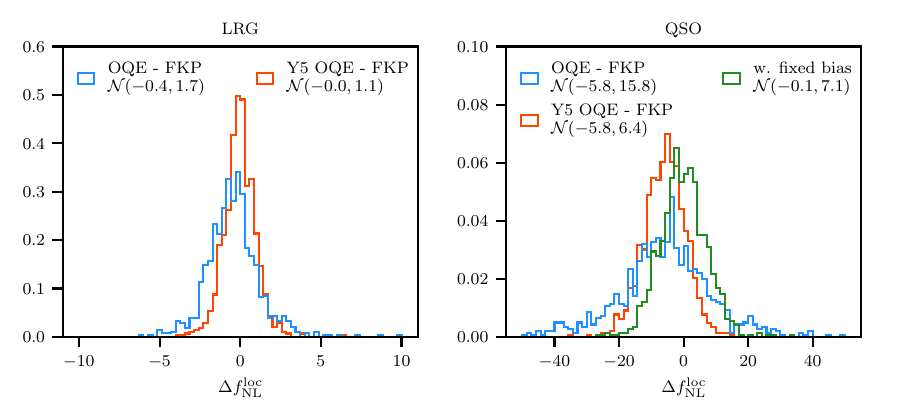}
    \caption{Normalised distribution from individual fit of the 1000 realizations of the difference between the best-fit value of $\fnl$ obtained with FKP or OQE weights. LRGs are on the left and QSOs on the right. Blue is for DR1, red for Y5 mocks and green is for DR1 but with the bias fixed during the fit. The mean and the standard deviation for each histogram are given in the legend.}
    \label{fig:oqe_vs_fkp}
\end{figure}

To investigate this effect, we fit individually the 1000 realizations with the different weighting schemes. First, we check that the standard deviation from the best fit value of $\fnl$ on the 1000 EZmocks is compatible with the errors given in \cref{tab:fit_ezmocks}. Then, the normalised distribution of the difference between the best fit value of $\fnl$ with the different weights are shown in \cref{fig:oqe_vs_fkp}, where the mean and the standard deviation of each distribution are displayed in the legend. 

The shift observed between the FKP and the OQE weights in \cref{tab:fit_ezmocks} ($-6$) is consistent with the mean ($-5.8$) of the distribution displayed in blue in \cref{fig:oqe_vs_fkp}. The shift does not disappear by increasing the data size (blue versus red histogram), however, the standard deviation becomes lower, meaning that the shift seems to be a real bias between the two weighting schemes.

The shift between the measurement of $\fnl$ with the two weighting schemes is lower than the statistical errors, and it is still the case for this first DESI data release ($\sim 0.5 \sigma$). However, as shown with the forecast for the Y5 data, this will not be the case with the increase of the data size in the upcoming DESI release. Thus, additional study will be required to avoid biaising the measurement. To investigate, we perform the fit with the linear bias $b_1$ fixed for the Y1 mocks and the discrepancy vanished as shown by the green histogram in \cref{fig:oqe_vs_fkp}. Hence, a better knowledge on $b_1$ could help to obtain an unbiased measurement of $\fnl$ with the OQE weighting scheme. This could be achieved by increasing $k_{\rm max}$, which, in turn, would require the use of much more complex model than \cref{eqn:pk_theo} as in \cite{DESI2024V}. We leave this analysis and improvement for future work.

This discrepancy appears also later when measuring $\fnl$ from the data. However, the majority of this discrepancy is due to a residual systematics in the lower redshift range of the QSO sample that is under-weighted by the OQE weighting scheme. The difference is $\Delta \fnl \sim 20$, so that the use of OQE weights is necessary to have an unbiased measurement, see \cref{sec:unblinded_analysis}. 


\subsection{Radial Integral Constraint} \label{sec:ric}
First, the \emph{global integral constraint} (GIC) \cite{DeMattia2019} is described in \cref{sec:gic} and we show that it can be neglected in this analysis.

\begin{figure}
    \centering
    \begin{subfigure}{0.48\textwidth}
        \centering
        \includegraphics[scale=0.98]{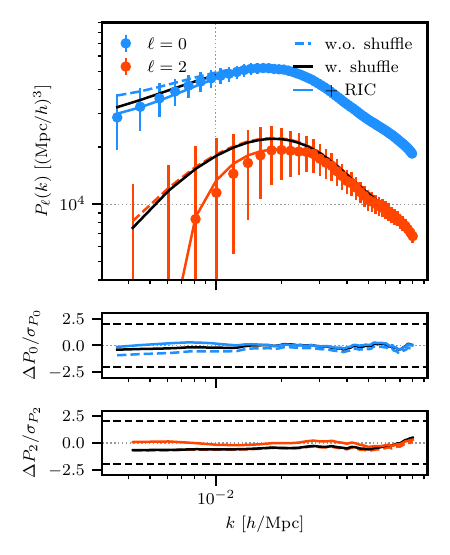}
        \caption{LRG ($0.4 < z < 1.1$) with FKP.}
    \end{subfigure}
    \begin{subfigure}{0.48\textwidth}
        \centering
        \includegraphics[scale=0.98]{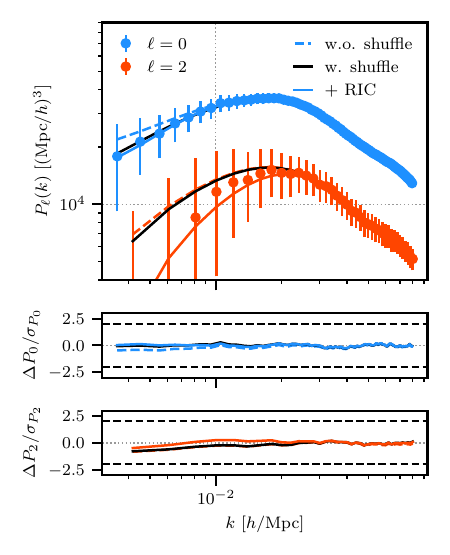}
        \caption{QSO ($0.8 < z < 3.1$) with FKP.}
    \end{subfigure}
    \caption{Multipoles of the mean power spectrum (dotted points), computed with FKP weights, over 100 realizations describing either the DESI DR1 LRGs (left) or QSOs (right) with the shuffling method applied. The errors are from the covariance matrix for the DR1 analysis. The colour dashed lines are the best fits from the realizations without the shuffling method. The black lines are the best fit from the realizations with the shuffling method but without taking into account the radial integral constraint contribution while the solid colour lines are the best fit taking into account this contribution. On the monopole, this contribution can be reproduced by decreasing the value of $\fnl$, thus needs to be corrected to measure this parameter correctly. Note that the quadrupole ($\ell=2$) has almost no constraining power on $\fnl$ such that even with the important lack of power at large-scales on the \emph{shuffled} quadrupole, this does not impact the value of $\fnl$.}
    \label{fig:ric}
\end{figure}

Up until now, the randoms of the EZmocks were generated in a box in three dimensions such that they already have their proper redshifts. We simply sampled them to match the desired redshift distribution. However, as explained in \cref{sec:observational_weights}, the redshift of the randoms are drawn directly from the data catalog using the so-called \emph{shuffling} method \cite{Ross2012}. By imprinting the data redshifts into the randoms, radial modes in the measured power spectrum are nulled leading to the so-called \emph{Radial Integral Constraint} (RIC) \cite{DeMattia2019}. To quantify the contribution of the effect, we apply the shuffling method on the first 100 EZmocks that we use. As illustrated in \cref{fig:ric} and in \cref{tab:ric}, the use of the shuffling method without any correction biases the measurement of $\fnl$ by $1\sigma$ of the statistical uncertainty.

In contrast to \cite{Mueller2021,Cagliari2023} which implement an additive correction to account for the radial constraint by taking simply the difference of the power spectrum between the mocks with and without  shuffling, we instead provide a multiplicative correction\footnote{Multiplicative because the convolved power spectrum is obtained by multiplying the window matrix to the theoretical prediction, as described in \cref{eqn:convolved_prediction}, and thus, any correction to the window matrix is propagated in a \emph{multiplicative} way in the convolved power spectrum.} by modifying the window function: 
\begin{equation}
    \mathcal{W} \rightarrow \mathcal{W} - \mathcal{W}^{\rm RIC}.
\end{equation}
Hence, the correction does not depend on the value of the power spectrum on which it is estimated. 

As described in Section~2.2 of \cite{DeMattia2019}, the contribution of the RIC has a similar shape as the global one with additional anisotropy and scale dependence coming in compared to \cref{eqn:wm_gic} such that it may be reasonable to look for:
\begin{equation} \label{eqn:wm_ric}
\left(\mathcal{W}^{\mathrm{RIC}}_{\ell \ell^\prime}\right)_{ij} = \dfrac{\left(\mathcal{W}_{\ell p} \right)_{im}}{\left(\mathcal{W}_{0 0} \right)_{00}} \left(f_{pq}\right)_{mn} \left(\mathcal{W}_{q \ell^\prime} \right)_{nj},
\end{equation}
with the summation runs over $p, q, n, m$. $\left(f_{pq}\right)_{mn}$ decreases rapidly with increasing $p, q$, e.g.:
\begin{equation} \label{eqn:param_f}
\left(f_{pq}\right)_{mn} = A_{pq} \, e^{-(k_n^{2} + k_m^{2}) / \sigma_{pq}^{2}},
\end{equation}
where $A_{pq}$ and $\sigma_{pq}$ are $2 \times (3 \times 3)$ unknown coefficients. Note that under this parametrization, one can retrieve the GIC contribution given in \cref{eqn:wm_gic} by setting $\left(f_{00}\right)_{00} = 1$ and 0 for the others. 

The coefficients $A_{pq}$ and $\sigma_{pq}$ can be estimated with the set of EZmocks with and without the shuffling method. The convolved power spectrum of this EZmocks with the shuffling method can be written as 
\begin{equation}
    \left(\hat{P}_{\ell}^{\rm shu}\right)_i = \left(\mathcal{W}_{\ell \ell^\prime} \right)_{ij} \left(\hat{P}_{\ell\prime}^{\rm box} \right)_j - \left(\mathcal{W}_{\ell \ell^\prime}^{\rm RIC} \right)_{ij} \left(\hat{P}_{\ell\prime}^{\rm box} \right)_j,
\end{equation}
where $\hat{P}_{\ell\prime}^{\rm box}$ is the power spectrum from the box used to build the cutsky mock \ie corresponds to one realization of the underlying $P_{\ell\prime}^{\rm theo}$ used to generate the mocks. The first term of the RHS is the observed power spectrum measured without the shuffling: 
\begin{equation}
\left(\hat{P}_{\ell}^{\mathrm{no \; shu}}\right)_{i}  = \left(\mathcal{W}_{\ell \ell^\prime} \right)_{ij} \left(\hat{P}_{\ell\prime}^{\rm box} \right)_j.
\end{equation}

Due to the large variance during the subsampling to go from the box to the cutsky, we do not want to compare the power spectrum from the box and the one from the cutsky. Fortunately, one can extract the window matrix from \cref{eqn:wm_ric} such that the RIC contribution can be re-written as 
\begin{equation}
    \left(\mathcal{W}_{\ell \ell^\prime}^{\rm RIC} \right)_{ij} \left(\hat{P}_{\ell\prime}^{\rm box} \right)_j = \left(\mathcal{W}_{\ell p} \right)_{im} \left(f_{pq}\right)_{mn} \left(\hat{P}_{q}^{\mathrm{no \; shu}}\right)_{n}.
\end{equation}
Note that $\left(\mathcal{W}_{0 0} \right)_{00}$ is a constant, and to simplify, we renormalise $\left(f_{pq}\right)_{mn}$ such that $\left(f_{pq}\right)_{mn} \rightarrow \left(f_{pq}\right)_{mn} / \left(\mathcal{W}_{0 0} \right)_{00}$ without changing the result.

Finally, the coefficients $A_{pq}$ and $\sigma_{pq}$ in $\left(f_{pq}\right)_{mn}$ can be estimated by minimising the sum over 100 independent realization of a standard $\chi^2$ defined for each realization by
\begin{equation}
\chi^2 = \left(\Delta_{\ell}\right)_{i}^{T} \left(C^{-1}_{\ell\ell^\prime}\right)_{ij} \left(\Delta_{\ell^{\prime}}\right)_{j},
\end{equation}
where $\left(C_{\ell\ell^\prime}\right)_{ij}$ is the covariance matrix and $\left(\Delta_{\ell}\right)_{i}$ is given by
\begin{equation}\label{eqn:ric_delta}
    \left(\Delta_{\ell}\right)_{i} = \left(\hat{P}_{\ell}^{\mathrm{shu}}\right)_{i} -  \left( \hat{P}_{\ell}^{\mathrm{no \; shu}}\right)_{i} + \left(\mathcal{W}_{\ell p} \right)_{im} \left(f_{pq}\right)_{mn} \left(\hat{P}_{q}^{\mathrm{no \; shu}}\right)_{n}.
\end{equation}
The minimization is performed with \texttt{iminuit} and is fitted independently for the FKP or OQE weights and for the different tracers. We use $\ell=0, 2, 4$ for FKP weights and only $\ell=0, 2$\footnote{We do not have computed the window matrix for $\ell=4$ in the OQE case. The contribution obtained from the minimization of $\ell=4$ in the FKP case could be neglected as well.} for OQE weights with $0.003~h\text{Mpc}^{-1}< k < 0.01~h\text{Mpc}^{-1}$. As a verification, the minimization was also performed only with the first 50 mocks and tested on the mean of the 50 others, and similar result were obtained.

Note that by definition the GIC is included in the RIC \cite{DeMattia2019}. However, in \cref{eqn:ric_delta}, we used only measured power spectra such that the GIC vanished in $\hat{P}_{\ell}^{\mathrm{shu}} -  \hat{P}_{\ell}^{\mathrm{no \; shu}}$ and cannot be modelled with this method. Fortunately, we show in \cref{sec:gic} that GIC is negligible for our analysis.

\Cref{fig:ric} shows the mean power spectrum over the 100 realizations for the LRGs and the QSOs using the shuffling method. The dashed lines are the best fits to the mean power spectrum without the shuffling method, illustrating the radial integral constraint contribution to both the monopole and the quadrupole. The contribution to the monopole can be easily reproduced by decreasing the value of $\fnl$, while the suppression of the power at large-scales in the quadrupole cannot. Adding the RIC correction enables us to measure $\fnl$ as shown in \cref{tab:ric} which compares the result of the best fit with and without the RIC correction to the one without the shuffling method. Not introducing this correction would bias $\fnl$ by $2/3 \sigma$. 

To validate this multiplicative correction, we also test the RIC correction on the mean of 30 mocks with a different power spectrum than the ones used to estimate this correction. For this reason, we applied the blinding procedure described in \cite{Chaussidon2024} with $f_{\rm NL}^{\rm blind} = 20$. As shown in the last row of \cref{tab:ric}, the correction performs well even if the shape of the power spectrum is different, validating the multiplicative correction proposed here\footnote{We also tested for this test the standard additive correction that is the option used in \cite{Mueller2021,Cagliari2023} and found consistent results.}. 


Finally, the covariance obtained from 100 realizations with the shuffling method is very similar to the one obtained from 100 realizations without it. Thus, in what follows, we always use the covariance estimated from 1000 realizations without the shuffling method as described in \cref{sec:covariance_matrix}. 

\begin{table} 
  \centering
  \caption[]{Results of the fits using the DR1 covariance matrix on the mean of the power spectrum with FKP weights, with and without the shuffling method and the RIC contribution, over 100 realizations for the LRGs and QSOs and over 30 realizations for the QSOs with the blinding applied ($f_{\rm NL}^{\rm blind} = 20$). The central values are best fit values from the \texttt{minuit} minimization while the errors are the $1\sigma$ credible intervals from the chains. Note that the RIC correction for the bottom row is computed from the mocks without the blinding (\ie the correction is the same for the second and bottom row), validating our multiplicative correction. The systematic error contribution is discussed in \cref{sec:systematic_errors}.} 
  \label{tab:ric}
    \begin{tabular}{lcccc}
    \toprule
        & parameter & no shuffle & shuffle & shuffle + RIC \\ \midrule \vspace{1mm}
    DR1 LRG & $\fnl$    & \cellcolor{green!30} $ 6_{-12}^{+17} $ & \cellcolor{red!30} $ -4_{-13}^{+16} $ & \cellcolor{green!30} $ 7_{-12}^{+17}$ \\ \vspace{1mm}
        & $b_1$        & $ 2.060_{-0.037}^{+0.032} $ & $ 2.069_{-0.036}^{+0.033}$ & $ 2.055_{-0.035}^{+0.031}$ \\ \vspace{1mm}
        & $s_{n,0}$    & $ 0.046_{-0.063}^{+0.062} $ & $ 0.058_{-0.070}^{+0.058}$ & $ 0.052_{-0.058}^{+0.064}$ \\ \vspace{1mm}
        & $\Sigma_s$   & $ 4.32_{-0.37}^{+0.52} $ & $ 4.57_{-0.38}^{+0.47}$ & $ 4.42_{-0.39}^{+0.48}$ \\ \midrule \vspace{1mm}
    DR1 QSO & $\fnl$ & \cellcolor{green!30} $ 3_{-15}^{+16} $ & \cellcolor{red!30} $ -7_{-15}^{+19} $ & \cellcolor{green!30} $ 6_{-16}^{+14} $ \\ \vspace{1mm}
        & $b_1$        & $ 2.438_{-0.049}^{+0.046} $ & $ 2.451_{-0.054}^{+0.048} $ & $ 2.431_{-0.045}^{+0.047} $ \\ \vspace{1mm}
        & $s_{n,0}$    & $ -0.002_{-0.054}^{+0.052} $ & $ -0.001_{-0.052}^{+0.058} $ & $ 0.004_{-0.050}^{+0.053} $ \\ \vspace{1mm}
        & $\Sigma_s$   & $ 2.67_{-0.78}^{+1.25} $ & $ 2.98_{-0.67}^{+1.20} $ & $ 2.87_{-0.77}^{+1.19} $ \\ \midrule \vspace{1mm}
    DR1 QSO with blinding & $\fnl - f_{\rm NL}^{\rm blind}$ & \cellcolor{green!30} $ 8_{-12}^{+13} $ & \cellcolor{red!30} $ 1_{-12}^{+16} $ & \cellcolor{green!30} $ 9_{-12}^{+14} $ \\ \vspace{1mm}
    (RIC from DR1 QSO)   & $b_1$        & $ 2.406_{-0.043}^{+0.042} $ & $ 2.410_{-0.048}^{+0.043} $ & $ 2.403_{-0.046}^{+0.038} $ \\ \vspace{1mm}
        & $s_{n,0}$    & $ -0.011_{-0.048}^{+0.052} $ & $ 0.006_{-0.053}^{+0.051} $ & $ 0.000_{-0.049}^{+0.049} $  \\ \vspace{1mm}
        & $\Sigma_s$   & $ 2.72_{-0.74}^{+1.31} $ & $ 2.99_{-0.65}^{+1.24} $ & $ 2.95_{-0.65}^{+1.26} $ \\ 
    \bottomrule
    \end{tabular}
\end{table}

\section{Imaging systematics: weights validation} \label{sec:blinded_analysis}
Imaging systematic mitigation aims to correct for the spurious density fluctuations in the angular distribution of the objects from the fluctuation of the imaging quality and foreground across the photometric survey used for the target selection. These fluctuations are illustrated in \cref{fig:LRG_systematic_plots} and in \cref{fig:QSO_systematic_plots} that show the relative density of the number of objects as a function of different templates where the black lines are for the sample not corrected for these dependences.

These systematics represent the most significant source of contamination in measuring the large-scale modes of the power spectrum. Over the past decade \cite{Vogeley1998,Scranton2002,Ross2011,Huterer2013}, mitigating these effects has been a major focus in both galaxy clustering analyses from spectroscopic surveys, as in eBOSS \cite{Rezaie2019, Kong2020, Ross2020, Rezaie2021}, and from photometric surveys as in the Dark Energy Survey (DES) \cite{Rodriguez-Monroy2022, Mena-Fernandez2024}. In \cref{sec:wsys_description}, we present the methodology used in DESI and through this paper to compute the imaging systematic weights. Then, in \cref{sec:validation_aic}, we use EZmocks to test the impact of different imaging mitigation weighting schemes on $\fnl$, and compute the angular integral constraint contribution to correct for the use of these weights. Finally, in \cref{sec:blinded_data}, we analyse the blinded data and validate the fiducial mitigation method. 

\subsection{Mitigation of the dependence on the imaging quality of the target selection} \label{sec:wsys_description}
\subsubsection{Default configuration in DESI} \label{sec:default_config}
In DESI \cite{DESI2024II}, we follow the commonly applied method based on \emph{template fitting} that was developed in the last major surveys \cite{Leistedt2013,Leistedt2016,Rezaie2019,Kitanidis2020,Weaverdyck2021}. This method provides a per-tracer correction weight $w_{\rm sys}$ calibrated from the observed variation of target density as a function of features that describe the imaging qualities. Recently, \cite{Rezaie2019, Rezaie2021} showed that this method could be improved by introducing some non-linearities between the template using supervised machine learning. Hence, in the following, the imaging weights $w_{\rm sys}$ are estimated at \texttt{HEALPix} level \cite{Gorski2005} using either a linear or a random forest-based regression with a k-fold training. Note that compared to the one in \cite{DESI2024II}, the linear regression here does not fit the data to binned statistics but rather the fluctuation at \texttt{HEALPix} level. All the weights are computed with \texttt{regressis}\footnote{\url{https://github.com/echaussidon/regressis}} as described in \cite{Chaussidon2022}.

Following \cite{DESI2024III, Kong2024} that assess the correlation between the target density and the different features, we consider only these 12 observational features\footnote{The creation of these feature maps is detailed in Appendix A of \cite{DESI2024III}. Some visualization of these maps can be found in Fig.4 of \cite{Chaussidon2022}.}: 
\begin{itemize}[noitemsep,topsep=1pt]
    \item Stellar density $[\rm{deg}^{-2}]$ is the density of point sources from Gaia DR2 \cite{GaiaCollaboration2018} in the magnitude range: $12 < \textsc{\small{PHOT\_G\_MEAN\_MAG}} < 17$.
    \item HI $[\rm{cm}^{-2}]$ is the hydrogen column density from the Effelsberg-Bonn HI Survey (EBHIS) and the third revision of the Galactic All-Sky Survey \cite{BenBekhti2016}.
    \item E(B-V) diﬀ GR / E(B-V) diﬀ RZ $[\rm{mag}]$: is the difference between the SFD E(B-V) \cite{Schlegel1998} and the E(B-V) determined from DESI stars spectra \cite{Zhou2024}. This new method from DESI data produce two values one based on $g-r$ and the other on $r-z$. Note, we are not using the standard E(B-V) map alone since it is strongly correlated to the large scale structure of the Universe via the Cosmic Infrared Background \cite{Chiang2019}.  
    \item PSF Depth $[1/\rm{nanomaggies}^2]$ (in $r$, $g$, $z$, $W1$, $W2$) is the 5-sigma point-source magnitude depth\footnote{For a $5\sigma$ point source detection limit in band $x$, $5/ \sqrt{x}$ gives the PSF Depth as flux in nanomaggies and $-2.5 \left( \log_{10}(5/ \sqrt{x}) - 9 \right)$ gives the corresponding magnitude (see \url{https://www.legacysurvey.org/dr9/catalogs/}).}.	
    \item Galaxy Depth $[1/\rm{nanomaggies}^2]$ (in $r$, $g$, $z$) is an alternative to PSF Depth. It measures the 5-sigma galaxy\footnote{(0.45" exp, round)}-source magnitude depth. It is only used instead of the corresponding PSF Depth. 
    \item PSF Size $[\rm{arcsec}]$ (in $r$, $g$, $z$): Inverse-noise-weighted average of the full width at half maximum of the point spread function, also called the delivered image quality.
\end{itemize}

As in \cite{DESI2024III}, the default configuration for the LRG sample is to compute the imaging weights in three different redshift bins ($0.4<z<0.6$, $0.6<z<0.8$ and $0.8<z<1.1$) and on three independent photometric regions (North, South (NGC), South (SGC) + DES)  with the following features: 
\begin{itemize}[noitemsep,topsep=1pt]
    \item $0.4<z<0.6$: Stellar density, HI, PSF Size $r$, Gal Depth $z$ / $r$, PSF Depth $W1$,
    \item $0.6<z<0.8$: Stellar density, HI, PSF Size $r$, Gal Depth $z$ / $g$, PSF Depth $W1$,
    \item $0.8<z<1.1$: Stellar density, HI, PSF Size $r$ / $z$, Gal Depth $z$, PSF Depth $W1$.
\end{itemize}
The default configuration for the QSO sample is also to compute the weights in three redshift bins ($0.8<z<1.3$, $1.3<z<2.1$ and $2.1<z<3.1$) and in three photometric regions (North, South (NGC) + South (SGC), DES) of but considering the same features in each bin:
\begin{itemize}[noitemsep,topsep=1pt]
    \item $0.8<z<1.3$ / $1.3<z<2.1$ / $2.1<z<3.1$: Stellar density, HI, E(B-V) diﬀ GR / RZ, PSF Depth $r$ / $g$ / $z$ / $W1$ / $W2$, PSF Size $g$ / $r$ / $z$.
\end{itemize}
These redshift bins were designed to match the redshift ranges of the sample used for the BAO or RSD measurements, except for the additional split for the QSOs at $z=1.3$.

This additional split was motivated by the lack of QSOs with small redshift ($z < 1.3$) in the regions where the PSF Depth is higher, as noted in \cite{Chaussidon2023}. Indeed, QSOs that are sufficiently close to us are increasingly identified as extended sources in regions with high PSF depth, leading to their rejection during the target selection.  This effect is limited to the low-$z$ end of the QSO sample and is not apparent when using the broad redshift bin of $0.8 < z < 2.1$. However, it contributes to an excess of power on large scales in the power spectrum if it is not properly addressed as shown in \cref{sec:appendix_impact_1.3_cut_wsys}.

In our template-fitting methodology, we assume that a template is fixed across the redshift bin (but allowed to vary between different bins). Hence, we are not able to model any redshift dependence inside a redshift bin. Note that this could also be useful for LRGs as found in \cite{Kong2024} . A more detailed analysis, which we leave for the future, might want to allow the template weights to vary within the redshift bins of individual tracers. 
\subsubsection{Test with other configurations}
\label{sec:other_config_ezmocks}
To assess the efficiency of the imaging systematic mitigation, we test several modifications of the default configuration. In particular, for the LRGs, we alternately adopt:
\begin{itemize}[noitemsep,topsep=1pt]
    \item \texttt{Default}: Default configuration, as described in \cref{sec:default_config}, computed either with a random forest using either $N_{\rm side}=128$ or $256$ or with a Linear regression using $N_{\rm side}=256$.
    \item \texttt{PSF Depth}: \texttt{Default} with a linear regression using $N_{\rm side}=256$, where the Gal Depth features are switched with the PSF Depth features.
    \item \texttt{Same Feature Zbin}: Using all the features for each redshift bins: Stellar density, HI, E(B-V) diﬀ GR / RZ, PSF Size $r$ / $z$, Gal Depth $z$ / $r$, PSF Depth $W1$. We test the linear regression using either $N_{\rm side}=128$ or $256$.
    \item \texttt{With DES}: as \texttt{Default}, but the regression is performed independently in (North, South, DES) instead of (North, South (NGC), all the SGC) with a linear regression using $N_{\rm side}=256$.
    \item  \texttt{4 regions}: as \texttt{Default}, but the regression is performed independently in (North, South (NGC), South (SGC), DES) instead of (North, South (NGC), all the SGC) with a linear regression using $N_{\rm side}=256$.
\end{itemize} 
For the QSOs, we test:
\begin{itemize}[noitemsep,topsep=1pt]
    \item \texttt{Default}: Default configuration computed either with a random forest with $N_{\rm side}=256$ or with a Linear regression using either $N_{\rm side}=128$ or $256$.
    \item \texttt{No PSF Size}: \texttt{Default} with a linear regression using $N_{\rm side}=128$ where the PSF size features in the $g$, $r$, $z$ bands are all removed.
    \item \texttt{No PSF Depth}: \texttt{Default} with a random forest regression using $N_{\rm side}=128$ where the PSF depth features in the $g$, $r$, $z$, $W1$ and $W2$ bands are all removed.
\end{itemize} 

\begin{figure}
    \hspace*{-17mm}
    \includegraphics[width=1.2\textwidth]{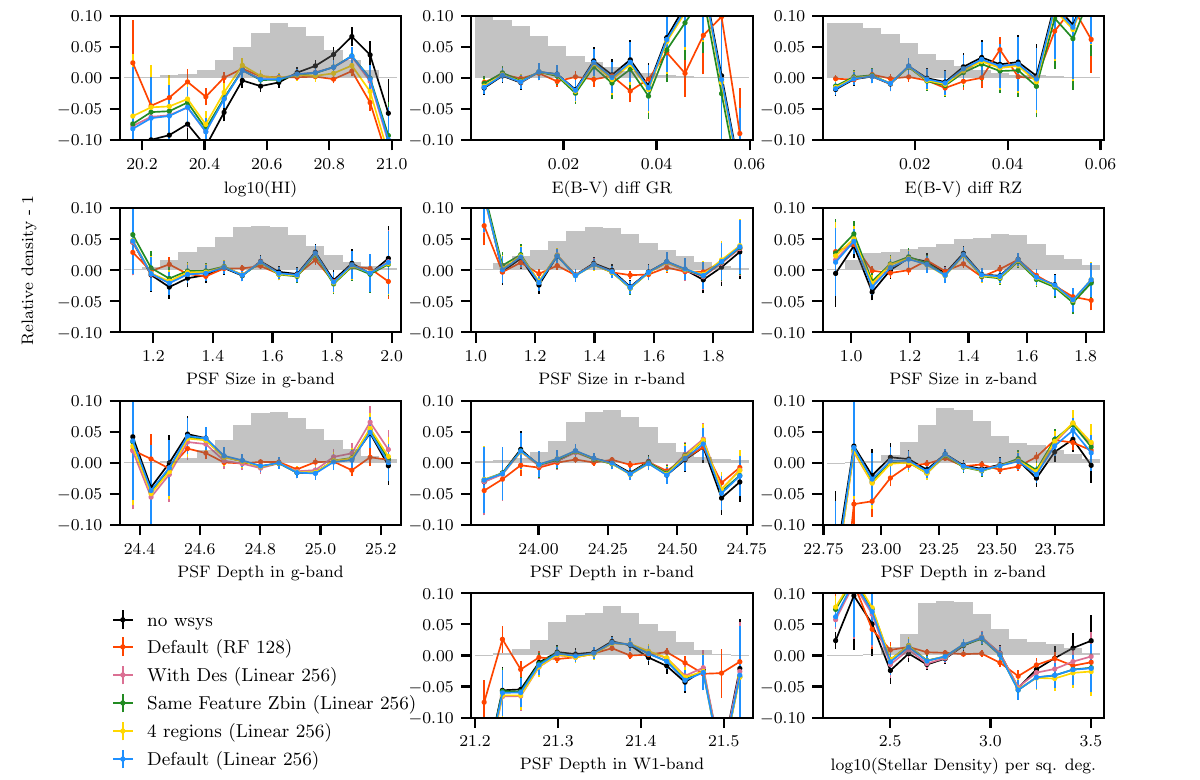}
    \centering
    \caption{Relative density centered around 0 as a function of the amplitude of several observational features of the LRGs (0.6 < z < 0.8) in the South (SGC) region. Back lines are without the imaging systematic correction while the different colours are for the different corrections (RF regression in red, Linear in blue/green). The histogram represents the fraction of objects in each bin for each observational feature and the error bars are the estimated standard deviation of the normalised density in each bin.}
    \label{fig:LRG_systematic_plots}
\end{figure}

\begin{figure}
    \hspace*{-17mm}
    \includegraphics[width=1.2\textwidth]{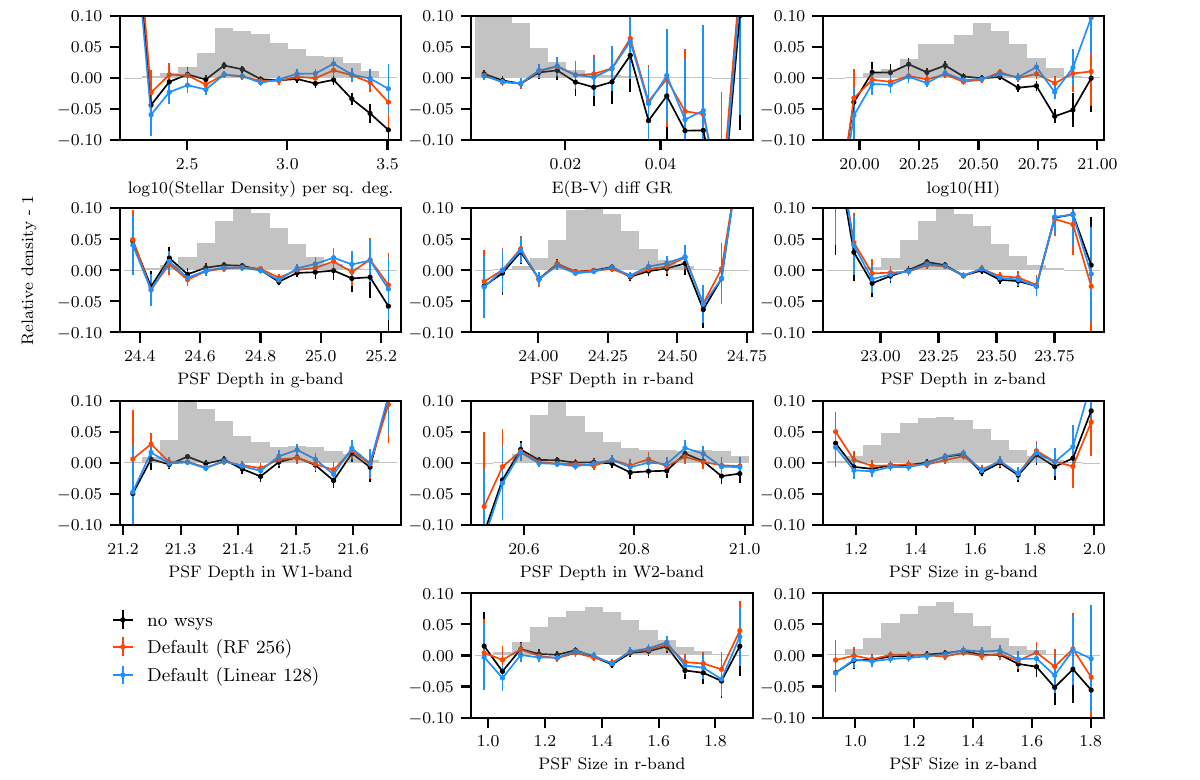}
    \centering
    \caption{Same as \cref{fig:LRG_systematic_plots} but for the QSOs (1.3 < z < 2.1) in the South (NGC).}
    \label{fig:QSO_systematic_plots}
\end{figure}

The efficiency of some of these variants is shown in \cref{fig:LRG_systematic_plots} for the LRG ($0.6 < z < 0.8$) sample in the South (SGC) region and in \cref{fig:QSO_systematic_plots} for the QSO ($1.3 < z < 2.1$) sample in the South (NGC) region. These plots show the relative density of the objects as a function of the amplitude of the templates corresponding to different features. The black lines are without any imaging systematic weights while the different colours are when we apply the different weights computed with the configurations explained above. Similar plots for the other regions and other redshift sub-samples were also used to assess the efficiency of the different corrections.

All the configurations using the linear regression, even if not shown in these figures for clarity, give very similar results. The most important difference appears when we use the non-linear regression (red lines) instead of the linear one (blue lines). However, as noted in \cite{Rezaie2021, Chaussidon2022, Rezaie2023}, non-linear regressions have more degrees of freedoms to "flatten" the black line such that it does not necessarily result in a better correction. In addition, this additional degree of freedom leads to a modification of the power spectrum at large scales as illustrated in the next section.  

\subsection{Angular Integral Constraint} \label{sec:validation_aic}
To compare the efficiency of the different imaging systematics, we first need to quantify their impacts on mocks without any contamination. Then, if necessary, we need correct for the angular integral constraint that appears due to the use of these weights.

\subsubsection{Illustration with the EZmocks}  \label{sec:validation_ezmocks_imaging}
As shown in \cite{Rezaie2023},  allowing too much flexibility with a neural network during regression results in significant removal of large-scale power in the monopole that strongly biases the measurement of $\fnl$. To address this, we first perform a null test, by using our set of EZmocks that does not contain any imaging systematic contamination. Doing so, the mock density of tracers (either QSOs or LRGs) is strictly uncorrelated (when averaging over many realizations) with the different imaging features presented in \cref{sec:wsys_description}. Hence, the imaging systematic mitigation \ie the computation of the per-tracer weights $w_{\rm sys}$ from linear or random forest regression, should not bias $\fnl$. Any observed impact on the power spectrum measurement is then attributable to the methodology itself and must be corrected to prevent bias in our results.

\begin{figure}
    \centering
    \includegraphics[scale=1]{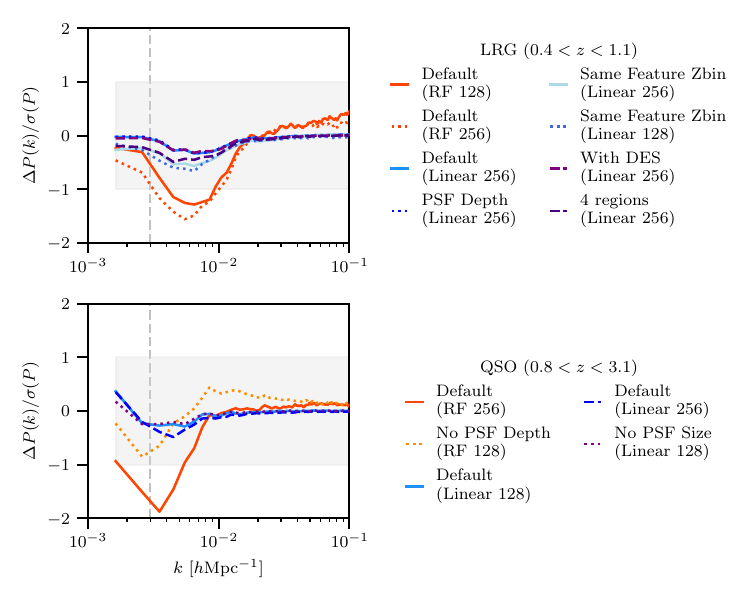}
    \caption{Relative difference  between the power spectrum monopoles obtained using FKP weights with and without imaging systematic weights to the DR1 statistical errors derived from the 1000 EZmocks. Each monopole that we show represents the mean computed from 30 realizations. LRGs are on the top and QSOs are on the bottom. The blue (\textit{resp.} red) lines are for linear (\textit{resp.} Random Forest) regressions while the specificity for each regression is explained in the text. The resulting values of $\fnl$ are displayed in \cref{fig:fits_null_test} and given in \cref{tab:fits_null_test}.}
    \label{fig:pk_null_test}
\end{figure}

From the different setups described above, we can compute the per-tracer correction weights $w_{\rm sys}$, to be used in the power spectrum estimator through \cref{eqn:observational_weights}. We measure the power spectrum monopoles as the mean computed over the first 30 EZmocks, for the LRGs and QSOs to isolate and quantify the impact of the imaging systematic weights\footnote{The imaging systematic weights are computed independently for each realization}. The impact on the monopole of these different configurations are shown in \cref{fig:pk_null_test} where we displayed the relative difference between the monopoles estimated with and without imaging systematic weights divided by the DR1 statistical errors. The regression using the random forest is displayed in red, while the linear regression in blue. From \cref{fig:pk_null_test}, it is clear that the random forest-based regression biases negatively the estimation of the monopole for either QSOs or LRGs compared to the linear regression, whose effect is smaller. This is because the regression has enough freedom to completely homogenise the angular distribution at a specific $N_{\rm side}$, nulling a lot of large cosmological angular modes. By removing physical modes on the large scales, random forest-based mitigation biases negatively the measurement of $\fnl$, while the linear regressions has a relatively small impact. 

\begin{figure}
    \centering
    \includegraphics[scale=1]{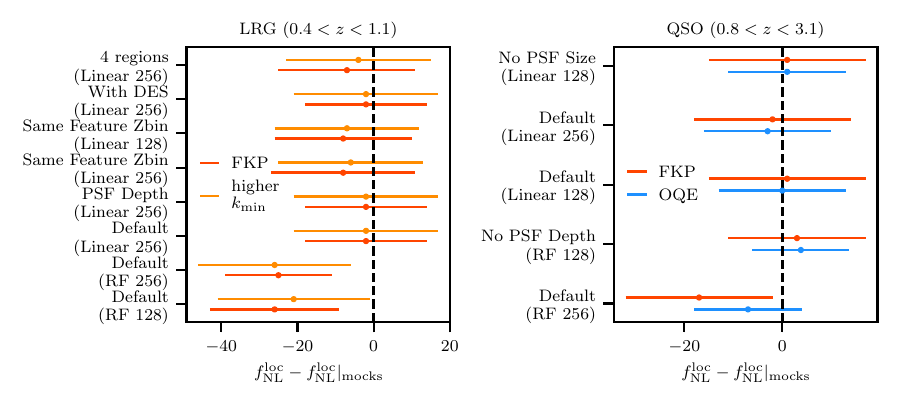}
    \caption{Best fit values of $\fnl$ with different imaging weights where the value from the EZmocks without weight is subtracted. Red whiskers denote the scenario of using FKP weights, while blue whiskers denote the OQE weights case. For the LRG, red is for $k_{\rm min} = 0.003~h\text{Mpc}^{-1}$ and orange for $k_{\rm min} = 0.006~h\text{Mpc}^{-1}$. The impact of the imaging weights on the power spectrum is shown in \cref{fig:pk_null_test}. The values are given in \cref{tab:fits_null_test}.}
    \label{fig:fits_null_test}
\end{figure}

To quantify the bias introduced on $\fnl$ measurement, we fit the mean of the power spectrum using the associated DR1 covariance matrix. The best fit for LRGs and QSOs using either FKP or OQE weights are displayed in \cref{fig:fits_null_test}. The Random Forest-based regression introduces an important negative bias in the estimation of $\fnl$. Indeed, by construction, these weights tend to flatten the angular density at the level of the difference pixels used during the regression, thus canceling modes that are in common between the different pixels. In the QSO case, the OQE weights help to prevent this effect by over-weighting the high-$z$ objects since at higher redshift the angular modes, nulled out by the use of imaging weights, are physically larger and so impact lower $k$'s. For the same reason, this effect is less important for the QSOs than for the LRGs.


Although less statistically significant than the random forest mitigation, this bias exists also in the case of the linear regression. One can reduce it by reducing the number of features used in the fit, as is already the case for the LRGs' \texttt{default} configuration compared to \texttt{Same Feature Zbin}. However, reducing the number of features can also reduce the efficiency of the weights by not including a feature that actually describes a remaining imaging systematic. A correction in this regard is proposed in \cref{sec:aic}.

To be conservative and have less significant correction of the effect illustrated in \cref{fig:pk_null_test}, we choose in the following as fiducial weight \texttt{Default (Linear 256)} for the LRGs and \texttt{Default (Linear 128)} for the QSOs. Establish the correction of this effect is the topic of the following section.


\subsubsection{Estimation of the Angular Integral Constraint} \label{sec:aic}
The suppression of power in \cref{fig:pk_null_test}, introduced by the imaging systematic weights, can be seen as an \emph{Angular Integral Constraint} (AIC) \cite{DeMattia2019}. Indeed, the imaging systematic weights aim to remove the angular fluctuations of the density at a pixel level, nulling out some angular modes and reducing the power at large scales. Hence, this effect can be taken into account by adding its contribution to the window matrix:
\begin{equation} \label{eqn:total_window}
    \mathcal{W} \rightarrow \mathcal{W} - \mathcal{W}^{\rm RIC} - \mathcal{W}^{\rm AIC},
\end{equation}
where $\mathcal{W}^{\rm AIC}$ can be estimated using the same method introduced in \cref{sec:ric}. We choose to use the same shape for the $\left(f_{p,q}\right)_{m,n}$ coefficients than for the RIC contribution. Note the $\mathcal{W}^{\rm AIC}$ is estimated with the realization of mocks without the shuffling method \ie without the RIC contribution, so that the AIC and RIC contributions is added linearly in the total window function. However, one can imagine, in the future, to model the two contributions simultaneously.

It is impossible to correctly quantify the efficiency of the imaging systematic mitigation on the power spectrum without taking into account the AIC correction. First, for the QSO and LRG fiducial weights that use linear regression, and second, for the weights that use random forest-based correction and either more features (for the LRGs) or a better $N_{\rm side}$ resolution (for the QSOs). Following the same methodology as in \cref{sec:ric}, the $\mathcal{W}^{\rm AIC}$ correction is estimated from the first 30 EZmock realizations on which the weighting scheme is applied.

\begin{figure}
    \hspace*{-15mm}
    \includegraphics[scale=1]{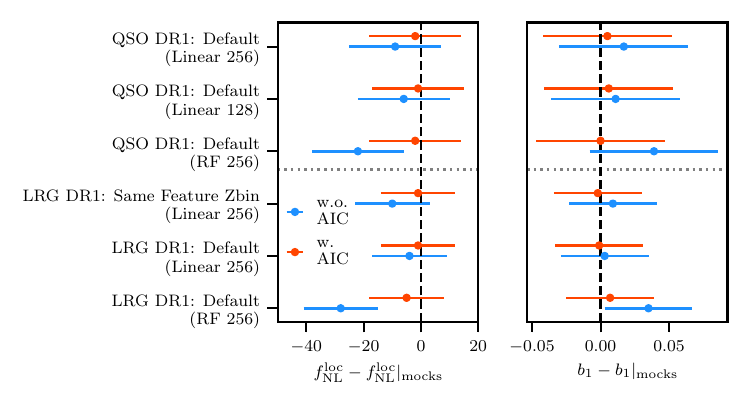}
    \centering
    \caption{Best fit values of ($\fnl$, $b_1$) without (blue) and with (red) the angular integral constraint contribution into the window function for different imaging weights where we have substracted the value from the EZmocks without imaging systematic weights. The values are given in \cref{tab:aic}. The systematic error contribution is discussed in \cref{sec:systematic_errors}.}
    \label{fig:aic}
\end{figure}

To test the impact of the AIC correction on parameter constraints, we fit the mean of 30 uncontaminated EZmock power spectra, each of them estimated with the imaging mitigation weighting schemes to be tested. The best fits using (red points) or not (blue points) the AIC correction in the parameter fit are given in \cref{fig:aic}. First, even if the contribution is very small ($\sim 0.4 \sigma$) for the linear-based weights, it exists, biasing the result, however, can be corrected. In all configurations, taking into account the AIC contribution enables us to recover the expected value of $\fnl$ that is measured from the realization without the weighting scheme applied (first column). Our correction is slightly worse for the random forest-based weights but is enough to quantify the efficiency of weights. 

In the following, we  neglect the impact of the AIC on the covariance matrix since it is too numerically expensive to run for the different $w_{\rm sys}$ configuration as many power spectra.

Note that the angular integral constraint is purely geometrical and does not reflect the efficiency of the imaging weights, meaning it can be reliably estimated using uncontaminated simulations. 

\subsubsection{\referee{Total window function}}
\referee{In this section, we summarize the total window function given in \cref{eqn:total_window}:
\begin{itemize}
    \item $\mathcal{W}$: standard window matrix that contains the wide-angle correction at first order as described in \cref{sec:window_function}.
    \item $\mathcal{W}^{\rm RIC}$: radial integral constraint contribution derived in \cref{sec:ric}.
    \item $\mathcal{W}^{\rm AIC}$: angular integral constraint contribution derived in \cref{sec:aic}. This contribution depends on the choice of the imaging systematic weights.
\end{itemize}
The computation of the three contributions depends on the choice of the weighting scheme to compute the power spectrum and is computed for the difference cases accordingly. In the following, all the fits presented used this total window function.}

\begin{figure}
    \centering
    \includegraphics[scale=1]{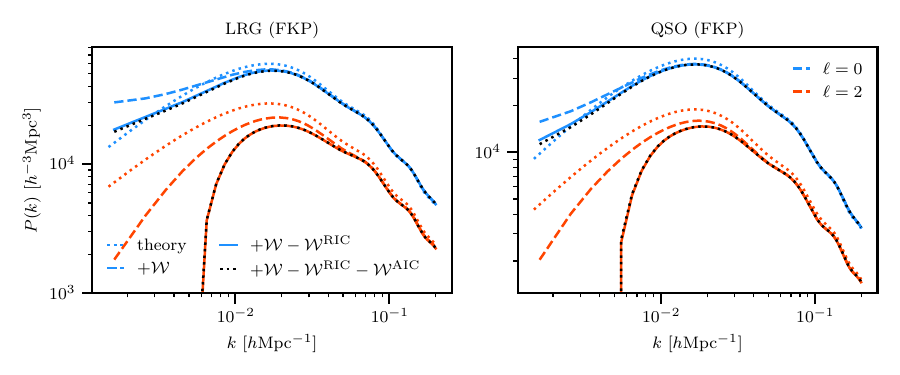}
    \caption{\referee{Impact of the window function and the different integral constraints on the monopole and the quadrupole for the LRG (left) and QSO (right) sample with the FKP weights. The theory without any window is in dotted color, while the window convolved theory is in dashed color, the window convolved theory corrected for the radial integral constraint is in full color, and the window convolved theory corrected for both the radial integral constraint and the angular integral constraint is in dotted black.}}
    \label{fig:wm_total}
\end{figure}

\referee{The impact of the window function and the different integral constraints are shown in \cref{fig:wm_total} for the fiducial choice used to analyse the LRG and QSO samples with FKP weights. As mentioned in \cref{sec:gic}, we do not include the global integral contribution that is negligible in our case. We choose in \cref{sec:internal_consistency}, imaging systematic weights that lead to a very tiny angular integral constraint contribution, such that the full color lines and the dotted black ones are very  similar.}

\subsection{Validation with the blinded data} \label{sec:blinded_data}
Following the analysis in \cite{Ross2024} and supported by the results on EZmocks (\cref{sec:validation_ezmocks_imaging}), our fiducial analyse uses the \texttt{Default (Linear 256)} weights for the LRGs and the \texttt{Default (Linear 128)} weights for the QSOs. In \cref{sec:consistency}, we check the internal consistency of the default configurations for LRG and QSO. From this, we identifiy possible remaining systematics that we explore through extended mitigation methods in \cref{sec:imaging_systematic_data_blind}. Finally, we check the robustness of the $\fnl$ measurement from blinded data in \cref{sec:internal_consistency}.

\subsubsection{Search for residual systematics} \label{sec:consistency}
First, we can assess the efficiency of the default configurations for imaging systematic mitigation by examining the compatibility between the blinded large-scale modes in the monopole, which are measured in different photometric regions and redshift bins. This is possible because, as shown in \cite{Chaussidon2024}, the blinding is not sensitive to the variation of the shot noise in the sample, and therefore is the same across the different redshift bins and photometric regions. Note also that RIC and AIC contributions to the window and the window matrix itself are different for each sub-sample either for the redshift or photometric region split, such that the very large scales should be different. Hence, the aim of this section is not to quantify the agreement of the different sub-samples, but only to look for any spurious excess of power at large scales resulting from a remaining systematic.

\begin{figure}
    \begin{subfigure}[]{\textwidth}
        \centering
        \includegraphics[]{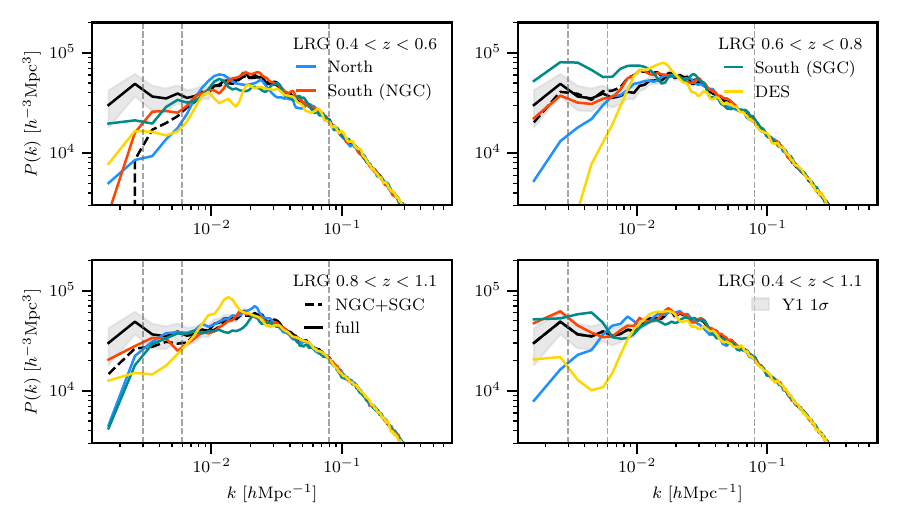}
        \caption{DESI DR1 LRG (blinded)}
        \label{fig:internal_consistency_LRG}
    \end{subfigure}
    \begin{subfigure}[]{\textwidth}
        \centering
        \includegraphics[]{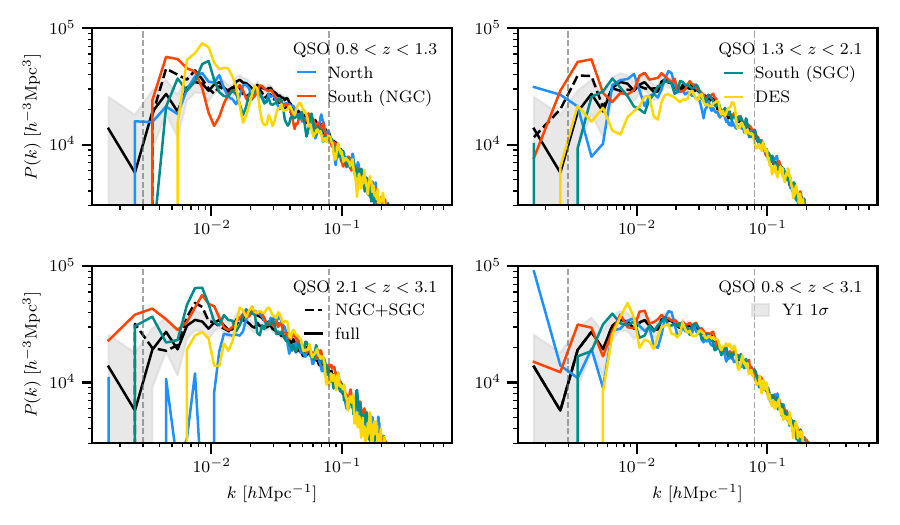}
        \caption{DESI DR1 QSO (blinded)}
        \label{fig:internal_consistency_QSO}
    \end{subfigure}
    \caption{Monopole of the power spectrum, measured with the fiducial $w_{\rm sys}$, of the \emph{blinded} LRG (top panels) and \emph{blinded} QSO (bottom panels) samples for different redshift bins and the different photometric regions (North in blue, South (NGC) in yellow, South (SGC) in green and the DES region in red). For each redshift bin, the black dashed line is the monopole measured on the entire footprint, while the solid black line is the monopole measured on the full redshift range and the entire footprint. The bottom right plot in each set of panels gives the monopole from the full redshift range on the different photometric regions. The gray region for each panel is the $1\sigma$ deviation from the EZmocks with the full redshift range and the entire footprint. The vertical grey dashed lines depict the scale range used for the fit.}
\end{figure}

In the following, we measure LRG and QSO blinded power spectrum monopoles separately in the different redshift bins detailed in \cref{sec:wsys_description}, as well as in the three different photometric regions, North, South (NGC) and South (SGC). As an indication of the statistical significance of the different photometric regions compared to the complete sample, the effective area for the DESI DR1 LRGs and QSOs are (in $\%$): 
\begin{itemize}[noitemsep,topsep=1pt]
    \item North: $(18, 20)$,
    \item South (NGC): $(48, 44)$,
    \item South (SGC): $(25, 27)$,
    \item DES: $(9, 9)$.
\end{itemize}
The DES region is the smallest region, while South (NGC) is the most important one. These differences in effective areas are visible in \cref{fig:angular_density_LRG_QSO}. To assess the statistical significance of each redshift bin, one can look at the redshift distribution shown in \cref{fig:redshift_distribution}. From this, we note that DES has a very low statistical significance compared to others. Then, using the fiducial imaging mitigation weighting schemes, these monopoles are shown in \cref{fig:internal_consistency_LRG} (resp. in \cref{fig:internal_consistency_QSO}) for the LRGs (resp. for the QSOs).

For the LRGs (\cref{fig:internal_consistency_LRG}), the shape of the monopole for the full sample (black line) exhibits a unusual shape at very large scales ($k \sim 0.003~h\text{Mpc}^{-1}$). We also note a relatively strong discrepancy at these scales between the monopole from the different photometric regions (bottom right panel). In addition, this discrepancy around these scales is the most important for the middle redshift bins (top right panel) where South (SGC) region shows an unexpected excess of power. This could be due to residual systematic contamination that the current default weight do not remove. For the two other redshift bins (top and bottom left panel), the monopoles agree with each other, and with the one from the entire footprint (black dashed lines), such that no clear remaining systematics appear in these bins.

In \cref{sec:imaging_systematic_data_blind}, we investigate different imaging weights to remove this excess of power. However, we note that increasing the minimal scale from $k_{\rm min} = 0.003~h\text{Mpc}^{-1}$ to $k_{\rm min} = 0.006~h\text{Mpc}^{-1}$ helps to reduce the discrepancy between the different photometric regions in \cref{fig:internal_consistency_LRG} (middle gray dashed lines) such that it can be a conservative approach.

In contrast to the case of LRGs, the results for the QSO monopole show a remarkable consistency between the different regions of the sky, for each redshift ranges (\cref{fig:internal_consistency_QSO}). In nearly each case shown in the Figure, the results agree within 1$\sigma$ of the monopole evaluated from the full footprint. We observe no spurious signals at large scales. We do observe some deviations in the highest redshift bins $2.1 < z < 3.1$ (lower left panel) for the North and the DES regions, however, this bin has much less statistical information than the full range and these two regions have a much smaller footprint and completeness compared to the South (NGC/SGC), such that the measurement is still compatible and does not raise any significant concern.

\subsubsection{Validation of the imaging systematic weights} \label{sec:imaging_systematic_data_blind}
In this section, we aim to validate the default weighting scheme as the best choice for correcting imaging systematics for the QSOs and to improve the correction on the LRGs. To quantitatively assess the \textit{full} imaging systematics mitigation procedure, we compare the measurement of $\fnl$ for the different imaging weight configurations with the corresponding AIC contribution that we compute for each different setups (see \cref{sec:aic}), rather than looking at the power spectrum level which do not inform us about due to the different integral constraints.

The blinded $\fnl$ constraints using the different imaging systematic weights are shown in \cref{fig:fits_blinded_test}. To assess the importance of the imaging weights, we also include the best fit when no imaging weights are used. Note that all fits, except when no weight is applied, incorporate the AIC contribution into the window matrix, requiring the computation of this contribution for each case. For the LRGs, we show the blinded constraints for two minimal scale cuts: $k_{\rm min} = 0.003~h\text{Mpc}^{-1}$ or $k_{\rm min} = 0.006~h\text{Mpc}^{-1}$.

We explore whether new regression methods and/or new set of templates can solve the remaining systematics highlighted in the previous subsection for the LRG sample. The relative density of LRGs ($0.6 < z < 0.8$) in South (SGC) region as a function of the most relevant observational features are displayed in \cref{fig:LRG_systematic_plots}. As mentioned in the introduction of this section, the aim of the imaging systematic weights is to mitigate the trend in the black lines by flattening them. We show the corrected densities using five different imaging mitigation weighting schemes: \texttt{Default (Linear 256)} in blue (our default mitigation method), \texttt{Default (RF 256)} in red, \texttt{Same Feature Zbin (Linear 256)} in green, \texttt{With Des (Linear 256)} in pink, and \texttt{4 regions (Linear 256)} in gold. While the first three configurations use the entire SGC footprint (South (SGC) + DES) for the regression, the fourth one uses the entire South footprint (South (NGC) + South (SGC)) and the last one only South (SGC). These different setups were tested with uncontaminated EZmocks, where the details can be found in \cref{sec:other_config_ezmocks}. Here, we show only one subregion of one redshift bin for simplicity, but the others have similar behavior.

\begin{figure}
    \centering
    \includegraphics[scale=1]{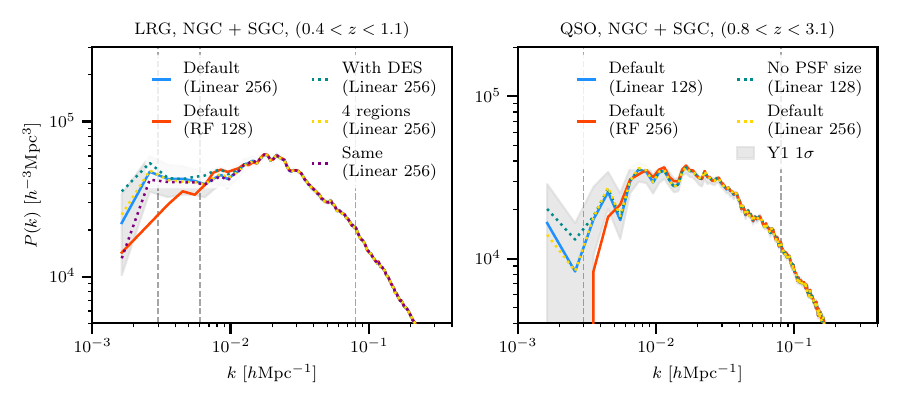}
    \caption{Monopole of the power spectrum measured on the \emph{blinded} data (LRG on the left and QSO on the right) with FKP weights and different imaging systematic weights. The gray region is the standard deviation from the 1000 EZmocks around the blue lines that are our fiducial choice. Since we need do include the angular integral constraint in the model, the very large-scale modes cannot be compared directly in this figure. The best fit values are displayed in \cref{fig:fits_blinded_test}.}
    \label{fig:wsys_impact_data}
\end{figure}

\begin{figure}
    \centering
    \includegraphics[scale=1]{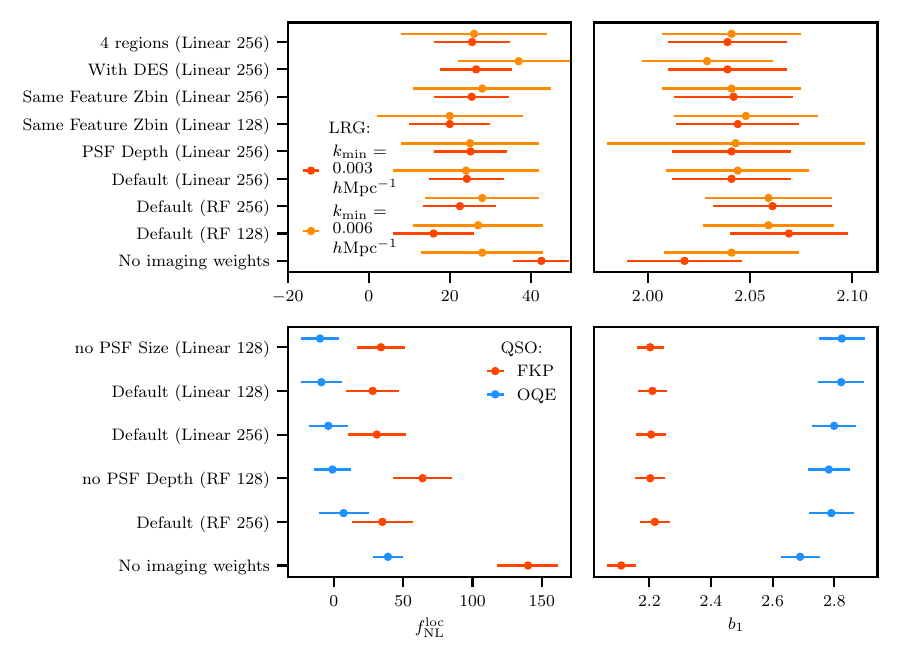}
    \caption{Best fit values of $(\fnl, b_1)$ for the DR1 \emph{blinded} LRGs with two values of $k_{\rm min}$ and for the \emph{blinded} QSOs using either FKP and OQE weights, for various configurations of imaging weights. All the fits include the radial and angular integral constraint contributions. The errors and the central values are from the minimization performed with \texttt{iminuit} and the covariance matrix is the one from the 1000 EZmocks. Note that the errors obtained on $\fnl$ for the blinded LRGs with $k_{\rm min}=0.003~h\mathrm{Mpc}^{-1}$ are too small compared to the expected on from the mean of EZmocks, see \cref{tab:fit_ezmocks}. The configuration chosen for the unblinded analysis is \texttt{Default (Linear 256)} for LRGs and \texttt{Default (Linear 128)} for QSOs. The table reporting the numbers are given in \cref{tab:fits_blinded_test}. The systematic error contribution is discussed in \cref{sec:systematic_errors}.}
    \label{fig:fits_blinded_test}
\end{figure}

Except for the RF weights, the four other (linear) corrections provide very similar results and have only a slight impact on the relative density although we have increased the number of templates or isolated the zone in the regression. The impact on the power spectrum of the full sample is shown in \cref{fig:wsys_impact_data} where only the RF weights appear to have a significant impact. However, as explained in \cref{sec:validation_aic}, one needs to take into account the angular integral constraint that is different for which weights. With this contribution the differences between the RF and the other linear weights in \cref{fig:LRG_systematic_plots} or in \cref{fig:wsys_impact_data} are mostly due to this contribution. Indeed, when fitting $\fnl$ with the AIC contribution, we find roughly the same amount of $\fnl$, see \cref{fig:fits_blinded_test}.

In particular, all the weights tested here provide only a minor correction and do not eliminate the apparent excess power at large scales described in \cref{sec:consistency} at very large scales. We note also that the errors on $\fnl$ obtained with $k_{\rm min} = 0.003~h\text{Mpc}^{-1}$ for the different weighting schemes  are smaller than expected from the mean of the EZmocks. This discrepancy may indicate the presence of a non-physical signal at these scales.

Hence, we resort to adopting a conservative approach to avoid any bias in our measurement by increasing the minimal scale from $k_{\rm min} = 0.003~h\text{Mpc}^{-1}$ to $k_{\rm min} = 0.006~h\text{Mpc}^{-1}$. This change improves the compatibility between the different regions and redshift bins and provides compatible errors between blinded data and simulations. As shown in \cref{sec:impact_imaging_systematic_weights}, at this new $k_{\rm min}$ the impact of imaging systematic weight is rather small. Note that with this new minimal scale cut in our LRG fit, the constraint on $\fnl$ is degraded by about $\sim 30 \%$ for the LRGs alone and $\sim 6 \%$ when  the LRGs are combined with the QSOs compared to the previous minimal scale cut.

Finally, the different configurations provide a consistent measurement of $\fnl$ with differences relative to the truth of $\Delta \fnl \sim 4$, and uncertainties $\sigma(\fnl)\sim 18$, illustrating the fact that the default configuration mitigates already all the effect that could be explained by our set of templates. Since none of the new configuration improve the statistical uncertainty of the measurement of $\fnl$, we decide to use \texttt{Default (Linear 256)} to be aligned with the recommendation of \cite{DESI2024II} and to avoid strong dependence on the AIC correction (see \cref{tab:fits_null_test}). In this way we reduce our measurement's sensitivity to a mismatch of the AIC estimation.

For the QSOs, we have conducted tests by using different sets of templates and/or regression methods. We showed in \cref{fig:QSO_systematic_plots} the uncorrected/corrected overdensities as a function of different imaging features for QSOs ($1.3 < z < 2.1$) in South (NGC) for the two configurations: \texttt{Default (Linear 256)} in blue (our default configuration) and \texttt{Default (RF 256)} in red. The impact of these weights on the power spectrum is shown in \cref{fig:wsys_impact_data}, while the measurement of $\fnl$ for the different configurations (again using their respective AIC corrections) is given in \cref{fig:fits_blinded_test}. The deficiency of power at large scales when using \texttt{Default (RF 256)} is not a sign of a better correction since one needs to include the AIC contribution, as explained in \cref{sec:validation_aic}. In this case, the use of \texttt{Default (RF 256)} leads to a measurement similar to \texttt{Default (Linear 256)}. We note that removing the PSF Depth templates increases a lot the value of $\fnl$ with the FKP weights. This is expected since these templates explain, for instance, the lack of true QSO at low-z which motivated the additional redshift split at $z=1.3$, see \cref{sec:appendix_impact_1.3_cut_wsys}. However, not using these templates has a very small impact for the measurement with OQE weights. That can be explained by the fact that these templates describe a systematic at low-$z$ that is under-weighted by the OQE weights and therefore does not impact the measurement.

Moreover, we found a tension between OQE and FKP results that cannot be explained by the small discrepancy found in \cref{sec:fkp_vs_oqe}. Again, since OQE down-weights the low-$z$ QSO sample compared to FKP, the tension can be explained by the presence of an unresolved systematic effect; we will address this point in the next section.

As for the LRGs, to reduce our sensitivity to the AIC correction, we decide to use \texttt{Default (Linear 256)}. This is different from \cite{DESI2024II} who use \texttt{Default (RF 256)}. 

\subsubsection{Robustness of the measurement} \label{sec:internal_consistency}
Before measuring $\fnl$ from the data without blinding, we assess the robustness of our analysis by testing several variations of our fiducial choices \ie by measuring $\fnl$:
\begin{itemize}[noitemsep,topsep=1pt]
    \item from NGC and SGC power spectra separately compared to the fiducial NGC+SGC measurements,
    \item using the tracers in smaller redshift intervals,
    \item fitting the power spectra in different $k$-ranges either by increasing $k_{\rm min}$ from $0.003$ (resp. 0.006) for QSOs (resp. for LRGs) to $0.008~h\text{Mpc}^{-1}$ or increasing $k_{\rm max}$ from $0.08$ to $0.1~h\text{Mpc}^{-1}$.
\end{itemize}
Note that we occasionally observe biases on $\fnl$ with the increasing $k_{\rm max}$ for the LRGs, where we may need to include perturbative theory instead of the linear power spectrum as well as non-linear galaxy bias model \cite{Desjacques2018}.

For each case, the window matrix $\mathcal{W}$, the RIC contribution $\mathcal{W}^{\rm RIC}$ and the AIC contribution $\mathcal{W}^{\rm AIC}$ are computed following the description given above in \cref{sec:aic}. For the separate NGC/SGC fits, the covariance is estimated with the EZmocks in the corresponding photometric region. However, the covariance for the two redshift sub-samples is approximated by the full sample covariance matrix, re-scaled by the ratio of the full and sub-sample effective volumes. The effective volume is computed with 
\begin{equation} \label{eqn:effective_volume}
    V_{\mathrm{eff}}=\int \left(\frac{n\left(z\right) P_{0}}{1+n\left(z\right) P_{0}}\right)^{2} \dd  V\left(z\right),
\end{equation}
where for simplicity we use a fixed value for $P(k, z) = P_0$ which we evaluate at the effective redshift and at the turnover scale of the power spectrum\footnote{Here, \referee{we use $P_{0} = 5\cdot10^4~h^{-3}\text{Mpc}^3$ for LRGs and $3\cdot10^4$} for the QSOs. Note these values are not the same as in \cite{DESI2024III}.}. It is computed at the effective redshift of the sample and at the turnover of the power spectrum. 

The effective volume of the different sub-samples is given in \cref{tab:effective_volume}. Hence, for the LRGs ($0.4 < z < 0.8$) (\textit{resp.} $0.6 < z < 1.1$), the full covariance matrix is rescaled by $\sim 2.1$ (resp. $\sim 1.2$). For the QSOs ($0.8 < z < 2.1$) (\textit{resp.} $1.6 < z < 3.1$), the full covariance matrix is rescaled by $\sim 1.3$ (resp. $\sim 1.8$). We also need to compute the specific effective redshift for these sub-samples; they are given in \cref{tab:effective_volume}. The result of the different posteriors over the $\fnl-b_1$ parameter space are displayed in \cref{fig:fit_blinded_data} (we give the values in the appendix, see \cref{tab:measurement_robustness}). 

\begin{table}
    \centering
    \caption[]{Effective volume computed with \cref{eqn:effective_volume} and effective redshift with \cref{eqn:effective_redshift} for the sub-sample of the LRGs and QSOs. For the OQE weights, we give only, for simplicity, the effective redshift for the monopole and $p=1.6$.}
    \label{tab:effective_volume}
    \begin{tabular}{lcccc}
      \toprule
      & & $V_{\mathrm{eff}}~[(\mathrm{Gpc}/h)^3]$ & $z_{\mathrm{eff}}$ (FKP) & $z_{\mathrm{eff}}$ (OQE - $\ell=0$) \\ 
      \midrule
      LRG & $0.4 < z < 1.1$ & $6.5$ & 0.733 & - \\
          & $0.4 < z < 0.8$ & $3.1$ & 0.601 & - \\
          & $0.6 < z < 1.1$ & $5.3$ & 0.831 & - \\
      \midrule
      QSO & $0.8 < z < 3.1$ & $8.3$ & 1.651 & 2.082 \\
          & $0.8 < z < 2.1$ & $6.4$ & 1.441 & 1.691 \\
          & $1.6 < z < 3.1$ & $4.6$ & 2.087 & 2.236 \\
      \bottomrule
    \end{tabular}
\end{table}

For the LRGs, we conducted the different robustness tests for the full sample ($0.4 < z < 1.1$) in \cref{fig:fit_data_blinded_lrg_fkp}. We find that the posterior is bimodal for the low-$z$ sub-sample ($0.4 < z < 0.8$) compared to the high-$z$ sample posterior which is Gaussian. This is not expected based on results from simulations. Such discrepancy between the low-$z$ and high-$z$ samples is most likely due to residual systematics in the mitigation of imagining systematics. In addition, we note that the result does not appear to be robust when decreasing $k_{\rm min}$.

Alternatively, we can focus on the high-$z$ sub-sample, whose robustness tests are repeated and displayed in \cref{fig:fit_data_blinded_lrg_fkp_high-z}. When repeating these tests, the fits are more robust when varying the fiducial configuration than for the full sample. Since the effective redshift of the high-$z$ sub-sample is higher, the linear bias is higher such that the constraining power on $\fnl$ remain relatively unchanged, even if the covariance matrix is properly rescaled by the ratio of the two effective volumes. Accordingly, one can restrict our baseline analysis to the high-$z$ sub-sample without degrading our measurement of $\fnl$, and in what follows, this is what we do to avoid any potential bias that could come from the low-$z$ sub-sample.

For the QSOs using FKP weights (see \cref{fig:fit_data_blinded_qso_fkp}), we find a good consistency between the NGC and SGC constraints, emphasising the robustness of our mitigation of imaging systematics in the different photometric regions. Moreover, the $k_{\rm min}$ and $k_{\rm max}$ robustness tests are fairly compatible, as our model of linear bias is consistent at these scales for QSOs. However, the low-$z$ ($0.8 < z < 2.1$) sub-sample and the high-$z$ ($0.8 < z < 3.1$) sub-sample prefer two different values of $\fnl$, even though they are statistically compatible, while the full sample prefers a value between the two. Since imaging systematics only add power, this slight tension on $\fnl$ may indicate a residual unknown systematic in the low-$z$ sample.

Fortunately, we can remove this potential systematic at low-$z$ thanks to the use of the OQE weights that drastically under-weight the low-$z$ object, see \cref{fig:optimal_weights}. The robustness tests for the OQE weights, see \cref{fig:fit_data_blinded_qso_oqe}, show excellent agreement even better than when we use the FKP weights, except for the low-$z$ sub-sample part that is now clearly biased. We note also that, the difference between $\fnl$ from FKP or OQE measurement is much more in agreement when considering only the high-$z$ sub-sample: $\Delta \fnl = 11$ that is statistically acceptable regarding\footnote{The high-$z$ sub-sample has a higher effective redshift and so a higher linear bias, such that the dispersion in \cref{fig:oqe_vs_fkp} is smaller for this specific configuration} \cref{fig:oqe_vs_fkp}, compared to $\Delta \fnl = 37$ when comparing the full sample. For these reasons, we consider OQE weights and the \texttt{Default (Linear 128)} mitigation option to analyse QSO unblinded data.

Moreover, for the QSO sample, we obtain in \cref{fig:fit_blinded_data} a lower bias than expected. We measured the linear bias as in \cref{sec:linear_bias} from the 2-point correlation function, and we also obtained a lower bias that is compatible to the one measured here. The problem seems to come from the blinded data itself, and we do not investigate this further since the bias obtained with the unblinded data is compatible with the one from eBOSS.

Finally, the decisions took in this section are based on the robustness tests from the \emph{blinded} data, such that are insensitive to any confirmation bias, and can be used as our fiducial confirmation in the following.


\begin{figure}
    \centering
    \begin{subfigure}[t]{0.48\textwidth}
        \centering
        \includegraphics[scale=0.97]{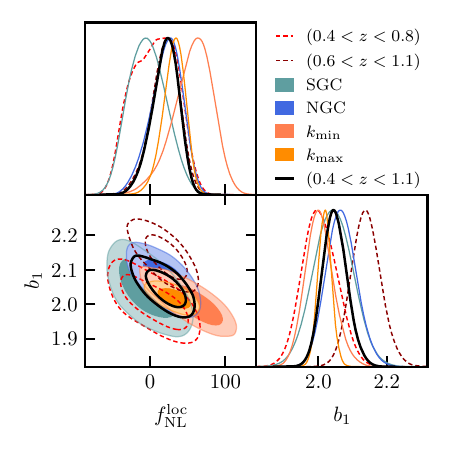}
        \caption{\emph{Blinded} LRG (FKP).}
        \label{fig:fit_data_blinded_lrg_fkp}
    \end{subfigure}
    \begin{subfigure}[t]{0.48\textwidth}
        \centering
        \includegraphics[scale=0.97]{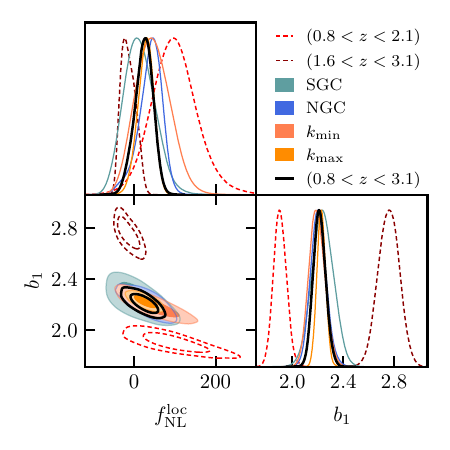}
        \caption{\emph{Blinded} QSO (FKP).}
        \label{fig:fit_data_blinded_qso_fkp}
    \end{subfigure}
    \begin{subfigure}[t]{0.48\textwidth}
        \centering
        \includegraphics[scale=0.97]{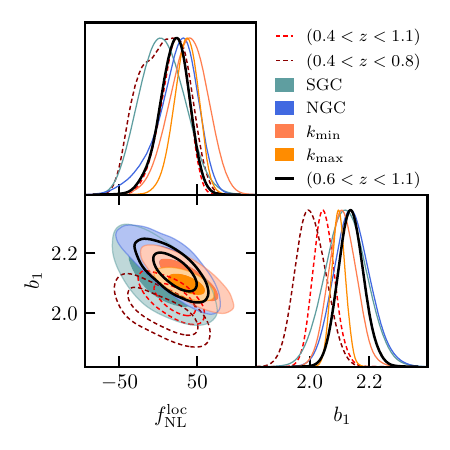}
        \caption{\emph{Blinded} LRG (FKP) high-$z$.}
        \label{fig:fit_data_blinded_lrg_fkp_high-z}
    \end{subfigure}
    \begin{subfigure}[t]{0.48\textwidth}
        \centering
        \includegraphics[scale=0.97]{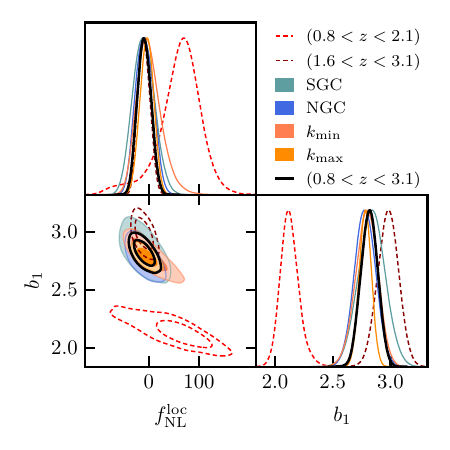}
        \caption{\emph{Blinded} QSO (OQE).}
        \label{fig:fit_data_blinded_qso_oqe}
    \end{subfigure}
    \caption{Posteriors in the $(\fnl, b_1)$ plane for the \emph{blind} analysis applied to real data with different modifications around our fiducial choices. We show QSO with  FKP (top right) and with OQE (bottom right), as well as the LRG (top left) and the high-$z$ ($0.6 < z < 1.1$) LRG (bottom left). In each panel, the fiducial choice is displayed in black, while the redshift sub-samples are with dashed lines, and in full colors the measurement using only a part of the footprint or with the modification of the fitting ranges. For the LRGs, the posteriors for the different redshift range are the same in (a) and (c). The parameters $s_{n,0}, \Sigma_s$ are free during the MCMC but are not shown for better visibility. The corresponding MAP values are given in \cref{tab:measurement_robustness}.}
    \label{fig:fit_blinded_data}
\end{figure}

\section{Primordial non-Gaussianity measurement} \label{sec:unblinded_analysis}
The model validation done in \cref{sec:model} and the data analysis performed with the blinding scheme in \cref{sec:blinded_analysis} lead to no relevant systematic bias in our methodology, see \cref{sec:systematic_errors} for the quantification of the systematic errors. Therefore, we can measure confidently $\fnl$ from the unblinded data of the DESI DR1 LRG and QSO samples. We remind the reader that the fiducial choices for our analysis are given in \cref{tab:fid_param}.

\begin{table}
    \centering
    \caption{Summary of the fiducial choices that we used to fit the unblinded LRG and QSO sample.}
    \label{tab:fid_param}
    \begin{tabular}{lcc}
        \toprule
        & LRG & QSO \\
        \midrule
        weighting scheme & FKP & OQE \\
        $z$ range & $0.6 - 1.1$ & $0.8 - 3.1$ \\
        $z_{\rm eff}$ & $0.831$ & $2.082$ \\
        $k_{\rm min}$ [$h\text{Mpc}^{-1}$] & 0.006 & 0.003 \\
        $k_{\rm max}$ [$h\text{Mpc}^{-1}$] & 0.08  & 0.08 \\
        $w_{\rm sys}$ (\cref{sec:default_config}) & \texttt{Default (Linear 256)} & \texttt{Default (Linear 128)} \\
        \bottomrule
    \end{tabular}
\end{table}

Before we proceed, we draw the reader attention to the fact that several versions of the clustering catalogues were built as described in \cite{DESI2024II}. The last version with the blinding is v1.2, and this is the version used for all the tests in \cref{sec:blinded_analysis}. Here, for the unblinded measurement, we are using v1.5, the last version available, and the one used for RSD measurement in \cite{DESI2024V}. The differences between the versions are described in Appendix B of \cite{DESI2024II}, and we note no relevant change for our analysis between the unblind catalogues from v1.2 and v1.5. 

\subsection{Consistency validation with the unblinded data}
\begin{figure}
    \centering
    \begin{subfigure}[t]{0.48\textwidth}
        \centering
        \includegraphics[scale=0.97]{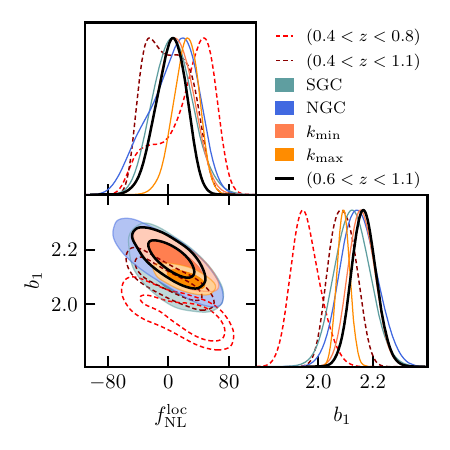}
        \caption{\textit{Unblinded} LRG (FKP) high-$z$.}
        \label{fig:fit_data_unblinded_lrg_fkp_high-z}
    \end{subfigure}
    \begin{subfigure}[t]{0.48\textwidth}
        \centering
        \includegraphics[scale=0.97]{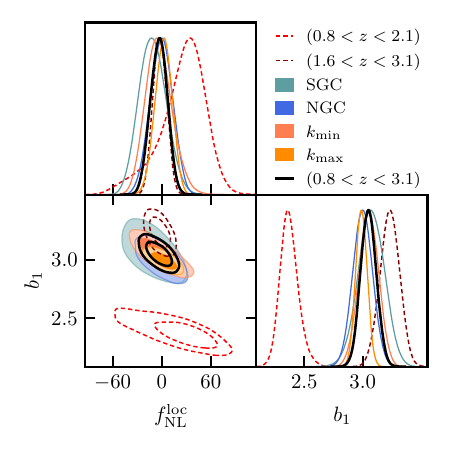}
        \caption{\textit{Unblinded} QSO (OQE).}
        \label{fig:fit_data_unblinded_qso_oqe}
    \end{subfigure}
    \caption{Analogous of \cref{fig:fit_blinded_data} but using the \emph{unblinded data}. The fiducial choices for our analysis, see \cref{tab:fid_param}, are displayed in black. The MAP values are given in \cref{tab:measurement_robustness_unblind}. Note that the linear bias $b_1$ is higher with the unblinded data.}
    \label{fig:fit_unblinded_data}
\end{figure}

We perform the same tests as in \cref{sec:internal_consistency} but with the unblinded data, to check the internal consistency of the data. The best fit values are given in \cref{tab:measurement_robustness_unblind}, and the associated posteriors for $(b1, \fnl)$ are given in \cref{fig:fit_unblinded_data} for the different samples and weighting scheme.

Similar to our conclusions raised with blinded measurements, there are no major differences for the QSO case displayed in \cref{fig:fit_data_unblinded_qso_oqe}, and the discrepancy between the low-$z$ and high-$z$ sample still exists. However, we note one unexpected minor change: the linear bias measured in the blinded catalogues is lower than that measured in the unblinded ones. Note that the blinding method described \cite{Chaussidon2024} does not create a such large bias modification, and we do not know what is the reason of this difference. This has a direct consequence on the error bars that we have in \cref{tab:measurement_robustness_unblind} compared to \cref{tab:measurement_robustness} as the errors are smaller.

We observe that the high-redshift QSO sample provides an unexpectedly stronger constraint on $\fnl$ relative to the use of the full sample with $\sigma(\fnl) = 8.4$ instead of $\sigma(\fnl) = 11.5$, and despite the use of the OQE weights. This result is consistent with what is seen in the blinded data (see \cref{tab:measurement_robustness_unblind}), although the errors were larger due to a lower-than-expected bias recovered in the blinded catalog. For this high-$z$ sample, the expected constraint from the mean of EZmocks is $\sigma(\fnl) = 10.5$ and from individual fits, $8.4$ is compatible. However, the aim of the OQE weights is to provide the optimal measurement on $\fnl$ by introducing a redshift dependence in the FKP weights and under-weight the subpart of the sample that does not really matter, such that they give the best constraints with the full sample compare to a subpart of it. Given this, we do not consider the high-redshift sample for our final constraint on $\fnl$.

For the LRGs, the high-$z$ sample displayed in \cref{fig:fit_data_unblinded_lrg_fkp_high-z} is still robust with respect to the different choice of analysis, except with the low-$z$ part of the sample. Moreover, the 1D posterior of the low-z LRG sample in $\fnl$ has a very non-Gaussian shape, and this may be responsible for the wide $\fnl$ posterior of the full LRG sample shown in red dashed in \cref{fig:fit_data_unblinded_lrg_fkp_high-z} which displays larger errors than the high-z sample alone. This unexpected mismatch between $b_1$ and $\fnl$ constraints between the low-$z$ and high-$z$ samples validates our choice to use the high-$z$ sample instead of the full sample. 

\begingroup 
\renewcommand{\arraystretch}{1.3} 
\begin{table}
    \centering
    \caption[]{Best-fit results with the \emph{unblinded} LRGs and QSOs for different variations of our fiducial analysis. The central values are best fit values from the \texttt{iminuit} minimization while the errors are the 1$\sigma$ credible intervals from the chains. The window functions used during these fits contain both RIC and AIC contributions and were recomputed for the different configurations when it was needed. We use $k_{\rm min}=0.006~h\text{Mpc}^{-1}$ for the LRGs and $0.003~h\text{Mpc}^{-1}$ for the QSOs. The posteriors are displayed in \cref{fig:fit_unblinded_data}.}
    \label{tab:measurement_robustness_unblind}
    \resizebox{0.99\textwidth}{!}{%
    \begin{tabular}{llccccc}
        \toprule
        &  & $f_{\rm NL}^{\mathrm{loc}}$ & $b_{1}$ & $s_{n, 0}$ & $\Sigma_{s}$ & $\chi_{\rm red}^2$ \\
        \midrule
        LRG (FKP) 
        & \cellcolor{green!30} \textbf{high-$z$ fiducial} & \cellcolor{green!30} ${6}_{-18}^{+22}$ & \cellcolor{green!30} ${2.166}_{-0.045}^{+0.045}$ & \cellcolor{green!30} ${-0.024}_{-0.075}^{+0.074}$ & \cellcolor{green!30} ${3.70}_{-0.53}^{+0.68}$ & \cellcolor{green!30} 1.24 \\
        $(0.6 < z < 1.1)$ & NGC                                  & ${19}_{-25}^{+39}$& ${2.132}_{-0.071}^{+0.062}$ & ${0.02}_{-0.100}^{+0.107}$ & ${3.84}_{-0.57}^{+0.96}$ & 0.93 \\
        & SGC                                  & ${4}_{-26}^{+25}$ & ${2.132}_{-0.069}^{+0.062}$ & ${0.05}_{-0.11}^{+0.11}$    & ${3.1}_{-0.90}^{+1.55}$ & 1.34 \\
        & $(0.4 < z < 0.8)$                   & ${46}_{-19}^{+45}$ & ${1.936}_{-0.061}^{+0.044}$ & ${0.014}_{-0.088}^{+0.105}$ & ${3.14}_{-0.61}^{+1.23}$ & 0.62 \\
        & $(0.4 < z < 1.1)$ & ${-25}_{-34}^{+27}$& ${2.124}_{-0.053}^{+0.047}$ & ${-0.101}_{-0.076}^{+0.081}$&${3.86}_{-0.42}^{+0.62}$ & 1.27 \\
        & $k_{\rm min}=0.008~h\text{Mpc}^{-1}$ & ${9}_{-22}^{+25}$ & ${2.161}_{-0.052}^{+0.045}$ & ${-0.017}_{-0.074}^{+0.084}$&${3.69}_{-0.54}^{+0.68}$ & 1.27 \\
        & $k_{\rm max} = 0.1~h\text{Mpc}^{-1}$ & ${25}_{-15}^{+19}$& ${2.092}_{-0.024}^{+0.024}$ & ${0.100}_{-0.029}^{+0.032}$ &${2.65}_{-0.36}^{+0.53}$ & 1.24 \\ 
        \midrule \midrule
        QSO (OQE)
        & \cellcolor{green!30} \textbf{fiducial} & \cellcolor{green!30} ${-2}_{-10}^{+11}$ & \cellcolor{green!30} ${3.048}_{-0.070}^{+0.064}$ & \cellcolor{green!30} ${-0.039}_{-0.068}^{+0.076}$ & \cellcolor{green!30} ${0.0}_{-1.41}^{+0.43}$ & \cellcolor{green!30} 1.18 \\
        $(0.8 < z < 3.1)$
        & NGC                              & ${2}_{-14}^{+14}$     & ${2.996}_{-0.082}^{+0.086}$ & ${-0.007}_{-0.086}^{+0.095}$ & ${0.0}_{-1.62}^{+0.49}$ & 1.19 \\
        & SGC                              & ${-12}_{-18}^{+15}$   & ${3.09 }_{-0.100}^{+0.123}$ & ${-0.02}_{-0.12}^{+0.12}$    & ${3.0}_{-1.7}^{+1.5}$ & 1.26 \\
        & $(0.8 < z < 2.1)$                & ${37}_{-18}^{+33}$ & ${2.353}_{-0.084}^{+0.067}$ & ${0.087}_{-0.079}^{+0.087}$ & ${0.0}_{-1.44}^{+0.43}$ & 0.61 \\
        & $(1.6 < z < 3.1)$                & ${-2.3}_{-8.4}^{+8.4}$ & ${3.236}_{-0.079}^{+0.088}$ & ${-0.055}_{-0.093}^{+0.090}$&${4.38}_{-0.88}^{+1.18}$ & 0.75\\
        & $k_{\rm min}=0.008~h\text{Mpc}^{-1}$& ${-5}_{-18}^{+14}$ & ${3.060}_{-0.080}^{+0.086}$ & ${-0.048}_{-0.085}^{+0.081}$ &${0.7}_{-1.44}^{+0.45}$ & 1.22 \\
        & $k_{\rm max} = 0.1~h\text{Mpc}^{-1}$& ${3}_{-10}^{+11}$ & ${2.987}_{-0.050}^{+0.048}$ & ${0.039}_{-0.041}^{+0.043}$  &${2.17}_{-0.73}^{+1.20}$& 1.14 \\
        \bottomrule        
    \end{tabular}%
    }
\end{table}
\endgroup 

\begin{figure}
    \centering
    \begin{subfigure}[t]{0.48\textwidth}
        \centering
        \includegraphics[scale=0.98]{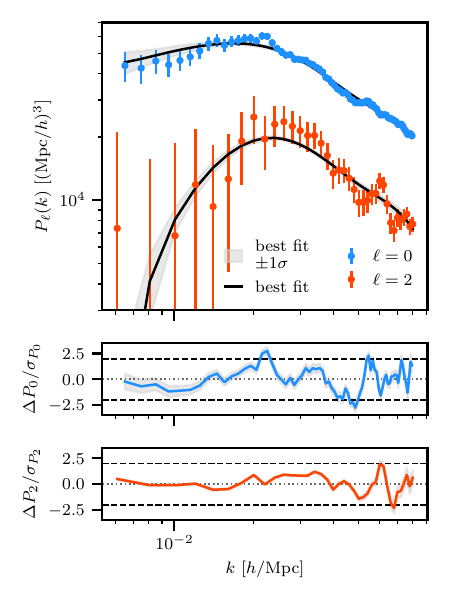}
        \caption{LRG (FKP) high-$z$.}
        \label{fig:bestfit_data_unblinded_lrg_fkp}
    \end{subfigure}
    \begin{subfigure}[t]{0.48\textwidth}
        \centering
        \includegraphics[scale=0.98]{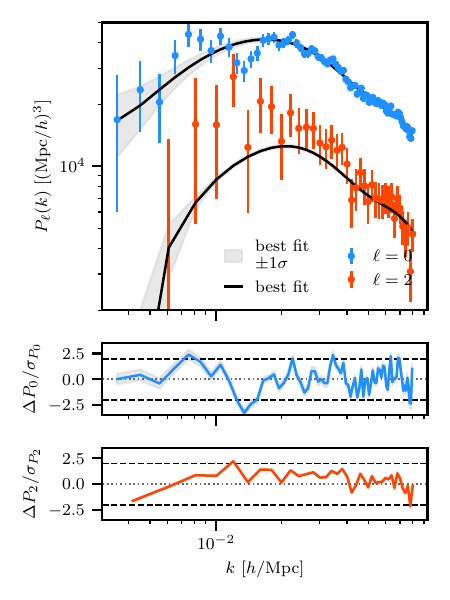}
        \caption{QSO (OQE).}
        \label{fig:bestfit_data_unblinded_qso_oqe}
    \end{subfigure}
    \caption{Monopole (blue) and quadrupole (red) of the unblinded DESI DR1 LRG high-$z$ (left) and QSO (right) with the best fit model (black). The errors are the standard deviation from the DR1 power-spectrum covariance matrix. The two lower panels give the relative difference normalised to the errors between the data and the best fit model. Gray regions around the best fit are the $1\sigma$ fluctuation of the model computed as the standard deviation of 1000 realizations of theory generated using the posteriors of the parameters.}
    \label{fig:bestfit_unblinded_data}
\end{figure}

The unblinded power spectrum for the LRG high-$z$ and the QSO (OQE), with their best fit that are given in \cref{tab:fit_unblind_png} (first block), are displayed in \cref{fig:bestfit_unblinded_data}. The gray regions around the best fit model (black lines) are computed as the standard deviation of 1000 realizations of the theory generated with the posteriors from the chains partially shown in \cref{fig:fit_unblinded_data} and centered around the best fit values given in \cref{tab:measurement_robustness_unblind}. This region illustrates the $1\sigma$ fluctuation where the model can be, although in this case, each bin $k$ is not independent.

Hence, the choice of the analysis done in \cref{sec:internal_consistency} remains viable, with the unblinded data requiring no further investigation. 


\subsection{Constraints on PNG with DESI DR1}
\begin{figure}
    \centering
    \includegraphics[scale=1.1]{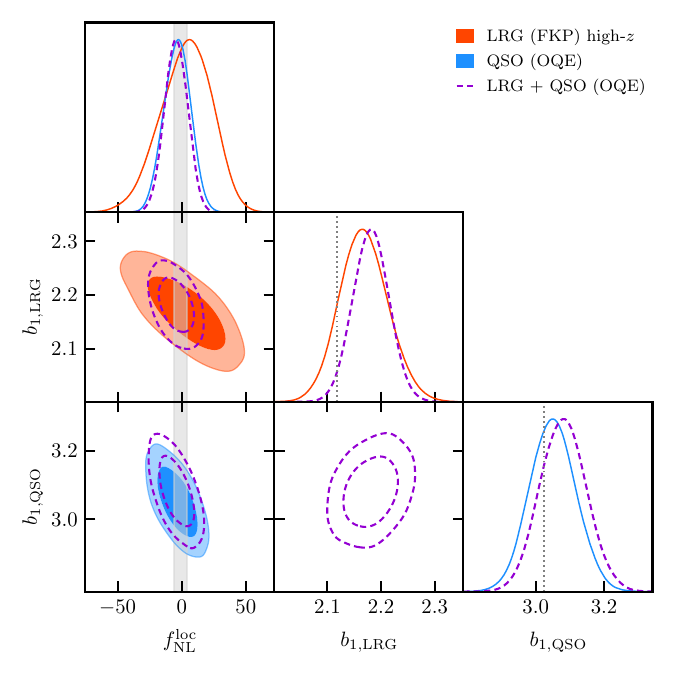}
    \caption{Posteriors in the $(b_1, \fnl)$ plane of our final constraints on PNG, obtained with unblinded DR1 LRG and QSO samples. The gray band is the constraint from Planck 2018. The MAP values are given in \cref{tab:fit_unblind_png}. The gray dotted lines are the values of $b_1$ for LRG and QSO samples measured from the monopole of the 2-point correlation function, see \cref{sec:linear_bias}. We do not consider the cross-covariance between the LRGs and QSOs.}
    \label{fig:fit_unblinded_final}
\end{figure}

The $\fnl$ constraints from the LRG and QSO samples: 
\begin{equation}
    \fnl = {6}_{-18}^{+22} \quad (68\%) \quad \text{[LRG]} \quad \text{and} \quad \fnl = {-2}_{-10}^{+11} \quad (68\%) \quad \text{[QSO]}
\end{equation}
are in good agreement such that one can combine them to improve the constraint on $\fnl$\footnote{As described in \cite{Barreira2020a}, a stellar mass selected sample leads to $p=0.55$, and thus, to $\fnl = 2_{-14}^{+15} \, (68\%)$ from the LRG sample.}. The best fit measurement and the posterior for the combination\footnote{Note that we combine the LRG high-$z$ sample even if it is not mentioned in the labels of \cref{fig:fit_unblinded_final} and in \cref{tab:fit_unblind_png}.} are shown in \cref{tab:fit_unblind_png} (second block) and in \cref{fig:fit_unblinded_final} in brown (\textit{resp.} in purple) for LRG high-$z$ (FKP) + QSO (FKP) (\textit{resp.} for LRG high-$z$ (FKP) + QSO(OQE)), as well as the summary of the independent measurements. Combining these two tracers leads to 
\begin{equation}
    \fnl = -3.6_{-9.1}^{+9.0} \quad (68\%) \quad \left[\text{LRG}+\text{QSO} \right],
\end{equation}
and improves the constraint by a factor of $10\%$ compared to QSO (OQE) alone. Moreover, we see that our combined results are compatible at the $<1\sigma$ level with $\fnl$ measurement from CMB data by \textit{Planck} in 2018 represented by the gray shaded region in \cref{fig:fit_unblinded_final}, given by $\fnl = -0.9 \pm 5.1$ at $68\%$ CL \cite{Akrami2020}. 

In this analysis, we have chosen a recent merger model for the QSO leading to $p=1.6$ since we do not have precise knowledge of $b_{\Phi}$, see \cite{Slosar2008, Breiding2024}. Adopting a more aggressive approach by setting $p=1.0$ leads to
\begin{equation}
    \fnl = 1.7_{-7.7}^{+8.4} \quad (68\%) \quad \left[\text{LRG } (p=1.0) + \text{QSO } (p=1.0) \right]
\end{equation}
when combining the LRG and the QSO sample, and reduces the $\fnl$ error bars by a factor of $11\%$. The gain for QSO (OQE) alone is about $10\%$ compared to using $p=1.6$, as shown in the corresponding row of \cref{tab:fit_unblind_png}. For the case of $p=1.0$, we recomputed the power spectra with the OQE weights (\cref{eqn:oqe_weights}), as well as the corresponding covariance matrix and the window matrix and its integral constraint contributions.

So far, we have built the bias model for $b_\Phi$ in our fitting procedure \cref{eqn:b_phi_with_p}, linked to the linear bias $b_1$. Without further information about $b_\Phi$, one can only measure directly the product $b_{\Phi} \times \fnl$ in \cref{eqn:pk_theo}, jointly with the linear bias $b_1$, which leads to 
\begin{equation}
    b_{\Phi} \fnl = {22}_{-63}^{+92} \quad (68\%) \quad \text{[LRG]} \quad \text{and} \quad b_{\Phi} \fnl = {-13}_{-56}^{+56} \quad (68\%) \quad \text{[QSO]}.
\end{equation}
The constraints with DESI DR1 data are given in the last row of \cref{tab:fit_unblind_png}, but the uncertainties remain too large to provide meaningful insights in the search for PNGs. Note that as already shown with the EZmocks (see \cref{tab:fit_ezmocks}), the errors on the recovered parameters for the OQE weights are equal than the ones obtained with FKP weights, since in the case of the OQE weighting scheme the higher effective redshift leads to a higher value of $b_\Phi$, thus increasing the errors on the combined parameter $b_\Phi \times \fnl$.

Since our combined constraints are compatible with $\fnl=0$ at the $<1\sigma$ level, we can proceed to a sanity check by comparing the unblinded errors to the ones obtained with EZmocks, where $\fnl = 0$ was used (see \cref{sec:validation_ezmocks}). We note that the errors derived from the fit with the unblinded QSOs are remarkably in agreement with the ones given in \cref{tab:fit_ezmocks} that we found during the fit of the mean on 1000 EZmocks. This comparison highlights that, at the scales over which we fit the data, our pipeline provides a fair description of the different statistical noises and accounts for the different systematic effects that appear in the real data. The LRG unblinded errors are larger than the EZmocks ones as expected, because we are considering here only the high-$z$ part of the sample. After re-normalising the EZmocks errors to the effective linear bias for the high-$z$ sample, the EZmocks and unblinded errors are compatible.



\begingroup 
\renewcommand{\arraystretch}{1.3} 
\begin{table}
    \centering
    \caption[]{Our final constraints on PNG, obtained with unblinded DR1 data. The central values are the best fit value from the \texttt{iminuit} minimization while the errors are the 1$\sigma$ credible interval from the chains. We use $k_{\rm min}=0.006~h\text{Mpc}^{-1}$ for the LRGs and  $0.003~h\text{Mpc}^{-1}$ for the QSOs. The posteriors are displayed in \cref{fig:fit_unblinded_final}.}
    \label{tab:fit_unblind_png}
    \resizebox{0.99\textwidth}{!}{%
    \begin{tabular}{lcc|cc|cc|c}
        \toprule
        \textbf{Individual ($\fnl$)}                   & $\fnl$ & \multicolumn{2}{c}{$b_{1}$} & \multicolumn{2}{c}{$s_{n, 0}$} & \multicolumn{2}{c}{$\Sigma_{s}$} \\
        \midrule
        \textbf{LRG (FKP) high-$z$} & ${6}_{-18}^{+22}$ & \multicolumn{2}{c}{${2.166}_{-0.045}^{+0.045}$ } & \multicolumn{2}{c}{${-0.024}_{-0.075}^{+0.074}$} & \multicolumn{2}{c}{${3.70}_{-0.53}^{+0.68}$} \\
        \textbf{QSO (OQE)} & ${-2}_{-10}^{+11}$ & \multicolumn{2}{c}{${3.048}_{-0.070}^{+0.064}$} & \multicolumn{2}{c}{${-0.039}_{-0.068}^{+0.076}$} & \multicolumn{2}{c}{${0.0}_{-1.41}^{+0.43}$} \\ 
        QSO (OQE) ($p=1.0$) & ${3.5}_{-7.4}^{+10.7}$ & \multicolumn{2}{c}{${2.792}_{-0.064}^{+0.059}$} & \multicolumn{2}{c}{${-0.038}_{-0.067}^{+0.064}$} & \multicolumn{2}{c}{${1.8}_{-1.41}^{+0.74}$}\\
        \bottomrule \toprule
        \textbf{Joint ($\fnl$)}          & $\fnl$ & \multicolumn{2}{c}{$b_1$} & \multicolumn{2}{c}{$s_{n, 0}$} & \multicolumn{2}{c}{$\Sigma_{s}$} \\
                  &        & {\footnotesize LRG} & {\footnotesize QSO} & {\footnotesize LRG} & {\footnotesize QSO} & {\footnotesize LRG} & {\footnotesize QSO}\\
        \midrule 
        \textbf{LRG + QSO (OQE)}   & ${-3.6}_{-9.1}^{+9.0}$ & {\footnotesize ${2.181}_{-0.033}^{+0.035}$} & {\footnotesize ${3.081}_{-0.070}^{+0.066}$} & {\footnotesize ${-0.051}_{-0.062}^{+0.063}$} & {\footnotesize ${-0.063}_{-0.078}^{+0.072}$} & {\footnotesize ${3.64}_{-0.47}^{+0.67}$} & {\footnotesize ${0.0}_{-1.36}^{+0.41}$} \\
        LRG + QSO (OQE) ($p=1.0$) & ${1.7}_{-7.7}^{+8.4}$ & {\footnotesize ${2.174}_{-0.032}^{+0.035}$} & {\footnotesize ${2.816}_{-0.059}^{+0.065}$} & {\footnotesize ${-0.042}_{-0.063}^{+0.063}$} & {\footnotesize ${-0.054}_{-0.073}^{+0.063}$} & {\footnotesize ${3.64}_{-0.52}^{+0.62}$} & {\footnotesize ${1.6}_{-1.45}^{+0.60}$} \\
        \bottomrule \toprule 
        \textbf{Individual ($b_{\Phi}\fnl)$} & $b_{\Phi}\fnl$ & \multicolumn{2}{c}{$b_1$} & \multicolumn{2}{c}{$s_{n, 0}$} & \multicolumn{2}{c}{$\Sigma_{s}$} \\
        \midrule
        LRG (FKP) high-$z$ & ${22}_{-63}^{+92}$ & \multicolumn{2}{c}{${2.166}_{-0.046}^{+0.045}$} & \multicolumn{2}{c}{${-0.024}_{-0.068}^{+0.082}$} & \multicolumn{2}{c}{${3.70}_{-0.49}^{+0.69}$} \\
        QSO (OQE) & ${-13}_{-56}^{+56}$ & \multicolumn{2}{c}{${3.036}_{-0.066}^{+0.072}$} & \multicolumn{2}{c}{${-0.002}_{-0.072}^{+0.070}$} & \multicolumn{2}{c}{${2.7}_{-0.89}^{+1.36}$} \\
        QSO (OQE) ($p=1.0$) & ${12}_{-47}^{+54}$ & \multicolumn{2}{c}{${2.800}_{-0.058}^{+0.058}$} & \multicolumn{2}{c}{${-0.039}_{-0.064}^{+0.058}$} & \multicolumn{2}{c}{${2.3}_{-1.0}^{+1.2}$} \\
        \bottomrule        
    \end{tabular}%
    }
\end{table} 
\endgroup

\subsection{Evaluation of systematic errors} \label{sec:systematic_errors}
The errors presented in this analysis exclude any contribution from systematic errors. In this section, we provide an estimate of these errors. They are summed up in \cref{tab:systematic_errors} where they are given as a percentage of the statistical error.

First, the geometrical description provided in \cref{sec:validation_ezmocks} recovers the parameters of interest within $\sim 0.3 \sigma$ in the worst case, see \cref{tab:fit_ezmocks}. Then, the systematic errors from the RIC contribution (\cref{tab:ric}) can be estimated by comparing the first and the last column of \cref{tab:ric} and can lead up to $0.2\sigma$ discrepancy. Similarly, for the AIC contribution (\cref{tab:aic}), the systematic error is about $0.06\sigma$, see \cref{tab:ric}, so that one can neglect these systematics that come from our theoretical model. Note that in the case of the LRG, the systematic is estimated from the fit up to $k_{\rm min} = 0.006 h\text{Mpc}^{-1}$, while the systematics from our model is for $k_{\rm min} = 0.003 h\text{Mpc}^{-1}$, such that they should be even smaller for this new $k_{\rm min}$.

Regarding the efficiency of the imaging weights, we adopted conservative cuts to guard against any residual effects though it weakens our final constraints. With the upcoming data releases and the reduction of the statistical uncertainty on $\fnl$, one will need to carefully assess the systematics in the future analysis. However, we can estimate it by looking the fluctuation in $\fnl$ as a function of the different imaging weights in \cref{tab:fits_blinded_test}, the imaging systematics contribute to $0.22\sigma$ for the LRGs and $0.38\sigma$ for the QSOs. Hence, the imaging errors are still the most important source of the systematics in our analysis.

Finally, assuming all these contributions independent, one can add them in quadrature such that 
\begin{equation}
    \sigma_{\rm tot} = \sqrt{\sigma_{\rm stat}^2 + \sigma_{\rm geo.}^2 + \sigma_{\rm RIC}^2 + \sigma_{\rm AIC}^2 + \sigma_{w_{\rm sys}}^2 }.
\end{equation}
Hence, in the worst case (\cref{tab:systematic_errors}), the total systematic error represents an increase of $8\%$ for LRGs and $12\%$ for QSOs over the statistical errors alone. We neglect them in this analysis, and further work may be required as the data sample size increases with the next release of DESI data.

\begin{table} 
    \centering
    \caption{Summary of the systematic error estimates in our analysis. All are given as a percentage of the associated statistical errors.}
    \label{tab:systematic_errors}
    \begin{tabular}{ccc}
        \toprule
        & LRG & QSO \\
        \midrule
        Geometry & $34\%$ & $26\%$ \\
        RIC           & $6\%$  & $20\%$ \\
        AIC           & $6\%$  & $6\%$  \\
        $w_{\rm sys}$ & $22\%$ & $38\%$ \\ 
        \bottomrule
    \end{tabular}
\end{table}
\section{Conclusions} \label{sec:conclusion}
In this work, we investigate the large-scale modes of the power spectrum from the largest LRG and QSO sample from the first DESI data release. We then proceed to measure the scale-dependent bias with DESI the luminous red galaxies and quasars, and obtain the first constraints on the PNG parameter $\fnl$ from DESI spectroscopic data.

We validate the power spectrum description that handles the geometrical effects used in DESI, and we develop an innovative method to include the radial integral constraint (RIC) and the angular integral constraint (AIC)  contributions into the power spectrum window matrix formalism. RIC appears due to the use of the shuffling method that consists of drawing the randoms redshifts from the data ones, while AIC stems from the geometrical impact inherent to the regression method used to correct for imaging systematics. This is the first time that both of these two contributions are handled in a multiplicative way for the analysis of the large scale modes of the power spectrum. In addition, we show that using the optimal quadratic weights as in \cite{Cagliari2023} improves our constraint on $\fnl$ and does not bias the signal if the linear bias is sufficiently well constrained.

We re-weight the QSO and LRG angular distribution to mitigate the dependence of the target selection on the different imaging properties of the DESI Legacy Imaging Surveys. The bias introduced by these weights, along with the imaging systematics themselves, was the most important systematic effect in the previous measurement using large-scale structures \cite{Mueller2021,Rezaie2023}. In this paper, our incorporation of the angular integral constraint (AIC) in the power spectrum window matrix enables us to not bias the measurement by using of standard imaging weights $w_{\rm sys}$. In addition, correctly handling this AIC contribution allows us to properly test the different imaging weights by looking at the impact of each weight on $\fnl$. 

Our $\fnl$ measurement is carried out, for the first time, with a fully blinded procedure that enables us to make the fiducial choices of our analysis without any confirmation bias. Specifically, we decide to use the QSOs ($0.8 < z < 3.1$) with the optimal quadratic weights from $k_{\rm min}=0.003~h\text{Mpc}^{-1}$ to $k_{\rm max}=0.08~h\text{Mpc}^{-1}$, and the high-$z$ part of the LRG sample ($0.6 < z < 1.1$) with FKP weights from $k_{\rm min}=0.006~h\text{Mpc}^{-1}$ to $k_{\rm max}=0.08~h\text{Mpc}^{-1}$. 

Combining both the DESI DR1 LRG and QSO samples, we find
\begin{equation}
  \fnl = \left\{
    \begin{array}{rl}
      -3.6_{-9.1}^{+9.0} & \quad (68\%) \quad \text{with } p_{\rm QSO} = 1.6\\ [1ex]
      1.7_{-7.7}^{+8.4}  & \quad (68\%) \quad \text{with } p_{\rm QSO} = 1.0
    \end{array} \right.
\end{equation}
leading to the tightest constraint to date using the large-scale structure and improving by a factor $\sim 2.3$ the previous one performed with the latest data release of eBOSS: $-23< \fnl <21$ \cite{Mueller2021,Cagliari2023}. Our new measurement, despite the significant reduction in error bars, is in agreement with the one from Planck 2018: $\fnl = -0.9 \pm 5.1$ at $68\%$ confidence \cite{Akrami2020}. For each tracer independently, we obtain
\begin{equation}
  \fnl = \left\{
    \begin{array}{rl}
      6_{-18}^{+22} & \quad (68\%) \quad \text{LRG only} \\ [1ex]
      -2_{-10}^{+11} & \quad (68\%) \quad \text{QSO only with } p_{\rm QSO} = 1.6 \\ [1ex]
      3.5_{-7.4}^{+10.7} & \quad (68\%) \quad \text{QSO only with } p_{\rm QSO} = 1.0
    \end{array} \right. . 
\end{equation}

This analysis could benefit from several improvements that we plan to implement in future analyses of DESI data. First, residual systematics in the LRG sample prevent us from using the full redshift sample and using larger scales up to $k_{\rm min} = 0.003~h\text{Mpc}^{-1}$. The upcoming DESI DR2 data with a larger LRG sample should help us investigate these systematics and correct them. Furthermore, to better handle imaging systematics, we plan to enhance our imaging mitigation method and allow their redshift dependence within individual bins. Such improvements should be feasible in the future, in the forthcoming DESI data releases.


For the DESI Y5 data, we forecast that with our current maximal scale $k_{\rm min} = 0.003~h\text{Mpc}^{-1}$ and by combining the full LRG and QSO samples, we should achieve $\sigma(\fnl) \sim 6.5$. Moreover, the geometrical model outlined here can be extended to larger scales, reaching  $k_{\rm min} = 0.001~h\text{Mpc}^{-1}$. We may need some additional validations for the window matrix computation and for the correct incorporation of the different integral constraint contributions into the window matrix. This larger maximal scale is expected to yield a 20–25$\%$ improvement in $\fnl$ constraints, resulting in  $\sigma(\fnl) \sim 5$. Thus, DESI data in the near future could approach the constraining power \referee{\citep{DESICollaboration2016, Sailer2021}} of the 2018 \textit{Planck} CMB results \citep{Planck18}. 



\section*{Data Availability}
Data from the plots in this paper are available on Zenodo\footnote{\url{https://doi.org/10.5281/zenodo.15185403}} as part of DESI’s Data Management Plan.
The data used in this analysis will be made public along with Data Release 1 of DESI planed in 2025\footnote{Details can be found here: \url{https://data.desi.lbl.gov/doc/releases/}}.
\acknowledgments
This material is based upon work supported by the U.S. Department of Energy (DOE), Office of Science, Office of High-Energy Physics, under Contract No. DE–AC02–05CH11231, and by the National Energy Research Scientific Computing Center, a DOE Office of Science User Facility under the same contract. Additional support for DESI was provided by the U.S. National Science Foundation (NSF), Division of Astronomical Sciences under Contract No. AST-0950945 to the NSF’s National Optical-Infrared Astronomy Research Laboratory; the Science and Technology Facilities Council of the United Kingdom; the Gordon and Betty Moore Foundation; the Heising-Simons Foundation; the French Alternative Energies and Atomic Energy Commission (CEA); the National Council of Science and Technology of Mexico (CONACYT); the Ministry of Science and Innovation of Spain (MICINN), and by the DESI Member Institutions: \url{https://www.desi.lbl.gov/collaborating-institutions}. Any opinions, findings, and conclusions or recommendations expressed in this material are those of the author(s) and do not necessarily reflect the views of the U. S. National Science Foundation, the U. S. Department of Energy, or any of the listed funding agencies.

The authors are honored to be permitted to conduct scientific research on Iolkam Du’ag (Kitt Peak), a mountain with particular significance to the Tohono O’odham Nation.

\bibliographystyle{biblio_style}
\bibliography{bibli}

\clearpage \newpage
\appendix
\section{Appendices}
\subsection{\referee{Impact of the fiber assignment}}\label{sec:fiber_assignment}

\begin{figure}
    \centering
    \includegraphics{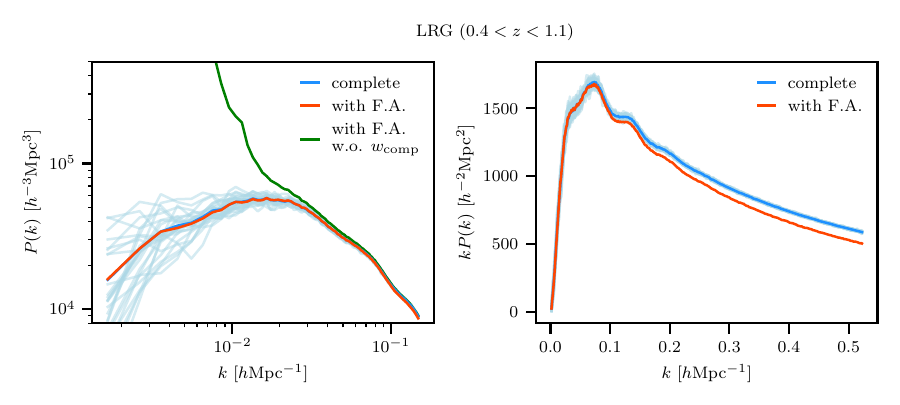}
    \caption{\referee{Impact of the DESI fiber assignment on the monopoles of mocks mimicking the DR1 LRG sample. The monopoles are the mean over 25 realizations. Blue lines are for the mocks without fiber assignment applied, reds are mocks with the fiber assignment applied and corrected by the completeness weights, while the green one is for mocks with fiber assignment applied but without the correction for completeness of the observations. For large scale study, the impact of the fiber assignment is therefore negligible.}}
\label{fig:impact_fa}
\end{figure}

\referee{As described in \cite{Schlafly2023}, the DESI survey follows a complex strategy with a limited amount of available fibers per observation, resulting in the fiber assignment step that could impact our cosmological measurement. The lack of available fibers has two effects: reducing the completeness of the observation (too many targets compared to the number of fibers) and reducing the number of closed pairs that could be observed (objects are too close to be reached by different fibers within the same observation). The first one is corrected by the completeness weights, $w_{\rm comp}$, as described in \cite{Ross2024}, while the second one, known as the \emph{fiber assignment impact at small scales}, can be modeled following \cite{Pinon2024}.}

\referee{\cref{fig:impact_fa} shows the impact of the fiber assignment on the monopole of the power spectrum computed from mocks that describe the DR1 LRG sample. These monopoles are the mean over 25 realizations. The mocks are the \texttt{AbacusSummit} mocks described in Section 11 of \cite{DESI2024II}. The blue lines are the monopoles when no fiber assignment is applied, known as complete mocks, while the green one is when the fiber assignment is applied and no weights are used to correct for its impact. However, once the completeness of the observation is corrected by $w_{\rm comp}$, the large-scale modes of the power spectrum are correctly recovered up to $k \sim 0.1~h\text{Mpc}^{-1}$. In addition, in our case, $\Sigma_s$ in the model \cref{eqn:pk_theo} can capture some residual effect if needed. Hence, any analysis using only large-scale modes of the power spectrum can neglect the fiber assignment impact if they correct for the completeness of the observation with $w_{\rm comp}$.}

\referee{Note that quasars have higher priority during the observation and are less dense, such that the impact of the fiber assignment is lower than for the LRGs. Naturally, we expect this effect to become less significant as the DESI survey continues to observe the sky and increases its completeness.}

\subsection{Impact of imaging systematic weights} \label{sec:impact_imaging_systematic_weights}
Despite the low stellar and extra-galactic contamination of the DESI galaxy clustering sample thanks to the spectroscopy, the fluctuations imprinted into the target density still play a major role at the large scales of the power spectrum. 

\begin{figure}
    \centering
    \includegraphics{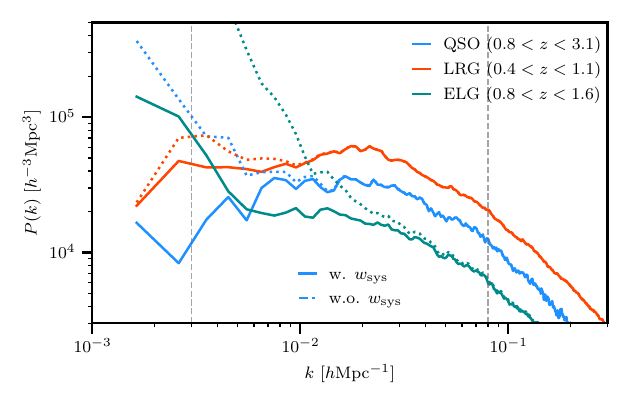}
    \caption{Monopoles of the DESI DR1 LRG, ELG, and QSO samples with the blinding scheme applied. Solid lines are with the imaging systematic weights, while the dotted ones are without.}
\label{fig:power_with_without_wsys}
\end{figure}

\Cref{fig:power_with_without_wsys} shows the monopole for the different dark time tracers of DESI, namely the Luminous Red Galaxies (LRG) with 0.4 < z < 1.1 \cite{Zhou2023}, Emission Line Galaxies (ELG) with 0.8 < z < 1.6 \cite{Raichoor2023} and the Quasars (QSO) with 0.8 < z < 3.1 \cite{Chaussidon2023}, with and without the imaging systematic weights that mitigate these spurious fluctuations.

The value of $\fnl$ measured without applying the imaging systematic weights for LRGs and QSOs was given in \cref{tab:fits_blinded_test}. Recall that, due to a residual systematic, which we are not able to correct in the LRG sample, we had to increased $k_{\rm min}$ from $0.003$ to $0.006~h\text{Mpc}^{-1}$ such that the imaging systematic weights have no impact on the monopole and on the $\fnl$ measurement.

\subsection{Linear bias for DESI LRGs and QSOs} \label{sec:linear_bias}
The optimal weighting scheme for measuring $\fnl$, detailed in \cref{sec:optimal_weight}, assumes a redshift evolution of the linear bias $b_1$. Hence, we provide here the measurements of the linear bias of the LRGs and QSOs in several redshift bins from the unblind DESI DR1 data.

The linear bias is measured in the monopole of the 2-point correlation function $\xi_{0}$ using only the scales: $30 \; h^{-1}\text{Mpc} < s < 80 \; h^{-1}\text{Mpc}$, so that one can only consider the simple Kaiser formula to take into account the redshift space distortion
\begin{equation}
    \xi_{0}(s) = \left(b_1^2 + \dfrac{2}{3}b_1f + \dfrac{1}{5}f^2\right) \xi_{\rm lin}(s),
\end{equation}
where $f$ is the growth rate and $\xi_{\rm lin}$ is the linear 2-points correlation function obtained from \texttt{Class}. Both of these quantities are evaluated assuming Planck 2018 cosmology \cite{Planck18} and at an effective redshift that is computed as in \cref{eqn:effective_redshift}. The covariance matrix that we use is estimated using the jackknife method with 128 sub-samples.

\begin{figure}
    \centering
    \includegraphics[scale=1]{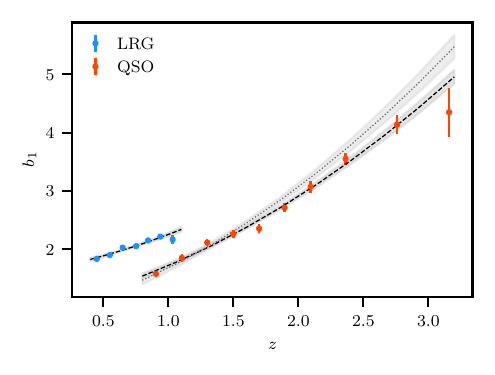}
    \caption{Redshift evolution of the linear bias $b_1$ as measured in the unblind DESI DR1 LRGs (blue) and QSOs (red). Black dashed lines are the best fit of \cref{eqn:appa_linear_bias_evolution} to these points, while the gray dotted line is the redshift evolution measured from the BOSS and eBOSS QSO sample \cite{Laurent2017}. Gray regions around the best fit are the 1$\sigma$ fluctuation of the model computed as the standard deviation of 1000 theory generated
    assuming Gaussian posteriors of the parameters with covariance matrix from the best fit.}
    \label{fig:bias_lrg_qso}
\end{figure}

The linear bias for the LRGs and QSOs is given in \cref{tab:bias_lrg_qso}, while these measurements are displayed in \cref{fig:bias_lrg_qso}. We propose a function to describe the redshift evolution and measure its parameters from the data; specifically,
\begin{equation} \label{eqn:appa_linear_bias_evolution}
    b_1(z) = a (1 + z)^2  + b,
\end{equation}
with $a, b = 0.209 \pm 0.025, 1.415 \pm 0.076$ for the DESI DR1 LRGs and $a, b = 0.237 \pm 0.010, 0.771 \pm 0.070$ for the DESI DR1 QSOs.

\begin{table}
    \centering  
    \caption{Measurements of the linear bias $b_1$ of the LRGs and QSOs from the unblind DESI DR1 data.} \label{tab:bias_lrg_qso}
    \begin{tabular}{lccc}
        \toprule
        Tracer  & $z_{\rm min}$ -- $z_{\rm max}$  & $z_{\rm eff}$ & $b_1$    \\ 
        \midrule
        LRG     & 0.4 -- 0.5 & 0.45 & $1.834 \pm 0.036$ \\
                & 0.5 -- 0.6 & 0.55 & $1.900 \pm 0.038$ \\
                & 0.6 -- 0.7 & 0.65 & $2.029 \pm 0.037$ \\
                & 0.7 -- 08  & 0.75 & $2.053 \pm 0.033$ \\
                & 0.8 -- 0.9 & 0.84 & $2.153 \pm 0.038$ \\
                & 0.9 -- 1.0 & 0.94 & $2.217 \pm 0.040$ \\
                & 1.0 -- 1.1 & 1.03 & $2.167 \pm 0.069$ \\ 
        \midrule
        QSO     & 0.8 -- 1.0 & 0.90 & $1.579 \pm 0.057$ \\
                & 1.0 -- 1.2 & 1.10 & $1.851 \pm 0.067$ \\
                & 1.2 -- 1.4 & 1.29 & $2.116 \pm 0.062$ \\
                & 1.4 -- 1.6 & 1.49 & $2.263 \pm 0.064$ \\
                & 1.6 -- 1.8 & 1.69 & $2.355 \pm 0.084$ \\
                & 1.8 -- 2.0 & 1.89 & $2.713 \pm 0.078$ \\
                & 2.0 -- 2.2 & 2.09 & $3.070 \pm 0.097$ \\
                & 2.2 -- 2.6 & 2.36 & $3.551 \pm 0.098$ \\
                & 2.6 -- 3.0 & 2.76 & $4.137 \pm 0.156$ \\
                & 3.0 -- 3.5 & 3.15 & $4.350 \pm 0.418$ \\
        \bottomrule
    \end{tabular}
\end{table}

Note that we can transform \cref{eqn:appa_linear_bias_evolution} to match the function used in \citep{Laurent2017}:
\begin{equation}
    b_1(z) = a_L \left[ (1 + z)^2 - 6.565 \right] + b_L,
\end{equation}
where $a_L, b_L = 0.237 \pm 0.010, 2.328 \pm 0.026$ for the DESI DR1 QSOs. For comparison, \cite{Laurent2017} found for the BOSS/eBOSS QSO sample a slightly higher bias at high redshift described by $a_L, b_L = 0.278 \pm 0.018, 2.393 \pm 0.042$.

\subsection{Impact of photometric region normalization on the power spectrum}\label{sec:appendix_pk_normalization}
As described in \cref{sec:pk_normalization}, one need to normalize South (NGC) to North when we compute the power spectrum on all the NGC as well as South (SGC) to the DES region for the full SGC part of the footprint. The normalization means that $\alpha$ in \cref{eqn:FKP_field} is set to match the corresponding data separately in each region. 

\begin{figure}
    \centering
    \includegraphics[scale=1]{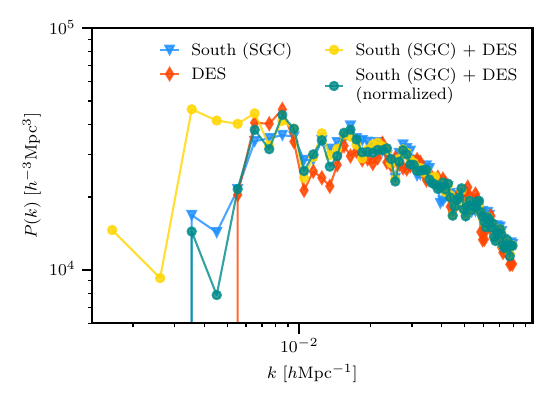}
    \caption{Comparison of the monopole of the \emph{blinded} QSO sample computed either on South (SGC) (blue), DES (red), South (SGC) + DES without the normalization (yellow) or with the normalization given in \cref{eqn:normalization_factor_region} (green).}
    \label{fig:south_des_pk}
\end{figure}

In particular, one needs to perform this renormalization between the South (SGC) and DES even though the two regions are from the same photometric survey. \Cref{fig:south_des_pk} shows the monopole for the blinded QSO sample without the normalization factor (yellow) that exhibits an excess of power at large scales compared with the monopole computed to the normalization factor given in \cref{eqn:normalization_factor_region} (green). Not accounting for this normalization would bias the measurement of primordial non-Gaussianity.

\subsection{Impact of quadrupole on PNG measurement} \label{sec:gain_with_quadrupole}
\Cref{tab:quadrupole_impact} shows the gain of including the quadrupole ($\ell=2$) during the fit. It leads only to few percent of improvement and can be neglected in this entire analysis, simplifying a lot the measurement with the OQE weights. However, for completeness, we include it during all the analysis.

\begin{table}
    \centering
    \caption{Result of the fit using the mean of the power spectrum over 1000 realizations including or not the quadrupole ($\ell=2$) with the
    corresponding covariance matrix for the LRGs and QSOs. Central values and the errors are both from the MCMC chains.}
    \label{tab:quadrupole_impact}
    \begin{tabular}{lcc|cc}
    \toprule
     & \multicolumn{2}{c}{with $\ell=2$} & \multicolumn{2}{c}{without $\ell=2$} \\
    \midrule  \vspace{1mm}
     & $\fnl$ & $b_1$ & $\fnl$ & $b_1$ \\ \midrule  \vspace{2mm}
    LRG Y1 (FKP) & ${5}_{-11}^{+18}$ & ${2.065}_{-0.037}^{+0.032}$ & ${7}_{-12}^{+18}$ & ${2.032}_{-0.044}^{+0.043}$ \\  \vspace{2mm}
    LRG Y5 (FKP) & ${3.8}_{-7.5}^{+10.7}$ & ${2.065}_{-0.024}^{+0.021}$ & ${2.5}_{-8.9}^{+9.8}$ & ${2.044}_{-0.044}^{+0.043}$ \\ 
    \midrule \vspace{1mm}
    QSO Y1 (FKP) & ${3.0}_{-13}^{+18}$ & ${2.440}_{-0.046}^{+0.050}$ & ${2}_{-14}^{+18}$ & ${2.427}_{-0.069}^{+0.060}$ \\  \vspace{2mm}
    QSO Y5 (FKP) & ${4.6}_{-9.6}^{+9.9}$ & ${2.435}_{-0.033}^{+0.030}$ & ${3}_{-10.0}^{+10.3}$ & ${2.417}_{-0.051}^{+0.050}$ \\ 
    \midrule \vspace{1mm}
    QSO Y1 (OQE) & ${-4}_{-10}^{+13}$ & ${3.088}_{-0.080}^{+0.064}$ & ${1}_{-10}^{+13}$ & ${3.009}_{-0.088}^{+0.080}$ \\  \vspace{2mm}
    QSO Y5 (OQE) & ${-1.4}_{-7.8}^{+7.8}$ & ${3.083}_{-0.049}^{+0.045}$ & ${1.8}_{-7.4}^{+8.7}$ & ${2.994}_{-0.062}^{+0.071}$ \\
    \bottomrule
    \end{tabular}
\end{table}

\subsection{Global Integral Constraint} \label{sec:gic}
As explained in \cite{Peacock1991, Nichol1992, Beutler2014, Wilson2017, DeMattia2019}, the real mean density of the tracer in the Universe, used in the FKP field \pcref{eqn:FKP_field}, is unknown and has to be estimated from our finite survey volume. Such an estimation suppresses  all the fluctuations with scales larger than the survey size such $P(k) \underset{k \to 0}{\longrightarrow} 0$. This effect is known as the \emph{global integral constraint} (GIC) and can be taken into account in the convolved prediction \pcref{eqn:convolved_prediction}:
\begin{equation} \label{eqn:gic_contribution}
    \left(P_{\ell}^{\rm obs}\right)_{i} = \left( \mathcal{W}_{\ell \ell^\prime} \right)_{ij} \left(P_{\ell^\prime}\right)_{j} - \left(\mathcal{W}^{\mathrm{GIC}}_{\ell \ell^\prime}\right)_{ij} \left(P_{\ell\prime}^{\mathrm{theo}}\right)_{j},
\end{equation}
where the summation runs over $\ell^\prime$ and $j$. The contribution of the global integral constraint $\left(\mathcal{W}^{\mathrm{GIC}}_{\ell \ell^\prime}\right)_{ij}$ is given by 
\begin{equation} \label{eqn:wm_gic}
    \left(\mathcal{W}^{\mathrm{GIC}}_{\ell \ell^\prime}\right)_{ij} = \left(\mathcal{W}_{\ell 0} \right)_{i0} / \left(\mathcal{W}_{0 0} \right)_{00} \left(\mathcal{W}_{0 \ell^\prime} \right)_{0j}.
\end{equation}

\begin{figure}
    \centering
    \begin{subfigure}{0.48\textwidth}
        \centering
        \includegraphics[scale=0.95]{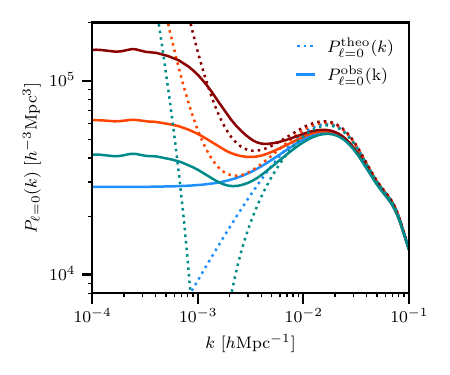}
        \caption{LRG ($0.4 < z < 1.1$) with FKP.}
    \end{subfigure}
    \begin{subfigure}{0.48\textwidth}
        \centering
        \includegraphics[scale=0.95]{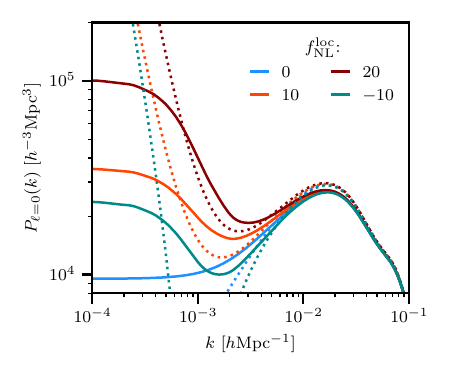}
        \caption{QSO ($0.8 < z < 3.1$) with FKP.}
    \end{subfigure}
    \caption{Monopole of the power spectrum for LRGs (left) and QSOs (right) for several values of $\fnl$ at very large-scales. The convolved (\textit{resp.} unconvolved) monopole is shown in solid (\textit{resp.} dashed) lines. The value of $P_{\ell=0}(k \rightarrow 0)$ depends of $\fnl$.}
    \label{fig:pk_0}
\end{figure}

The GIC contribution in \cref{eqn:gic_contribution} is directly proportional to $P_{\ell=0}(k \rightarrow 0)$ \cite{DeMattia2019} such that this contribution depends on the value of $\fnl$ used to evaluate the theory. The convolved power spectrum for LRGs and QSOs at very large scales ($k < 10^{-3}~[h/\mathrm{Mpc}]$) is displayed \cref{fig:pk_0}. The convolved monopole of the power spectrum converges at very large scales to a non-zero value such that the contribution of the GIC is 4 (\textit{resp.} 10) times larger for a situation with $\fnl=20$ compared to $\fnl=0$.

\begin{figure}
    \centering
    \begin{subfigure}{0.485\textwidth}
        \centering
        \includegraphics[scale=0.90]{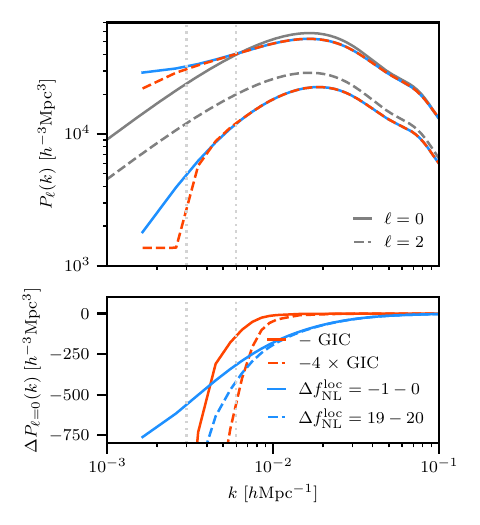}
        \caption{LRG ($0.4 < z < 1.1$) with FKP.}
    \end{subfigure}
    \begin{subfigure}{0.485\textwidth}
        \centering
        \includegraphics[scale=0.90]{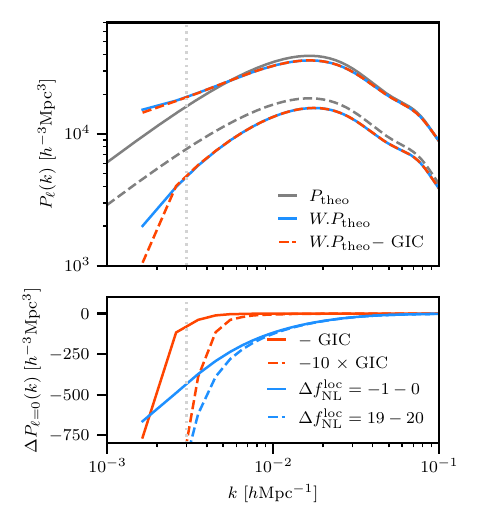}
        \caption{QSO ($0.8 < z < 3.1$) with FKP.}
    \end{subfigure}
    \caption{Top panel: Convolved power spectrum with the global integral constraint contribution (red dashed lines) on the monopole and the quadrupole for the LRGs (left) and the QSOs (right) compared to the convolved power spectrum given by \cref{eqn:convolved_prediction} (blue lines). The gray lines are the unconvolved monopoles. Bottom panel: Comparison between the GIC contribution for $\fnl = 0$ (\textit{resp.} $\fnl=20$) in red solid (\textit{resp.} dashed) lines and the amplitude of the signal of $\Delta \fnl = - 1$ at $\fnl=0$ (\textit{resp.} $\fnl=20$) in solid (\textit{resp.} dashed) blue lines.}
    \label{fig:gic}
\end{figure}

The GIC contribution is shown in \cref{fig:gic} for the DESI DR1 sample. As expected, the GIC contribution is larger for the LRGs since they probe a smaller volume of the Universe than the QSOs. In the QSO case, the GIC contribution, even for an underlying value of $\fnl=20$ is negligible compared to the fluctuation of the signal due to a modification of $\Delta \fnl = 1$ for the scales of interest ($k > 0.003 ~[h/\mathrm{Mpc}]$). For the LRGs, the contribution becomes comparable to the fluctuation of the signal due to a modification of $\Delta \fnl = 1$ for the largest scales. As shown in the following, the expected sensitivity for $\fnl$ with this sample is about $14.5$ so that one can neglect this impact, especially because this contribution impacts only the largest scales that are the one with the most statistical uncertainty, see \pcref{fig:data_vs_ezmocks}. Since these scales are also very sensitive to the residual imaging systematics, such that one need for the data to increase our fiducial value (see the following section, \cref{sec:validation_ezmocks}) of $k_{\rm min}$ to $0.006~[h/\mathrm{Mpc}]$ where the contribution becomes, as for the QSOs, negligible compare the fluctuation of the signal for $\Delta \fnl = 1$. 

Since, this contribution does not contribute significantly to our measurement, we do not include its contribution to the window matrix $\mathcal{W}$. Note that this contribution is lower due to the increase of the survey size of the upcoming DESI data release, such that one can certainly still neglect it even in light of the reduction in errors associated with this new data.

\subsection{Target selection dependence on the PSF detection at low-z.} \label{sec:appendix_impact_1.3_cut_wsys}
\begin{figure}
    \centering
    \includegraphics[scale=1]{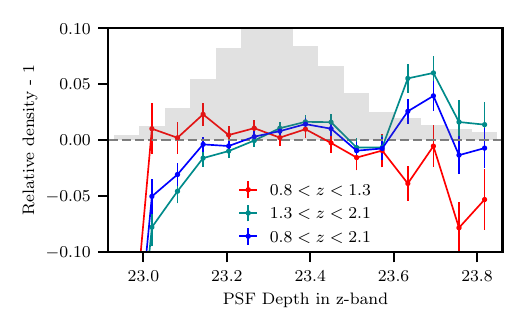}
    \caption{Relative density of the QSO sample for different redshift subsamples as a function of PSF Depth $z$ in the South (NGC+SGC) region. The histogram represents the fraction of objects in each bin of the observational feature and the error bars are the estimated standard deviation of the normalized density in each bin.}
    \label{fig:qso_new_cut}
\end{figure}

As noted in \cite{Chaussidon2023} (Fig. 11), the QSO target selection requires objects to be classified as point sources, which leads to the exclusion of many low-redshift quasars classified as extended sources as their host galaxies were resolved. Improved photometry exacerbates this issue, as more host galaxies are resolved, meaning the number of low-redshift quasars affected more so than the number of their higher-redshift counterparts. This behavior is demonstrated in \cref{fig:qso_new_cut}, where the low-redshift sample ($0.8 < z < 1.3$) in red shows a distinct pattern compared to the mid-redshift sample ($1.3 < z < 2.1$) in green. The choice $z=1.3$ is motivated by comparing the relative density as a function of the PSF Depth $z$ of several redshift bins.

The figure highlights the dependence on PSF depth in the z-band, the deepest band in the Legacy Surveys which drives object morphology classification. The distinct behavior of the low-redshift sample is masked when analysing the full range ($0.8 < z < 2.1$) and cannot be properly addressed with imaging systematic weights computed across the entire sample. However, the small fraction of low-redshift objects ($23\%$ of the data) has a significant impact on the large-scale modes of the power spectrum if not properly accounted for.

\begin{figure}
    \centering
    \includegraphics[scale=1]{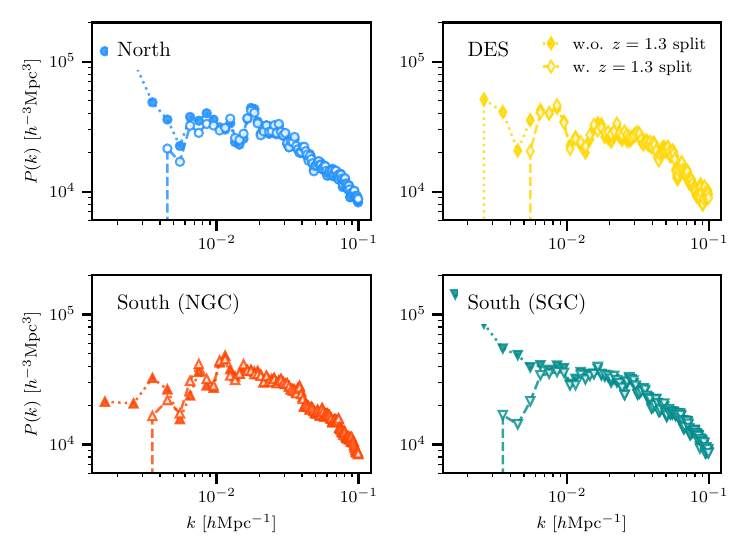}
    \caption{Comparison of the monopole of the \emph{blinded} QSO sample with (empty marker) or without (solid marker) the additional split at $z=1.3$ to compute the imaging systematic weights.}
    \label{fig:impact_1.3_cut_wsys}
\end{figure}

\Cref{fig:impact_1.3_cut_wsys} compares the monopole across different photometric regions: using a single weight for the full sample (solid markers) versus separate weights (open markers) for the two redshift ranges, 0.8 < z < 1.3 and 1.3 < z < 2.1. Without these tailored weights, it becomes impossible to correctly measure the large-scale modes of the power spectrum.

We note that this improvement in imaging systematic weights does not yet account for any redshift dependence within the sub-samples. Addressing this limitation is left to future work, expected with the next DESI data release.

\subsection{Additional tables}
In this appendix, we include several tables that were used to generate figures in the main text and that we may need for the assessment of systematic errors.

\begingroup 
\renewcommand{\arraystretch}{1.1} 
\begin{table}
    \centering
    \caption[]{Best fit results on the mocks for the LRGs and QSOs (mean over 10 realizations) using either FKP and OQE weights, for various configurations of imaging weights with the radial integral constraint but without the angular integral constraint. The errors are from the minimization performed with \texttt{iminuit}. The $\fnl$ values are plotted in \cref{fig:fits_null_test} for visualization.}
    \label{tab:fits_null_test}
    \resizebox{0.99\textwidth}{!}{%
    \begin{tabular}{llcccc}
        \toprule
        & & $f_{\rm NL}^{\mathrm{loc}}$ & $b_{1}$ & $s_{n, 0}$ & $\Sigma_{s}$   \\
        \midrule
        LRG (FKP)  & No weight & $4 \pm 13$ & $2.066 \pm 0.032$ & $0.044 \pm 0.061$ & $4.32 \pm 0.44$ \\ 
        \small{$k_{\rm min} = 0.003~h\text{Mpc}^{-1}$}& Default (RF 128) & $-24 \pm 17$ & $2.102 \pm 0.038$ & $0.009 \pm 0.068$ & $4.58 \pm 0.41$ \\
                   & Default (RF 256) & $-22 \pm 14$ & $2.103 \pm 0.036$ & $0.003 \pm 0.065$ & $4.61 \pm 0.41$ \\ 
                   & \texttt{Default (Linear 256)} & $1 \pm 16$ & $2.068 \pm 0.035$ & $0.044 \pm 0.063$ & $4.31 \pm 0.44$ \\ 
                   & PSF Depth (Linear 256) & $1 \pm 16$ & $2.068 \pm 0.034$ & $0.045 \pm 0.063$ & $4.31 \pm 0.44$ \\ 
                   & Same Feature Zbin (Linear 128) & $-5 \pm 19$ & $2.075 \pm 0.038$ & $0.038 \pm 0.067$ & $4.31 \pm 0.44$ \\ 
                   & Same Feature Zbin (Linear 256) & $-6 \pm 18$ & $2.076 \pm 0.037$ & $0.035 \pm 0.066$ & $4.33 \pm 0.44$ \\ 
                   & With DES (Linear 256) & $0. \pm 16$ & $2.069 \pm 0.035$ & $0.044 \pm 0.063$ & $4.31 \pm 0.44$ \\ 
                   & 4 regions (Linear 256) & $-4 \pm 18$ & $2.074 \pm 0.037$ & $0.038 \pm 0.066$ & $4.32 \pm 0.44$ \\ 
        \midrule
        LRG (FKP)   & No weight & $7 \pm 18$ & $2.062 \pm 0.038$ & $0.049 \pm 0.067$ & $4.31 \pm 0.44$ \\ 
        \small{$k_{\rm min} = 0.006~h\text{Mpc}^{-1}$} & Default (RF 128) & $-19 \pm 20$ & $2.096 \pm 0.040$ & $0.016 \pm 0.070$ & $4.57 \pm 0.42$ \\ 
                    & Default (RF 256) & $-23 \pm 20$ & $2.104 \pm 0.041$ & $0.001 \pm 0.070$ & $4.60 \pm 0.41$ \\ 
                    & \texttt{Default (Linear 256)} & $1 \pm 19$ & $2.070 \pm 0.039$ & $0.042 \pm 0.068$ & $4.32 \pm 0.44$ \\ 
                    & PSF Depth (Linear 256) & $1 \pm 19$ & $2.070 \pm 0.039$ & $0.042 \pm 0.068$ & $4.32 \pm 0.44$ \\ 
                    & Same Feature Zbin (Linear 128) & $-4 \pm 19$ & $2.075 \pm 0.039$ & $0.037 \pm 0.069$ & $4.32 \pm 0.44$ \\ 
                    & Same Feature Zbin (Linear 256) & $-5 \pm 20$ & $2.076 \pm 0.039$ & $0.036 \pm 0.069$ & $4.33 \pm 0.44$ \\ 
                    & With DES (Linear 256) & $1 \pm 19$ & $2.069 \pm 0.039$ & $0.042 \pm 0.068$ & $4.32 \pm 0.44$ \\ 
                    & 4 regions (Linear 256) & $-2 \pm 19$ & $2.073 \pm 0.039$ & $0.039 \pm 0.068$ & $4.33 \pm 0.44$ \\ 
        \midrule \midrule
        QSO (FKP) & No weight & $6 \pm 16$ & $2.428 \pm 0.047$ & $0.007 \pm 0.053$ & $2.71 \pm 0.87$ \\ 
                  & Default (RF 256) & $-16 \pm 15$ & $2.467 \pm 0.048$ & $-0.017 \pm 0.053$ & $3.19 \pm 0.74$ \\ 
                  & No PSF Depth (RF 128) & $5 \pm 15$ & $2.431 \pm 0.046$ & $-0.001 \pm 0.052$ & $3.20 \pm 0.75$ \\ 
                  & \texttt{Default (Linear 128)} & $0 \pm 17$ & $2.439 \pm 0.049$ & $0.000 \pm 0.054$ & $2.75 \pm 0.85$ \\ 
                  & Default (Linear 256) & $-3 \pm 17$ & $2.445 \pm 0.050$ & $-0.004 \pm 0.054$ & $2.74 \pm 0.86$ \\ 
                  & No PSF Size (Linear 128) & $1 \pm 17$ & $2.437 \pm 0.049$ & $0.001 \pm 0.053$ & $2.75 \pm 0.85$ \\ 
        \midrule
        QSO (OQE) & No weight & $0 \pm 12$ & $3.062 \pm 0.071$ & $-0.020 \pm 0.074$ & $0.0 \pm 9.2$ \\ 
                  & Default (RF 256) & $-10 \pm 11$ & $3.101 \pm 0.069$ & $-0.054 \pm 0.073$ & $0.0 \pm 7.9$ \\ 
                  & No PSF Depth (RF 128) & $4.0 \pm 9.9$ & $3.050 \pm 0.066$ & $-0.029 \pm 0.070$ & $0.0 \pm 6.9$ \\ 
                  & \texttt{Default (Linear 128)} & $-3 \pm 13$ & $3.071 \pm 0.075$ & $-0.027 \pm 0.077$ & $0.0 \pm 6.3$ \\ 
                  & Default (Linear 256) & $-6 \pm 14$ & $3.085 \pm 0.079$ & $-0.035 \pm 0.079$ & $0.0 \pm 5.2$ \\ 
                  & No PSF Size (Linear 128) & $-2 \pm 13$ & $3.070 \pm 0.074$ & $-0.026 \pm 0.076$ & $0.0 \pm 9.1$ \\
        \bottomrule
    \end{tabular}%
    }
\end{table}
\endgroup

\begin{table}
    \centering
    \caption[]{Result of the fit on the mocks using the DR1 covariance matrix on the mean of the power spectrum with FKP weights over 30 realizations for the LRGs and QSOs without (first column) / with (second column) the imaging systematic weights and adding the angular integral constraint contribution into the window function (third column). The central values and errors are from the \texttt{minuit} minimization. The first column is comparable to the result from \cref{tab:fit_ezmocks}. These values are sum up in \cref{fig:aic}} 
    \label{tab:aic}
      \begin{tabular}{lcccc}
        \toprule
                             & params     & \cellcolor{green!30} w.o. $w_{sys}$    & \cellcolor{red!30} w. $w_{sys}$      & \cellcolor{green!30} w. $w_{sys}$ + AIC \\
        \midrule
        DR1 LRG (FKP)        & $\fnl$     & $4 \pm 13$        & $-24 \pm 17$       & $-1 \pm 19$        \\ 
        Default              & $b_1$      & $2.067 \pm 0.032$ & $2.102 \pm 0.038$  & $2.074 \pm 0.037$  \\
        (RF 256)             & $s_{n,0}$  & $0.042 \pm 0.061$ & $0.009 \pm 0.068$  & $0.026 \pm 0.065$  \\
                             & $\Sigma_s$ & $4.32 \pm 0.44$   & $4.58 \pm 0.41$    & $4.50 \pm 0.43$    \\ 
        \midrule
        Default              & $\fnl$     & $4 \pm 13$        & $0. \pm 16$        & $3 \pm 14$         \\
        (Linear 256)         & $b_1$      & $2.067 \pm 0.032$ & $2.070 \pm 0.035$  & $2.066 \pm 0.033$  \\ 
                             & $s_{n,0}$  & $0.042 \pm 0.061$ & $0.042 \pm 0.063$  & $0.044 \pm 0.062$  \\
                             & $\Sigma_s$ & $4.32 \pm 0.44$   & $4.30 \pm 0.44$    & $4.31 \pm 0.44$    \\ 
        \midrule
        Same Feature Zbin    & $\fnl$     & $4 \pm 13$        & $-6 \pm 19$        & $3 \pm 15$         \\
        (Linear 256)         & $b_1$      & $2.067 \pm 0.032$ & $2.076 \pm 0.038$  & $2.065 \pm 0.033$  \\
                             & $s_{n,0}$  & $0.042 \pm 0.061$ & $0.035 \pm 0.067$  & $0.047 \pm 0.062$  \\
                             & $\Sigma_s$ & $4.32 \pm 0.44$   & $4.31 \pm 0.44$    & $4.33 \pm 0.44$    \\
        \midrule \midrule
        DR1 QSO (FKP)        & $\fnl$     & $6 \pm 16$        & $-16 \pm 15$       & $4 \pm 16$         \\
        Default              & $b_1$      & $2.428 \pm 0.047$ & $2.467 \pm 0.048$  & $2.428 \pm 0.048$  \\ 
        (RF 256)             & $s_{n,0}$  & $0.007 \pm 0.053$ & $-0.017 \pm 0.053$ & $0.013 \pm 0.053$  \\
                             & $\Sigma_s$ & $2.71 \pm 0.87$   & $3.19 \pm 0.74$    & $2.91 \pm 0.81$    \\
        \midrule
        Default              & $\fnl$     & $6 \pm 16$        & $0. \pm 17$        & $5 \pm 16$         \\
        (Linear 128)         & $b_1$      & $2.428 \pm 0.047$ & $2.439 \pm 0.049$  & $2.434 \pm 0.048$  \\ 
                             & $s_{n,0}$  & $0.007 \pm 0.053$ & $0.000 \pm 0.054$  & $-0.003 \pm 0.053$ \\
                             & $\Sigma_s$ & $2.71 \pm 0.87$   & $2.75 \pm 0.85$    & $2.42 \pm 0.94$    \\
        \midrule
        Default              & $\fnl$     & $6 \pm 16$        & $-3 \pm 17$        & $4 \pm 17$        \\ 
        (Linear 256)         & $b_1$      & $2.428 \pm 0.047$ & $2.445 \pm 0.050$  & $2.433 \pm 0.049$ \\
                             & $s_{n,0}$  & $0.007 \pm 0.053$ & $-0.004 \pm 0.054$ & $0.001 \pm 0.053$ \\ 
                             & $\Sigma_s$ & $2.71 \pm 0.87$   & $2.74 \pm 0.86$    & $2.55 \pm 0.90$   \\ 
        \midrule
      \end{tabular}
  \end{table}

\begingroup 
\renewcommand{\arraystretch}{1.1} 
\begin{table}
    \centering
    \caption[]{Values used for \cref{fig:fits_blinded_test}. Best fit results for the DR1 \emph{blinded} LRGs and \emph{blinded} QSOs using either FKP ot OQE weights, for various configurations of imaging weights. All the fits include the radial and angular integral constraint contributions. The errors and the central values are from the minimization performed with \texttt{iminuit} and the covariance matrix is the one from the 1000 EZmocks. The green rows are the default choices for the analysis on the unblinded data.}
    \label{tab:fits_blinded_test}
    \resizebox{0.99\textwidth}{!}{%
    \begin{tabular}{llcccc}
        \toprule
        &        features (method)            & $f_{\rm NL}^{\mathrm{loc}}$ & $b_{1}$ & $s_{n, 0}$ & $\Sigma_{s}$        \\
        \midrule
        LRG (FKP)
        & No imaging weights & $42.6 \pm 6.9$ & $2.018 \pm 0.028$ & $0.125 \pm 0.056$ & $4.50 \pm 0.43$ \\
        \small{$k_{\rm min} = 0.003~h\text{Mpc}^{-1}$}
        & Default (RF 128) & $16 \pm 10$ & $2.069 \pm 0.029$ & $0.036 \pm 0.058$ & $4.53 \pm 0.41$ \\ 
        & Default (RF 256) & $22.5 \pm 9.0$ & $2.061 \pm 0.029$ & $0.059 \pm 0.058$ & $4.81 \pm 0.40$ \\ 
        & Default (Linear 256) & $24.2 \pm 9.3$ & $2.041 \pm 0.029$ & $0.089 \pm 0.057$ & $4.52 \pm 0.42$ \\ 
        & PSF Depth (Linear 256) & $25.1 \pm 9.1$ & $2.041 \pm 0.029$ & $0.090 \pm 0.057$ & $4.52 \pm 0.42$ \\ 
        & Same Feature Zbin (Linear 128) & $20. \pm 10$ & $2.044 \pm 0.030$ & $0.086 \pm 0.058$ & $4.56 \pm 0.42$ \\ 
        & Same Feature Zbin (Linear 256) & $25.4 \pm 9.3$ & $2.042 \pm 0.029$ & $0.088 \pm 0.057$ & $4.56 \pm 0.42$ \\ 
        & With DES (Linear 256) & $26.5 \pm 8.9$ & $2.039 \pm 0.029$ & $0.097 \pm 0.057$ & $4.61 \pm 0.42$ \\ 
        & 4 regions (Linear 256) & $25.5 \pm 9.3$ & $2.039 \pm 0.029$ & $0.092 \pm 0.057$ & $4.53 \pm 0.42$ \\ 
        \midrule
        LRG (FKP)
        & No imaging weights     & $28 \pm 15$ & $2.041 \pm 0.033$ & $0.090 \pm 0.061$ & $4.57 \pm 0.42$ \\ 
        \small{$k_{\rm min} = 0.006~h\text{Mpc}^{-1}$}
        & Default (RF 128) & $27 \pm 16$ & $2.059 \pm 0.032$ & $0.046 \pm 0.060$ & $4.48 \pm 0.42$ \\ 
        & Default (RF 256) & $28 \pm 14$ & $2.059 \pm 0.031$ & $0.061 \pm 0.060$ & $4.82 \pm 0.40$ \\ 
        & \cellcolor{green!30} Default (Linear 256) & $24 \pm 18$ & $2.044 \pm 0.035$ & $0.086 \pm 0.063$ & $4.54 \pm 0.42$ \\ 
        & PSF Depth (Linear 256) & $25 \pm 17$ & $2.043 \pm 0.035$ & $0.087 \pm 0.063$ & $4.54 \pm 0.42$ \\ 
        & Same Feature Zbin (Linear 128) & $20. \pm 18$ & $2.048 \pm 0.035$ & $0.081 \pm 0.063$ & $4.58 \pm 0.42$ \\ 
        & Same Feature Zbin (Linear 256) & $28 \pm 17$ & $2.041 \pm 0.034$ & $0.089 \pm 0.062$ & $4.57 \pm 0.42$ \\ 
        & With DES (Linear 256) & $37 \pm 15$ & $2.029 \pm 0.032$ & $0.108 \pm 0.060$ & $4.60 \pm 0.42$ \\ 
        & 4 regions (Linear 256) & $26 \pm 18$ & $2.041 \pm 0.034$ & $0.090 \pm 0.062$ & $4.55 \pm 0.42$ \\
        \midrule \midrule
        QSO (FKP)
        & No imaging weights & $140. \pm 17$ & $2.109 \pm 0.039$ & $0.274 \pm 0.044$ & $7.61 \pm 0.50$ \\ 
        \small{$k_{\rm min} = 0.003~h\text{Mpc}^{-1}$}
        & Default (RF 256) & $35 \pm 19$ & $2.218 \pm 0.044$ & $0.213 \pm 0.049$ & $7.58 \pm 0.47$ \\ 
        & no PSF Depth (RF 128) & $64 \pm 21$ & $2.203 \pm 0.042$ & $0.204 \pm 0.046$ & $7.42 \pm 0.48$ \\ 
        & Default (Linear 256) & $31 \pm 21$ & $2.206 \pm 0.046$ & $0.204 \pm 0.049$ & $7.37 \pm 0.48$ \\ 
        & \cellcolor{green!30} Default (Linear 128) & $28 \pm 22$ & $2.210 \pm 0.048$ & $0.196 \pm 0.050$ & $7.23 \pm 0.48$ \\ 
        & no PSF Size (Linear 128) & $34 \pm 22$ & $2.203 \pm 0.047$ & $0.208 \pm 0.050$ & $7.35 \pm 0.48$ \\ 
        \midrule
        QSO (OQE)
        & No imaging weights & $39 \pm 11$ & $2.689 \pm 0.061$ & $0.241 \pm 0.063$ & $7.50 \pm 0.63$ \\ 
        \small{$k_{\rm min} = 0.003~h\text{Mpc}^{-1}$}
        & Default (RF 256) & $7 \pm 18$ & $2.790 \pm 0.078$ & $0.169 \pm 0.073$ & $7.68 \pm 0.60$ \\ 
        & no PSF Depth (RF 128) & $-1 \pm 13$ & $2.782 \pm 0.069$ & $0.190 \pm 0.069$ & $7.77 \pm 0.61$ \\ 
        & Default (Linear 256) & $-4 \pm 14$ & $2.799 \pm 0.073$ & $0.171 \pm 0.071$ & $7.85 \pm 0.60$ \\ 
        & \cellcolor{green!30} Default (Linear 128) & $-9 \pm 15$ & $2.822 \pm 0.075$ & $0.147 \pm 0.072$ & $7.62 \pm 0.60$ \\ 
        & no PSF Size (Linear 128) & $-10. \pm 14$ & $2.824 \pm 0.074$ & $0.148 \pm 0.071$ & $7.67 \pm 0.60$ \\ 
        \bottomrule
    \end{tabular}%
    }
\end{table}
\endgroup

\begingroup 
\renewcommand{\arraystretch}{1.3} 
\begin{table}
    \centering
    \caption[]{Best-fit results from the \emph{blinded} LRGs and QSOs for different variations of our fiducial analysis. The central values are best fit value from the \texttt{iminuit} minimization while the errors are the 1$\sigma$ credible intervals from the chains. The window functions used during these fits contain both RIC and AIC contributions and were recomputed for the different configurations when it was needed. We use $k_{\rm min}=0.006~h\text{Mpc}^{-1}$ for the LRGs and  $0.003~h\text{Mpc}^{-1}$ for the QSOs. The posteriors are displayed in \cref{fig:fit_blinded_data}.}
    \label{tab:measurement_robustness}
    \begin{tabular}{llcccc}
        \toprule
        &  & $f_{\rm NL}^{\mathrm{loc}}$ & $b_{1}$ & $s_{n, 0}$ & $\Sigma_{s}$ \\
        \midrule
        LRG (FKP)
        & fiducial                             & ${24}_{-15}^{+19}$ & ${2.044}_{-0.039}^{+0.032}$ & ${0.086}_{-0.063}^{+0.066}$ &${4.54}_{-0.42}^{+0.44}$\\
        $(0.4 < z < 1.1)$
        & NGC                                  & ${27}_{-16}^{+25}$ & ${2.062}_{-0.046}^{+0.043}$ & ${0.054}_{-0.082}^{+0.080}$ &${4.86}_{-0.48}^{+0.57}$\\
        & SGC                                  & ${-7}_{-26}^{+25}$ & ${2.051}_{-0.057}^{+0.057}$ & ${0.08}_{-0.11}^{+0.10}$    &${3.81}_{-0.67}^{+1.13}$\\
        & $(0.4 < z < 0.8)$                    & ${16}_{-28}^{+33}$ & ${1.993}_{-0.051}^{+0.050}$ & ${-0.01}_{-0.087}^{+0.106}$ &${3.63}_{-0.52}^{+1.00}$\\
        & $(0.6 < z < 1.1)$                    & ${22}_{-17}^{+20}$ & ${2.137}_{-0.044}^{+0.041}$ & ${0.081}_{-0.068}^{+0.076}$ &${4.72}_{-0.45}^{+0.51}$\\
        & $k_{\rm min}=0.008~h\text{Mpc}^{-1}$ & ${64}_{-19}^{+33}$ & ${1.994}_{-0.043}^{+0.038}$ & ${0.153}_{-0.061}^{+0.076}$ &${4.47}_{-0.43}^{+0.44}$\\
        & $k_{\rm max}=0.1~h\text{Mpc}^{-1}$   & ${34}_{-13}^{+16}$ & ${2.017}_{-0.020}^{+0.019}$ & ${0.100}_{-0.026}^{+0.029}$ &${3.38}_{-0.30}^{+0.31}$\\
        \midrule
        LRG (FKP)
        & NGC                                  & ${32}_{-19}^{+31}$ & ${2.133}_{-0.067}^{+0.053}$ & $0.084_{-0.092}^{+0.107}$  & ${4.95}_{-0.54}^{+0.62}$\\
        $(0.6 < z < 1.1)$
        & SGC                                  & ${2}_{-30}^{+27}$ & ${2.128}_{-0.064}^{+0.074}$ & ${0.10}_{-0.13}^{+0.10}$   &${4.08}_{-0.77}^{+1.19}$\\
        & $k_{\rm min}=0.008~h\text{Mpc}^{-1}$ & ${41}_{-24}^{+25}$ & ${2.109}_{-0.046}^{+0.047}$ & ${0.117}_{-0.079}^{+0.072}$&${4.65}_{-0.44}^{+0.56}$\\
        & $k_{\rm max} = 0.1~h\text{Mpc}^{-1}$ & ${37}_{-15}^{+16}$ & ${2.096}_{-0.025}^{+0.022}$ & ${0.107}_{-0.029}^{+0.031}$&${2.94}_{-0.36}^{+0.43}$\\ 
        \midrule \midrule 
        QSO (FKP)
        & fiducial                             & ${28}_{-22}^{+23}$ & ${2.210}_{-0.054}^{+0.044}$ & ${0.197}_{-0.050}^{+0.050}$ &${7.23}_{-0.47}^{+0.51}$\\
        $(0.8 < z < 3.1)$
        & NGC                                  & ${47}_{-21}^{+33}$ & ${2.199}_{-0.059}^{+0.062}$ & ${0.212}_{-0.061}^{+0.063}$ &${7.11}_{-0.65}^{+0.67}$\\
        & SGC                                  & ${12}_{-40}^{+36}$ & ${2.247}_{-0.086}^{+0.084}$ & ${0.183}_{-0.077}^{+0.095}$ &${7.23}_{-0.68}^{+0.79}$\\
        & $(0.8 < z < 2.1)$                    & ${101}_{-55}^{+53}$ & ${1.905}_{-0.055}^{+0.047}$ & ${0.276}_{-0.054}^{+0.055}$&${6.90}_{-0.59}^{+0.67}$\\
        & $(1.6 < z < 3.1)$                    & ${-22}_{-22}^{+15}$ & ${2.783}_{-0.082}^{+0.083}$ & ${0.178}_{-0.067}^{+0.079}$&${9.16}_{-0.43}^{+0.72}$\\
        & $k_{\rm min}=0.008~h\text{Mpc}^{-1}$ & ${44}_{-43}^{+42}$ & ${2.195}_{-0.066}^{+0.059}$ & ${0.206}_{-0.053}^{+0.060}$ &${7.20}_{-0.49}^{+0.51}$\\
        & $k_{\rm max} = 0.1~h\text{Mpc}^{-1}$ & ${32}_{-16}^{+23}$ & ${2.225}_{-0.033}^{+0.034}$ & ${0.144}_{-0.029}^{+0.030}$ &${6.20}_{-0.37}^{+0.36}$\\
        \midrule
        QSO (OQE)
        & fiducial                             & ${-9}_{-15}^{+13}$ & ${2.822}_{-0.083}^{+0.064}$ & ${0.147}_{-0.069}^{+0.073}$ &${7.62}_{-0.58}^{+0.64}$\\
        $(0.8 < z < 3.1)$
        & NGC                                  & ${-8}_{-18}^{+18}$ & ${2.783}_{-0.088}^{+0.094}$ & ${0.164}_{-0.092}^{+0.085}$ &${7.34}_{-0.77}^{+0.84}$\\
        & SGC                                  & ${-11}_{-23}^{+20}$& ${2.84}_{-0.11}^{+0.12}$    & ${0.19}_{-0.11}^{+0.11}$ &${9.29}_{-0.31}^{+1.07}$\\
        & $(0.8 < z < 2.1)$                    & ${72}_{-30.}^{+43}$ & ${2.119}_{-0.087}^{+0.068}$ & ${0.283}_{-0.077}^{+0.079}$&${6.29}_{-0.81}^{+0.88}$\\
        & $(1.6 < z < 3.1)$                    & ${-11}_{-14}^{+10}$ & ${2.986}_{-0.086}^{+0.096}$ &${0.125}_{-0.080}^{+0.092}$&${9.22}_{-0.38}^{+0.84}$\\
        & $k_{\rm min}=0.008~h\text{Mpc}^{-1}$ & ${-4}_{-29}^{+18}$ & ${2.811}_{-0.082}^{+0.108}$ & ${0.155}_{-0.076}^{+0.089}$ &${7.67}_{-0.54}^{+0.69}$\\
        & $k_{\rm max} = 0.1~h\text{Mpc}^{-1}$ & ${-3}_{-14}^{+13}$ & ${2.783}_{-0.050}^{+0.054}$ & ${0.167}_{-0.041}^{+0.042}$ &${6.90}_{-0.45}^{+0.47}$\\
        \bottomrule        
    \end{tabular}%
\end{table}
\endgroup

\clearpage

\section{Author Affiliations}
\label{sec:affiliations}

\noindent \hangindent=.5cm $^{1}${Lawrence Berkeley National Laboratory, 1 Cyclotron Road, Berkeley, CA 94720, USA}

\noindent \hangindent=.5cm $^{2}${IRFU, CEA, Universit\'{e} Paris-Saclay, F-91191 Gif-sur-Yvette, France}

\noindent \hangindent=.5cm $^{3}${Center for Cosmology and AstroParticle Physics, The Ohio State University, 191 West Woodruff Avenue, Columbus, OH 43210, USA}

\noindent \hangindent=.5cm $^{4}${Department of Astronomy, The Ohio State University, 4055 McPherson Laboratory, 140 W 18th Avenue, Columbus, OH 43210, USA}

\noindent \hangindent=.5cm $^{5}${The Ohio State University, Columbus, 43210 OH, USA}

\noindent \hangindent=.5cm $^{6}${Physics Dept., Boston University, 590 Commonwealth Avenue, Boston, MA 02215, USA}

\noindent \hangindent=.5cm $^{7}${Dipartimento di Fisica ``Aldo Pontremoli'', Universit\`a degli Studi di Milano, Via Celoria 16, I-20133 Milano, Italy}

\noindent \hangindent=.5cm $^{8}${Department of Physics \& Astronomy, University College London, Gower Street, London, WC1E 6BT, UK}

\noindent \hangindent=.5cm $^{9}${Instituto de F\'{\i}sica, Universidad Nacional Aut\'{o}noma de M\'{e}xico,  Circuito de la Investigaci\'{o}n Cient\'{\i}fica, Ciudad Universitaria, Cd. de M\'{e}xico  C.~P.~04510,  M\'{e}xico}

\noindent \hangindent=.5cm $^{10}${University of California, Berkeley, 110 Sproul Hall \#5800 Berkeley, CA 94720, USA}

\noindent \hangindent=.5cm $^{11}${Institut de F\'{i}sica d’Altes Energies (IFAE), The Barcelona Institute of Science and Technology, Edifici Cn, Campus UAB, 08193, Bellaterra (Barcelona), Spain}

\noindent \hangindent=.5cm $^{12}${Departamento de F\'isica, Universidad de los Andes, Cra. 1 No. 18A-10, Edificio Ip, CP 111711, Bogot\'a, Colombia}

\noindent \hangindent=.5cm $^{13}${Observatorio Astron\'omico, Universidad de los Andes, Cra. 1 No. 18A-10, Edificio H, CP 111711 Bogot\'a, Colombia}

\noindent \hangindent=.5cm $^{14}${Institut d'Estudis Espacials de Catalunya (IEEC), c/ Esteve Terradas 1, Edifici RDIT, Campus PMT-UPC, 08860 Castelldefels, Spain}

\noindent \hangindent=.5cm $^{15}${Institute of Cosmology and Gravitation, University of Portsmouth, Dennis Sciama Building, Portsmouth, PO1 3FX, UK}

\noindent \hangindent=.5cm $^{16}${Institute of Space Sciences, ICE-CSIC, Campus UAB, Carrer de Can Magrans s/n, 08913 Bellaterra, Barcelona, Spain}

\noindent \hangindent=.5cm $^{17}${Departament de F\'{\i}sica Qu\`{a}ntica i Astrof\'{\i}sica, Universitat de Barcelona, Mart\'{\i} i Franqu\`{e}s 1, E08028 Barcelona, Spain}

\noindent \hangindent=.5cm $^{18}${Institut de Ci\`encies del Cosmos (ICCUB), Universitat de Barcelona (UB), c. Mart\'i i Franqu\`es, 1, 08028 Barcelona, Spain.}

\noindent \hangindent=.5cm $^{19}${Fermi National Accelerator Laboratory, PO Box 500, Batavia, IL 60510, USA}

\noindent \hangindent=.5cm $^{20}${Department of Physics, The Ohio State University, 191 West Woodruff Avenue, Columbus, OH 43210, USA}

\noindent \hangindent=.5cm $^{21}${School of Mathematics and Physics, University of Queensland, Brisbane, QLD 4072, Australia}

\noindent \hangindent=.5cm $^{22}${Department of Physics, University of Michigan, 450 Church Street, Ann Arbor, MI 48109, USA}

\noindent \hangindent=.5cm $^{23}${University of Michigan, 500 S. State Street, Ann Arbor, MI 48109, USA}

\noindent \hangindent=.5cm $^{24}${Department of Physics, Southern Methodist University, 3215 Daniel Avenue, Dallas, TX 75275, USA}

\noindent \hangindent=.5cm $^{25}${Department of Physics and Astronomy, University of California, Irvine, 92697, USA}

\noindent \hangindent=.5cm $^{26}${Sorbonne Universit\'{e}, CNRS/IN2P3, Laboratoire de Physique Nucl\'{e}aire et de Hautes Energies (LPNHE), FR-75005 Paris, France}

\noindent \hangindent=.5cm $^{27}${Departament de F\'{i}sica, Serra H\'{u}nter, Universitat Aut\`{o}noma de Barcelona, 08193 Bellaterra (Barcelona), Spain}

\noindent \hangindent=.5cm $^{28}${NSF NOIRLab, 950 N. Cherry Ave., Tucson, AZ 85719, USA}

\noindent \hangindent=.5cm $^{29}${Instituci\'{o} Catalana de Recerca i Estudis Avan\c{c}ats, Passeig de Llu\'{\i}s Companys, 23, 08010 Barcelona, Spain}

\noindent \hangindent=.5cm $^{30}${Department of Physics and Astronomy, Siena College, 515 Loudon Road, Loudonville, NY 12211, USA}

\noindent \hangindent=.5cm $^{31}${Department of Physics \& Astronomy and Pittsburgh Particle Physics, Astrophysics, and Cosmology Center (PITT PACC), University of Pittsburgh, 3941 O'Hara Street, Pittsburgh, PA 15260, USA}

\noindent \hangindent=.5cm $^{32}${Departamento de F\'{\i}sica, DCI-Campus Le\'{o}n, Universidad de Guanajuato, Loma del Bosque 103, Le\'{o}n, Guanajuato C.~P.~37150, M\'{e}xico.}

\noindent \hangindent=.5cm $^{33}${Instituto Avanzado de Cosmolog\'{\i}a A.~C., San Marcos 11 - Atenas 202. Magdalena Contreras. Ciudad de M\'{e}xico C.~P.~10720, M\'{e}xico}

\noindent \hangindent=.5cm $^{34}${Department of Physics and Astronomy, University of Waterloo, 200 University Ave W, Waterloo, ON N2L 3G1, Canada}

\noindent \hangindent=.5cm $^{35}${Perimeter Institute for Theoretical Physics, 31 Caroline St. North, Waterloo, ON N2L 2Y5, Canada}

\noindent \hangindent=.5cm $^{36}${Waterloo Centre for Astrophysics, University of Waterloo, 200 University Ave W, Waterloo, ON N2L 3G1, Canada}

\noindent \hangindent=.5cm $^{37}${Instituto de Astrof\'{i}sica de Andaluc\'{i}a (CSIC), Glorieta de la Astronom\'{i}a, s/n, E-18008 Granada, Spain}

\noindent \hangindent=.5cm $^{38}${Departament de F\'isica, EEBE, Universitat Polit\`ecnica de Catalunya, c/Eduard Maristany 10, 08930 Barcelona, Spain}

\noindent \hangindent=.5cm $^{39}${Universit\'e Clermont-Auvergne, CNRS, LPCA, 63000 Clermont-Ferrand,France}

\noindent \hangindent=.5cm $^{40}${Department of Physics and Astronomy, Sejong University, 209 Neungdong-ro, Gwangjin-gu, Seoul 05006, Republic of Korea}

\noindent \hangindent=.5cm $^{41}${CIEMAT, Avenida Complutense 40, E-28040 Madrid, Spain}

\noindent \hangindent=.5cm $^{42}${Department of Physics \& Astronomy, Ohio University, 139 University Terrace, Athens, OH 45701, USA}

\noindent \hangindent=.5cm $^{43}${Department of Astronomy, Tsinghua University, 30 Shuangqing Road, Haidian District, Beijing, China, 100190}

\noindent \hangindent=.5cm $^{44}${National Astronomical Observatories, Chinese Academy of Sciences, A20 Datun Rd., Chaoyang District, Beijing, 100012, P.R. China}

\end{document}